%% file: paper.tex
\pdfoutput=1

\newcommand*{\ATLASTEMPLATEPATH}{atlaslatex/}
\newcommand*{\ATLASLATEXPATH}{\ATLASTEMPLATEPATH latex/}
\newcommand*{\ATLASLOGOSPATH}{\ATLASTEMPLATEPATH logos/}
\documentclass[cernpreprint,texlive=2016,txfonts,UKenglish]{\ATLASLATEXPATH atlasdoc}
\usepackage[backend=bibtex, block=space]{\ATLASLATEXPATH atlaspackage}
\usepackage{\ATLASLATEXPATH atlasbiblatex}
\usepackage{\ATLASLATEXPATH atlascontribute}
\usepackage{\ATLASLATEXPATH atlasphysics}
\graphicspath{{\ATLASLOGOSPATH}{figures/}}
\usepackage{paper-defs}
\usepackage{paper-results}
\usepackage{xstring}
\usepackage{multirow}
\usepackage{adjustbox}
\usepackage{dcolumn}
\usepackage[nice]{nicefrac}
\usepackage{hyphenat}
\allowdisplaybreaks

\input{paper-metadata}

\hypersetup{pdftitle={ATLAS document},pdfauthor={The ATLAS Collaboration}}

\begin{document}
\maketitle

\section{Introduction}
\label{sec:intro}

The top quark is the heaviest known fundamental particle, 
making the measurement of its production and decay kinematic properties an important probe of physical processes beyond the Standard Model (SM).
Within the SM, the top quark decays predominantly
through the electroweak interaction to an on-shell $W$ boson and a
$b$-quark.
Due to its large mass~\cite{ATLAS:2014wva}, its lifetime
${\mathcal O} (10^{-25}~\text{s})$ is smaller than its hadronisation time-scale
${\mathcal O} (10^{-24}~\text{s})$, allowing this quark to be studied as a free
quark. 
Since the top-quark lifetime is also shorter than the depolarisation
timescale ${\mathcal O} (10^{-21}~\text{s})$~\cite{Bigi:1986jk}
and the $W$ boson is produced on-shell in the top-quark decay, the top-quark spin
information is directly transferred to its decay products.
Comparing angular measurements of the decay products of polarised top
quarks with precise SM predictions provides a unique way to study the
non-SM couplings in the \Wtb\ vertex~\cite{Jezabek:1994zv}.  The
normalised triple-differential cross-section (to be defined
in \Section{sec:triplediffrate}) is the joint probability distribution
in all three of the angles determining the kinematics of the decay
$t\rightarrow Wb$ from a polarised initial state. Its analysis is the most complete investigation of the dynamics of top-quark decay undertaken to date. 

\begin{figure}[!htb]
  \centering
  \centerline{\makebox{
      \subfloat[5FS (2$\rightarrow$2 process)]{\includegraphics[width=2.5in]{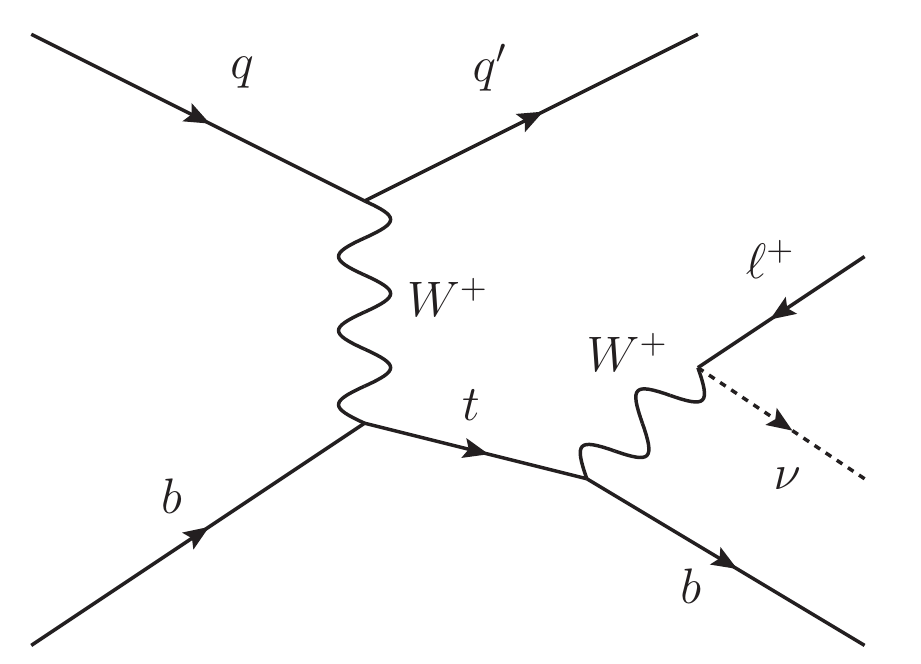}}
      \subfloat[4FS (2$\rightarrow$3 process)]{\includegraphics[width=2.5in]{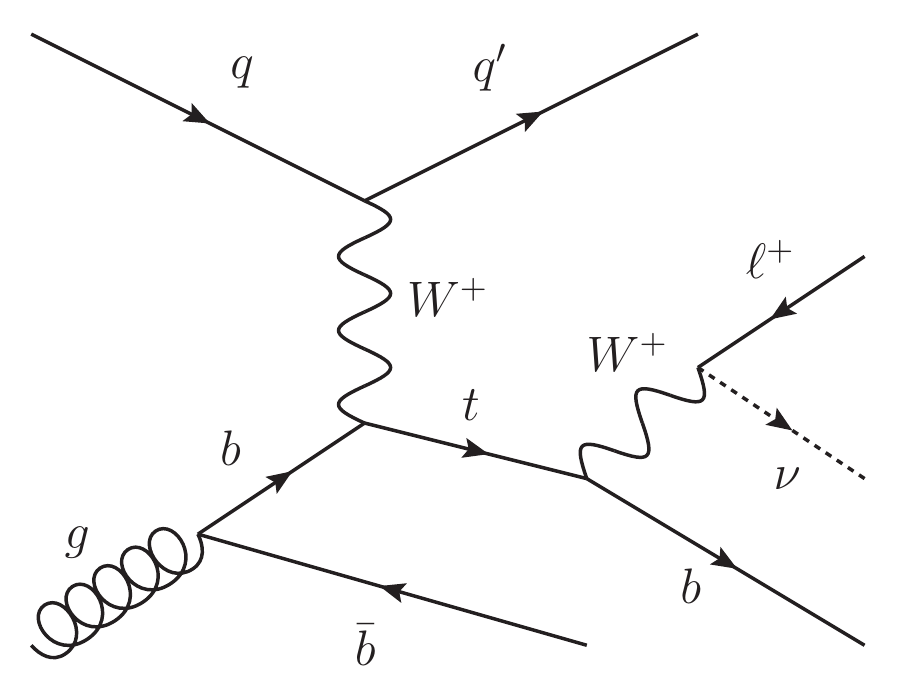}}
  }}
  \caption{Representative LO Feynman diagrams for \tch\ single-top-quark production and decay. Here $q$ represents a $u$- or $\bar{d}$-quark, and $q^\prime$ represents 
            (a) a $d$- or $\bar{u}$-quark, respectively, in which the initial
            $b$-quark arises from a sea $b$-quark in the 5FS or
            $2\to2$ process, or (b) a gluon splitting into a 
            $b\bar{b}$ pair in the 4FS or $2\to3$ process.}
\label{fig:tchannel}
\end{figure}

At hadron colliders, top quarks are produced predominantly in pairs
(\ttbar) via the flavour-conserving strong interaction, while an
alternative process produces single top quarks through the electroweak
interaction. Although the \ttbar\ production cross-section is larger than that of single-top-quark production, top quarks are produced unpolarised because of
parity conservation in quantum chromodynamics (QCD)~\cite{Mahlon:1995zn}, contrary to what happens for single
top quarks.
At the Large Hadron Collider (LHC)~\cite{Evans:2008zzb}, in proton--proton ($pp$) collision
data, the \tch\ is the dominant process for producing single top
quarks used for the measurements presented in this paper. 
\Figure{fig:tchannel} shows the
two representative leading-order (LO) Feynman diagrams for \tch\ single-top-quark
production. 
In these two diagrams, a light-flavour quark $q$ (i.e. $u$- or $\bar{d}$-quark) from one of the colliding protons
interacts with a $b$-quark by exchanging a virtual $W$ boson, producing a top quark $t$ and a
recoiling light-flavour quark $q'$, called the spectator quark.
The $b$-quark comes either directly from another colliding proton in the
five-flavour scheme (5FS) or $2\to2$ process (a) or from a gluon
splitting in the four-flavour scheme\footnote{In the
  5FS the $b$-quarks are treated as massless in the parton distribution functions, while in the 4FS, the parton distribution functions only contain parton distributions for the quarks
  lighter than the $b$-quark and $b$-quarks are
  treated as massive.} 
(4FS) or $2\to3$ process (b). In $pp$ collisions at \cmenergy, the predicted
\tch\ production cross-section using the 5FS is
87.8$^{+3.4}_{-1.9}$~pb~\cite{Kidonakis:2011wy}, calculated at next-to-leading order
(NLO) in QCD with resummed next-to-next-to-leading logarithmic (NNLL)
accuracy, and called approximate next-to-next-to-leading order (NNLO) in
the following. The calculation assumes a top-quark mass of 172.5~\GeV\ and uses the MSTW2008
NNLO~\cite{Martin:2009iq, Martin:2009bu} parton distribution
function (PDF) set. The uncertainties
correspond to the sum in quadrature of the uncertainty obtained from
the MSTW2008 NNLO PDF set at the 90\% confidence level (CL) and the factorisation and
renormalisation scale uncertainties.

As a consequence of the vector--axial (V$-$A) form
of the \Wtb\ vertex in the SM, the spin of single top quarks in \tch\ production is predominantly
aligned along the direction of the spectator-quark
momentum~\cite{Mahlon:1996pn}. 

Probes of new physics phenomena affecting the production or decay of the top quark can be parameterised with a series of effective couplings at each vertex~\cite{Li:1990qf,Kane:1991bg}; 
in the \tch\ single-top-quark production, both production and decay proceed through the \Wtb\ vertex, and thus are sensitive to the same set of 
effective couplings.

New physics can be described by an effective
Lagrangian, $\Lagr_{\mathrm{eff}}$,
represented by dimension-five and dimension-six operators in the framework of effective field theory~\cite{Buchmuller:1985jz, Zhang:2010dr}
\begin{equation*}
  \Lagr_{\mathrm{eff}} = {\cal L}_{\text{SM}} + \frac{1}{\Lambda_{\text{NP}}}{\cal L}_5 +\frac{1}{\Lambda_{\text{NP}}^2} {\cal L}_6 + \cdots \,,
  \label{eq:EffLagr}
\end{equation*}
where ${\cal L}_{\text{SM}}$ represents the SM Lagrangian of dimension
four,  ${\cal L}_5$ and ${\cal L}_6$ represent the contributions from dimension-five and dimension-six operators invariant under the SM gauge
symmetry, and $\Lambda_{\text{NP}}$ 
is a new physics scale chosen such that higher-dimension operators are sufficiently suppressed by higher powers of $\Lambda_{\text{NP}}$.
Of the standardised set of  
operators reported in Ref.~\cite{Buchmuller:1985jz}, only four operators, which are dimension six, contribute independently to the \Wtb\ vertex at LO, 
allowing these terms to be analysed separately from the rest of the full set of possible operators.
In a general Lorentz-covariant Lagrangian, expressed by
Refs.~\cite{Li:1990qf,Kane:1991bg}, corrections to the vertex are absorbed into four
non-renormalisable effective complex couplings called anomalous
couplings:
\begin{equation*}
  \Lagr_{\mathrm{eff}} = - \frac{g}{\sqrt{2}}{\overline{b}}\gamma^\mu \left( \vl \ProjL + \vr \ProjR \right) tW^-_\mu - 
  \frac{g}{\sqrt{2}}{\overline{b}}\frac{i\sigma^{\mu\nu}q_{\nu}}{\mW}
  \left( \gl \ProjL + \gr \ProjR \right) tW^-_\mu + \text{h.c.} \,,
\label{eq:WtbLagr}
\end{equation*}
where the four complex effective couplings $\vlr$, $\glr$ can be
identified with the dimension-six operators' Wilson coefficients~\cite{Wilson:1969zs}.
Here, $g$ is the weak coupling constant, 
and \mw\ and $q_{\nu}$ are the mass and the four-momentum of the $W$ boson. The terms $\ProjLR\equiv \left( 1\mp\gamma^5 \right)/2$ 
are the left- and right-handed projection operators and $\sigma^{\mu\nu}=i[\gamma^{\mu},\gamma^{\nu}]/2$. The terms $\vlr$ and $\glr$ are the left- 
and right-handed vector and tensor complex couplings, respectively. In
the SM at LO, all coupling constants vanish, except $\vl=\Vtb$, which is a quark-mixing element in the 
Cabibbo--Kobayashi--Maskawa (CKM) matrix. Deviations from these values would provide hints of physics 
beyond the SM, and furthermore, complex values could imply that the top-quark decay has
a CP-violating component~\cite{delAguila:2002nf,AguilarSaavedra:2006fy,AguilarSaavedra:2008zc,AguilarSaavedra:2009mx,AguilarSaavedra:2010nx}.

Indirect constraints on \vl, \vr, \gl, and \gr\ were
obtained~\cite{Grzadkowski:2008mf, Cao:2015doa} from precision
measurements of $B$-meson decays. 
These results yield constraints in a six-dimensional space of operator coefficients, where four of them correspond to \Wtb\ couplings.
Considering one coefficient at a time results in very tight
constraints on a particular combination of \vr\ and \gl,
but if several coefficients are allowed to move simultaneously, then individual bounds are not possible.
Very tight constraints on CP-violating interactions have been derived from measurements of electric dipole moments~\cite{Cirigliano:2016njn}.
Those constraints also depend on combinations of couplings, and in a global fit~\cite{Cirigliano:2016nyn}, cannot constrain $\imgr$ better than direct measurements, as are presented here.
Measurements of the $W$ boson
helicity fractions in top-quark decays~\cite{Aad:2012ky,
  Chatrchyan:2013jna, Aaboud:2016hsq, Khachatryan:2016fky,Khachatryan:2014vma} are
sensitive to the magnitude of combinations of anomalous couplings,
which are assumed to be purely real, corresponding to the
CP-conserving case. These measurements can only place limits
on combinations of couplings, and thus the quoted limits on individual
couplings depend on the assumptions made about other couplings while
\vl\ is fixed to the SM value of one. More stringent limits are
set either in these analyses on $\Re{\gr}$ by considering the measurements of the \tch\ single-top-quark production cross-section~\cite{Aad:2014fwa, Chatrchyan:2012ep, Khachatryan:2014iya} or by performing a global fit considering
the most precise measurements of the $W$ boson helicity fractions at
the LHC combined with measurements of single-top-quark production
cross-sections for different centre-of-mass energies at the LHC and
Tevatron~\cite{Birman:2016jhg}. Direct searches for anomalous
couplings in \tch\ single-top-quark events set limits simultaneously on
either both $\Re{\gr/\vl}$ and $\Im{\gr/\vl}$~\cite{Aad:2015yem, Aaboud:2017aqp}, or on
pairs of couplings~\cite{Khachatryan:2016sib}.  
In both cases, analyses assume
SM values for the other anomalous couplings.

The goal of this analysis is to simultaneously constrain the full space of 
parameters governing the \Wtb\ vertex using the triple-differential angular decay rate of single
top quarks produced in the \tch\, as discussed in \Section{sec:triplediffrate}, in which the $W$ boson from the top quark subsequently decays leptonically. 
Conceptually, this is a measurement of each of
the anomalous coupling parameters \vlr\ and \glr\ plus the 
polarisation $P$ of the top quark, with a full covariance matrix;
however, any likelihood function derived from the triple-differential 
decay rate possesses invariances and/or parameter
space boundaries lying quite near to the SM point.  
Therefore, contours are presented instead, with only $\Re{\gr/\vl}$ and 
$\Im{\gr/\vl}$ showing approximate elliptical contours and therefore
admitting point estimation. The anomalous couplings \vr, \gl\ and \gr\ are
allowed to be complex and the measurements shown require no assumptions to
be made regarding the other anomalous couplings.
The analysis is carried out in a Fourier-dual space of coefficients
in an angular expansion~\cite{Boudreau:2013yna,Boudreau:2016pdi}. This
method is chosen because it permits an analytic deconvolution of
detector effects including both resolution and efficiency, while
permitting a simultaneous determination of the real and imaginary parts
of all of the anomalous couplings at the \Wtb\ vertex, in addition to
the polarisation of the top quark produced in the \tch.

This paper is organised as
follows. \Section{sec:triplediffrate} defines the
coordinate system and parameterisation used in the measurement and the
triple-differential formalism applied to polarised single top quarks. \Section{sec:detector} 
gives a short description of the ATLAS detector,
then \Section{sec:data_mc_samples} describes the data samples as well
as the simulated event samples used to predict properties of the \tch\ 
signal and background processes. \Section{sec:object_selection}
describes the event reconstruction for the identification of \tch\ events,
while \Section{sec:event_selection} presents the criteria to define
the signal region as well as the control and validation regions. The
procedures for modelling background processes are reported
in \Section{sec:background_estimation}. The event yields and
angular distributions comparing the predictions and the observed data
are shown in \Section{sec:eventyields_and_controlplots}. \Section{sec:analysis}
describes the efficiency, resolution, and background 
models used to translate the distribution of true \tch\ signal events
to the distribution of reconstructed signal and background events, and how the parameters 
of the model are estimated. \Section{sec:systematics} quantifies the
sources of uncertainty important in this measurement. \Section{sec:results} presents the resulting 
central value and covariance matrix for the model parameters and the
ratios $\Re{\gr/\vl}$ and $\Im{\gr/\vl}$, and the conclusions are
given in \Section{sec:conclusion}.

\section{Triple-differential decay rate of polarised single top quarks}
\label{sec:triplediffrate}

\begin{figure}[!htbp]
  \centerline{\makebox{\includegraphics[width=3.0in]{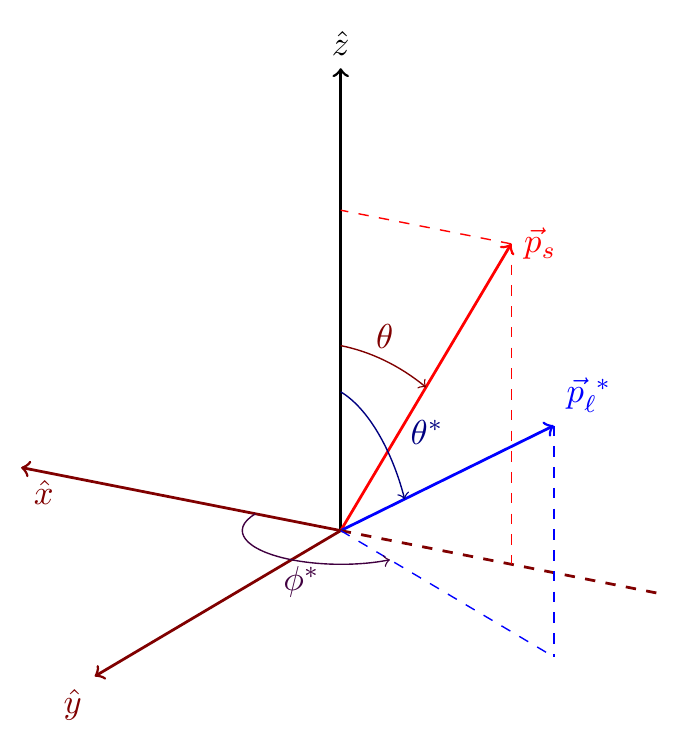}}}
  \caption{Definition of the right-handed coordinate system with $\hat{x}$, $\hat{y}$, and $\hat{z}$ defined as shown from the momentum directions of 
    the $W$ boson, $\Wdir \equiv \hat{z}$, and the spectator quark, $\Sdir$ with $\hat{y}=\Sdir\times\Wdir$, in the top-quark rest frame. The angles $\thetaS$ and $\phiS$ indicate 
    the direction of the lepton momentum, $\Ldir$, while the
    angle $\theta$ indicates the direction of the spectator-quark momentum, $\Sdir$, in this coordinate system.
    \label{fig:axes}}
\end{figure}

An event-specific coordinate system is defined for analysing the decay
of the top quark in its rest frame, using the directions of the
spectator quark $q^\prime$ that recoils against the top quark, 
the $W$ boson from the top-quark decay, and
the lepton $\ell$ ($e$, $\mu$ or $\tau$) from the $W$ boson decay, in the final state depicted in
\Figure{fig:axes}. The $\hat{z}$-axis is chosen along the direction of the
$W$ boson momentum, $\Wmom$, or equivalently along the direction opposite to the $b$-quark momentum, boosted into the top-quark rest frame,
$\hat{z}\equiv\Wdir={\Wmom}/{|\Wmom|}$. The reconstruction of the $W$
boson and top quark is discussed in \Section{sec:event_selection}.
As mentioned before,
the spin of single top quarks, 
\topspin, in \tch\ production is predominantly
aligned along the direction of the spectator-quark
momentum, $\Smom$, in the top-quark rest frame,
$\Sdir = {\Smom}/{|\Smom|}$~\cite{Mahlon:1996pn}. If this quark defines the spin-analysing direction, the degree of 
polarisation is shown in Refs.~\cite{Mahlon:1999gz,Schwienhorst:2010je,Jezabek:1994zv} to be $P\equiv \Sdir \cdot \topspin / |\topspin| \approx \PSM$ at \cmenergy\ 
for SM couplings. 
A three-dimensional right-handed coordinate system is defined from the $\Wdir$--$\Sdir$ plane and the perpendicular direction, with $\hat{y}=\Sdir\times\Wdir$ 
and $\hat{x}=\hat{y}\times\Wdir$. In this coordinate system, the
direction of the lepton momentum, $\Lmom$, in the $W$ boson rest frame,
$\Ldir = \Lmom/|\Lmom|$, is specified by the polar angle $\thetaS$ and the
azimuthal angle $\phiS$. The third angle $\theta$ is defined as the
angle between $\Sdir$ and $\Wdir$.
The angle $\thetaS$ is the same angle used to measure the $W$ boson helicity fractions in top-quark
decays~\cite{Aad:2012ky, Chatrchyan:2013jna, Aaboud:2016hsq, Khachatryan:2016fky,Khachatryan:2014vma}.

These three angles, $\theta$, $\thetaS$, and $\phiS$, arise as a natural choice for measuring a 
triple-differential distribution for the decay of the top quark, where the $W$ boson subsequently decays leptonically.
The $t\to W b$ transition is determined by four
helicity amplitudes, $A_{\lambda_W,\lambda_b}$, where $\lambda_W$ and $\lambda_b$ are the
helicities of the $W$ boson and the $b$-quark, respectively~\cite{Boudreau:2013yna}. For
$\lambda_b = \nicefrac{1}{2}$, only the $W$ boson helicities $\lambda_W = 1, 0$ are possible, while for $\lambda_b = -\nicefrac{1}{2}$,
$\lambda_W = -1, 0$ are possible. The angular dependence of these
transition amplitudes is given in Ref.~\cite{Boudreau:2013yna}. At LO
and neglecting the $b$-quark mass, the helicity amplitudes have a
simple dependence on the anomalous couplings. Up to a common
proportionality constant, the magnitudes can be expressed as
\begin{eqnarray*}
  \left|\ARR \right|^2  &\propto& 2 \left| x_{W}\vr - \gl\right|^2 ,\nonumber \\
  \left |\AZR \right|^2 &\propto& \left|\vr-x_{W}\gl\right|^2,\nonumber \\
  \left| \ALL \right|^2 &\propto& 2 \left| x_{W}\vl - \gr\right|^2 ,\nonumber \\
  \left |\AZL \right|^2 &\propto&  \left | \vl - x_{W} \gr\right|^2 ,
                            \label{eq:amplitudesAC}
\end{eqnarray*}
where $x_{W} = \mW/\mt$. The relative phases between $\ARR$ and $\AZR$ and between $\ALL$ and $\AZL$ are determined by the relative phases between \vr\ and \gl\ and between \vl\ and \gr, respectively.

From the four helicity amplitudes, three fractions can be independently determined.  
In addition,  the interference allows two relative phases between amplitudes to be experimentally determined.  
These are called the \emph{generalised helicity fractions and phases}~\cite{Boudreau:2013yna, Aad:2015yem}:
\begin{itemize}
\item $\fu$, the fraction of decays containing transversely polarised $W$ bosons,
  \begin{equation*}
    \fu =\frac {\left|A_{1,\frac{1}{2}}\right|^2 + \left|A_{-1,-\frac{1}{2}}\right|^2} {\left|A_{1,\frac{1}{2}}\right|^2 + \left|A_{-1,-\frac{1}{2}}\right|^2  + \left|A_{0,\frac{1}{2}}\right|^2 + \left|A_{0,-\frac{1}{2}}\right|^2},
    \label{eq:paramDef1}
  \end{equation*}

\item $\fup$, the fraction of $b$-quarks that are right-handed in events with transversely polarised $W$ bosons,
 \begin{equation*}
   \fup = \frac {\left|A_{1,\frac{1}{2}}\right|^2} {\left|A_{1,\frac{1}{2}}\right|^2 + \left|A_{-1,-\frac{1}{2}}\right|^2},
    \label{eq:paramDef2}
 \end{equation*}

\item $\fzp$, the fraction of $b$-quarks that are right-handed in events with longitudinally polarised $W$ bosons,
  \begin{equation*}
    \fzp = \frac {\left|A_{0,\frac{1}{2}}\right|^2} {\left|A_{0,\frac{1}{2}}\right|^2 + \left|A_{0,-\frac{1}{2}}\right|^2},
    \label{eq:paramDef3}
  \end{equation*}

\item $\delp$, the phase between amplitudes for longitudinally polarised and transversely polarised $W$ bosons recoiling against right-handed $b$-quarks,
  \begin{equation*}
    \delp = \operatorname{arg} \left(A_{1,\frac{1}{2}}\, A_{0,\frac{1}{2}}^* \right),
    \label{eq:paramDef4}
  \end{equation*}

\item $\delm$, the phase between amplitudes for longitudinally polarised and transversely polarised $W$ bosons recoiling against left-handed $b$-quarks,
  \begin{equation*}
    \delm = \operatorname{arg} \left(A_{-1,-\frac{1}{2}}\, A_{0,-\frac{1}{2}}^* \right).
    \label{eq:paramDef5}
  \end{equation*}

\end{itemize}

The fractions \fu\ and \fup\ are related to the quantities $\Fp$, $\Fz$, and $\Fm$ determined by
measurements of the $W$ boson helicity fractions in top-quark decays~\cite{Aad:2012ky,
  Chatrchyan:2013jna, Aaboud:2016hsq, Khachatryan:2016fky,Khachatryan:2014vma}, with
$\Fp=\fu\fup$, $\Fz=1-\fu$, and $\Fm=\fu (1-\fup)$. The fraction \fzp\
is previously unmeasured.

For convenience in what follows, $\params$ is defined as $\params
\equiv \left \{ \fu, \fup, \fzp, \delp, \delm \right\}$. 
From these five experimental observables, plus the relationships between the helicity amplitudes and the anomalous couplings, 
one can obtain constraints on all the couplings simultaneously.
Additionally, the top-quark polarisation, $P$, is considered separately
from \params\ because it depends on the production of the top quark,
rather than on its decay. 

At LO, the helicity amplitudes, and hence \params\ can be expressed as functions of the couplings and the parton masses~\cite{Fischer:2001gp, AguilarSaavedra:2010nx}.
Using SM couplings and  $\mb=4.95~\GeV$, $\mt =172.5~\GeV$, and
$\mW=80.399~\GeV$ with the derived analytic expressions 
for \params, the expected values are 
\begin{equation*}
  \fu = \fuSM, \qquad \fup =\fupSM, \qquad \fzp =\fzpSM, \qquad \delp = \delm = \delmSM.
  \label{eq:paramSMVals}
\end{equation*}
Calculations at  NNLO~\cite{Czarnecki:2010gb} predict \fu = \fuNNLO,
and \fup = \fupNNLO, where the largest part of the uncertainty in \fu\
comes from the experimental uncertainty of the top-quark mass, while for \fup\  it arises from uncertainties in $\alpha_\mathrm{s}$ and the $b$-quark mass.
An NNLO prediction does not yet exist for \fzp, but NLO calculations~\cite{Fischer:2001gp} yield a value \fzpNLO .  

In Refs.~\cite{Boudreau:2013yna,Boudreau:2016pdi} it is shown that the
Jacob--Wick helicity formalism~\cite{Jacob:1959at,Richman:1984gh}
applied to the decay of polarised top quarks in \tch\ production leads to the following expression for the
triple-differential decay rate for polarised top quarks in terms of
the three angles ($\theta$, $\thetaS$, and $\phiS$) and the top-quark
polarisation,

\begin{eqnarray}
  \varrho(\theta,\thetaS,\phiS; P) = \frac{1}{N}\frac{\mathrm{d}^3N}{\mathrm{d}(\cos{\theta}) \mathrm{d}\Omega^*}
  &=& \frac{1}{8\pi} \Bigg\{ \frac{3}{4}\left|A_{1,\frac{1}{2}}\right|^2 (1+P\cos\theta)(1+\cos\thetaS)^2 \Bigg.\nonumber\\
  &+& \frac{3}{4}\left|A_{-1,-\frac{1}{2}}\right|^2 (1-P\cos\theta)(1-\cos\thetaS)^2\nonumber\\
  &+& \frac{3}{2}\left(\left|A_{0,\frac{1}{2}}\right|^2(1-P\cos\theta) + \left|A_{0,-\frac{1}{2}}\right|^2(1+P\cos\theta)\right)\sin^2\thetaS\nonumber\\
  &-& \frac{3\sqrt{2}}{2} P\sin\theta\sin\thetaS (1+\cos\thetaS) \, \Re{{e^{i\phiS}A_{1,\frac{1}{2}}\,A_{0,\frac{1}{2}}^*}}\nonumber\\
  &-& \Bigg. \frac{3\sqrt{2}}{2} P\sin\theta\sin\thetaS (1-\cos\thetaS) \, \Re{{e^{-i\phiS}A_{-1,-\frac{1}{2}}\,A_{0,-\frac{1}{2}}^*}} \Bigg\}\nonumber\\
  &=& \sum_{k=0}^{1} \sum_{l=0}^{2} \sum_{m=-k}^{k} a_{k,l,m} \M{k,l,m}\,,
      \label{eq:tripleDiff}
\end{eqnarray}

where $\mathrm{d}\Omega^* \equiv \mathrm{d}(\cos{\theta^*})\mathrm{d}\phi^*$ (see
\Figure{fig:axes}). The $a_{k,l,m}$ represent the angular coefficients to
be determined and $\M{k,l,m}$ are orthonormal functions over the
three angles defined by the product of two spherical harmonics, $Y_k^m(\theta, 0)$ and $Y_l^m(\thetaS, \phiS)$,
\begin{equation*}
  \M{k,l,m} = \sqrt{2\pi}Y_k^m(\theta, 0)Y_l^m(\thetaS, \phiS).
\end{equation*}

The properties of these $M$-functions are detailed in Ref.~\cite{Boudreau:2016pdi}.
The restriction to $k \leq 1$ and
$l \leq 2$ in \Equation{eq:tripleDiff} is caused by the
allowed spin states of the initial- and final-state fermions and the
vector boson at the weak vertex.

Only nine of the angular coefficients $a_{k,l,m}$, not taking into
account $a_{0,0,0}$, which is constrained by normalisation ($|\ARR|^2 +
|\AZR|^2 + |\ALL|^2 + |\AZL|^2 = 1$), are non-zero and can be
parameterised in terms of the generalised helicity fractions and
phases.

The non-zero angular coefficients $a_{k,l,m} (\params; P)$ are:
\begin{eqnarray}
  a_{0,0,0}~& = & \frac{1}{\sqrt{8\pi}}\,,\nonumber \\
  a_{0,1,0}~& = & \frac{\sqrt{3}}{\sqrt{8\pi}} \fu \left( \fup - \frac{1}{2} \right) \,,\nonumber \\
  a_{0,2,0}~& = & \frac{1}{\sqrt{40\pi}}\left(\frac{3}{2}\fu - 1 \right) \,,\nonumber \\
  a_{1,0,0} & = & +P\frac{1}{\sqrt{24\pi}}\left(\fu(2\fup-1) + (1-\fu) (1-2\fzp) \right) \,,\nonumber \\
  a_{1,1,0} & = & +P\frac{1}{\sqrt{32\pi}} \fu \,,\nonumber \\
  a_{1,2,0} & = & +P\frac{1}{\sqrt{480\pi}}\left(\fu (2\fup-1) - 2 (1-\fu) (1-2\fzp)\right) \,,\nonumber \\
  a_{1,1,1}  & = & (a_{1,1,-1})^* = -P \frac{1}{\sqrt{16\pi}} \sqrt{\fu(1-\fu)} \left \{  \sqrt{\fup\fzp} \, \mathrm{e}^{i\delp} +  \sqrt{(1-\fup)(1-\fzp)} \, \mathrm{e}^{-i\delm} \right \}\,,\nonumber \\
  a_{1,2,1}  & = & (a_{1,2,-1})^* = -P \frac{1}{\sqrt{80\pi}} \sqrt{\fu(1-\fu)} \left \{  \sqrt{\fup\fzp} \, \mathrm{e}^{i\delp} -  \sqrt{(1-\fup)(1-\fzp)} \, \mathrm{e}^{-i\delm} \right \}\,,
  \label{eq:physicsCoefficients}
\end{eqnarray}

where $(a_{k,l,m})^*$ represents a complex conjugate. All the other angular coefficients are zero in top-quark decays. 

Coefficients of $M$-functions can also be determined from data.
In \Section{sec:analysis}, techniques are discussed for measuring those coefficients, 
how  to deconvolve them to obtain the coefficients presented here, 
and hence the parameters $\params$ and $P$. 

\section{ATLAS detector}
\label{sec:detector}

The ATLAS detector~\cite{Aad:2008zzm} consists of a set of
sub-detector systems, cylindrical in the central region and planar in the two 
endcap regions, that covers almost the full solid angle around the
interaction point (IP).\footnote{ATLAS uses a right-handed coordinate
  system with its origin at the nominal IP in the centre of the
  detector and the $z$-axis along the beam pipe. The $x$-axis points
  from the IP to the centre of the LHC ring, and the $y$-axis points
  upward. Cylindrical coordinates $(r,\phi)$ are used in the
  transverse plane, $\phi$ being the azimuthal angle around the
  $z$-axis. The pseudorapidity is defined in terms of the polar angle
  $\theta$ as $\eta=-\ln\tan(\theta/2)$. The transverse momentum and
  energy are defined as $\pT = p~\text{sin}~\theta$ and $E_{\text{T}}
  = E~\text{sin}~\theta$, respectively. The $\Delta R$ is the distance
  defined as $\Delta R = \sqrt{(\Delta\eta)^2 + (\Delta\phi)^2}$.}
ATLAS is composed of an inner detector (ID) for tracking close to the
IP, surrounded by a superconducting solenoid providing a 2~T 
axial magnetic field, electromagnetic (EM) and hadronic calorimeters,
and a muon spectrometer (MS). The ID consists of a silicon pixel
detector, a silicon micro-strip detector, providing tracking
information within pseudorapidity $|\eta| < 2.5$, and a straw-tube
transition radiation tracker that covers $|\eta| < 2.0$. The central
EM calorimeter is a lead and liquid-argon (LAr) sampling calorimeter
with high granularity, and is divided into a barrel region that covers
$|\eta| < 1.5$ and endcap regions that cover $1.4 < |\eta| < 3.2$. A
steel/scintillator tile calorimeter provides hadronic energy
measurements in the central range of $|\eta| < 1.7$. The endcap ($1.5
< |\eta| < 3.2$) and forward regions ($3.1 < |\eta| < 4.9$) are
instrumented with LAr calorimeters for both the EM and hadronic energy
measurements. The MS consists of three large superconducting toroid
magnets with eight coils each, a system of trigger chambers covering
$|\eta| < 2.4$, and precision tracking chambers covering $|\eta| <
2.7$. The ATLAS detector employs a three-level trigger
system~\cite{ATLAS:2016qun}, used to select events to be recorded for
offline analysis. The first-level trigger is hardware-based,
implemented in custom-built electronics and it uses a subset of the
detector information to reduce the physical event rate from 40~MHz to
at most 75~kHz. The second-level trigger and the final event filter,
collectively referred to as the high-level trigger (HLT), are
software-based and together reduce the event rate to about 400 Hz.

\section{Data and simulation samples}
\label{sec:data_mc_samples}

The analysis is performed using data from $pp$ collisions delivered by the LHC in 2012 at \cmenergy\ and recorded by the ATLAS detector. Stringent detector and data quality requirements were applied, resulting in a data sample corresponding to a total integrated luminosity of \lumiInInvFb~\cite{Aaboud:2016hhf}. The events were selected by single-lepton\footnote{Henceforth, ``lepton'' indicates electron or muon, and does not include $\tau$ leptons.}
triggers~\cite{ATLAS:2016qun,Aad:2014sca}, imposing at the HLT a threshold of 24~\GeV\ on 
the transverse energy (\ET) of electrons and on the transverse momentum (\pT) of muons, along with isolation requirements. To recover efficiency for
higher-\pT\ leptons, the isolated lepton  triggers were complemented by triggers without isolation 
requirements, but with a threshold raised to 60~\GeV\ for electrons and to 36~\GeV\ for muons.

Samples of events generated using Monte Carlo (MC) simulations were produced using different event generators interfaced to various parton showering (PS) and hadronisation generators.
Minimum-bias events simulated with the \pythia8 generator (ver. 8.1)~\cite{Sjostrand:2007gs} were overlaid to model the effect of multiple $pp$ collisions per bunch crossing (pile-up). The distribution of the average number of pile-up interactions in the simulation is reweighted to match the corresponding distribution in data, which has an average of 21~\cite{Aaboud:2016hhf}. The events were processed using the same reconstruction and analysis chain as for data events.

Single-top-quark \tch\ events were generated with the NLO \powheg\ generator (rev. 2556)~\cite{Frixione:2007vw} with the CT10f4~\cite{Lai:2010vv} PDF set, using the 4FS for the matrix-element (ME) calculations~\cite{Frederix:2012dh}. The renormalisation and factorisation scales were set to $\mu_{\text{R}}^2 = \mu_{\text{F}}^2 = 16 ( m_{b}^2 + p_{{\mathrm T},b}^2 )$, where $\mb$ is the mass of the $b$-quark and $p_{{\mathrm T},b}$ is the transverse momentum of the $b$-quark from the initial gluon splitting. Top quarks were decayed using \madspin~\cite{Artoisenet:2012st}, which preserves all spin correlations. Additional \tch\ samples were produced with the LO \protos\ generator (ver. 2.2b)~\cite{AguilarSaavedra:2008gt} using the CTEQ6L1 PDF set~\cite{Pumplin:2002vw} within the 4FS. Thus in addition to a SM sample, samples with anomalous couplings enabled in both the production and the decay vertices were produced using the \protos\ generator, varying simultaneously \vl\ with either $\revr \in [0.25, 0.50]$, $\regr \in [-0.26, 0.18]$ or $\imgr \in [-0.23, 0.23]$, such that the top-quark width was invariant. The factorisation scale was set to $\mu_{\text{F}}^2 = - p^2_W$ for the spectator quark and $\mu_{\text{F}}^2 = p^2_{\bar{b}} + m^2_b$ for the gluon, where $p_W$ and $p_{\bar{b}}$ are the three-momenta of the exchanged $W$ boson and of the $\bar{b}$-quark originating from the gluon splitting (the spectator $\bar{b}$-quark), respectively. 
In order to compare different LO generators, another sample of signal events was produced with the multi-leg LO \acermc\ generator (ver. 3.8)~\cite{Kersevan:2004yg} using the CTEQ6L1 PDF set. This generator incorporates both 4FS and 5FS, featuring an automated procedure to remove the overlap in phase space between the two schemes~\cite{Kersevan:2006fq}. The factorisation and renormalisation scales were set to $\mu_{\text{F}} = \mu_{\text{R}}=\mt=172.5~\GeV$.

In this analysis, all simulated signal event samples are normalised using the production cross-section mentioned in \Section{sec:intro}. Simulation samples produced with \powheg\ are used for predicting the acceptance and the template shape of the \tch\ signal. To estimate the efficiency and resolution models, the simulation samples in which parton-level information is well defined, i.e.~those produced with either \protos\ or \acermc, are used.

Samples of simulated events for \ttbar\ production and electroweak production of single top quarks in the associated \Wt\ and \sch\ were produced using the NLO \powheg\ generator (rev. 2819, rev. 3026) coupled with the CT10~\cite{Lai:2010vv} PDF set. The \textit{t-} and \sch\ processes do not interfere even at NLO in QCD and are thus well defined with that precision~\cite{Willenbrock:1986cr}. For \Wt\ associated production, the diagram removal scheme is used to eliminate overlaps between this process and \ttbar\ production at NLO. In the \ttbar\ sample, the resummation damping factor\footnote{The resummation damping factor, $h_{\text{damp}}$, is one of the parameters controlling the ME/PS matching in \powhegmodel\ and effectively regulates the high-\pT\ gluon radiation. In the used \powheg\ revision, $h_{\text{damp}} = \infty$ was the default value.} $h_{\text{damp}}$ was set to the top-quark mass~\cite{ATL-PHYS-PUB-2015-002}. An additional \ttbar\ sample with anomalous couplings enabled in the decay vertex was produced using the \protos\ generator (ver. 2.2) coupled with the CTEQ6L1 PDF set. This sample is used to take into account the dependence of \ttbar\ background upon the value of the anomalous couplings.

For all simulated event samples mentioned above, the PS, hadronisation and underlying event (UE) were added using \pythia~(ver. 6.426, ver. 6.427)~\cite{Sjostrand:2006za} with the Perugia 2011C set of tuned parameters (P2011C tune)~\cite{Skands:2010ak} and the CTEQ6L1 PDF set. The \tauola~\cite{Jadach:1990mz} program and the \photos~\cite{Golonka:2005pn} algorithm were used to properly simulate decays of polarised $\tau$ leptons including spin correlations and to generate quantum electrodynamics (QED) radiative corrections in decays to account for photon radiation. All these processes were simulated assuming a top-quark mass of 172.5~\GeV, and the decay of the top quark was assumed to be 100\% $t\rightarrow Wb$.

For estimating the \tch\ and \ttbar\ generator modelling uncertainties, additional samples were produced using alternative generators or parameter variations. For studying the top-quark mass dependence, supplementary single-top-quark and \ttbar\ simulated event samples with different top-quark masses were generated. These topics are further discussed in \Section{sec:systematics} and \Section{sec:results}, respectively.
    
Vector-boson production in association with jets was simulated using the multi-leg LO \sherpa\ generator~(ver. 1.4.1)~\cite{Gleisberg:2008ta} 
with its own parameter tune and the CT10 PDF set. Thus, $W$+jets and $Z$+jets events with up to four additional partons were generated and the contributions of $W/Z$+light-jets and $W/Z$+heavy-jets ($W/Z$+$bb$, $W/Z$+$cc$, $W/Z$+$c$) were simulated separately. \sherpa\ was also used to generate the hard 
process, but also for the PS, hadronisation and the UE, using the CKKW method~\cite{Hoeche:2009rj} to remove overlaps 
between the partonic configurations generated by the ME and by the PS. Samples of diboson events ($WW$, $WZ$, and $ZZ$), containing up to three additional partons where at least one of the bosons decays leptonically, were also produced using the \sherpa\ generator~(ver. 1.4.1) with the CT10 PDF set.

All baseline simulated event samples were passed through the full simulation of the ATLAS detector~\cite{Aad:2010ah} based on the GEANT4 framework~\cite{Agostinelli:2002hh}
while \protos\ simulated event samples and alternative samples used to estimate systematic uncertainties were processed through a faster simulation using the \afII\ framework~\cite{Richter-Was:1998iea}.

\section{Event reconstruction}
\label{sec:object_selection}

Electron candidates are reconstructed from isolated energy deposits in
the EM calorimeter associated with ID tracks fulfilling strict quality
requirements~\cite{Aad:2014fxa}. These electrons are required to
satisfy $\ET = E_{\text{cluster}}/\sin(\theta_{\text{track}}) >
25~\GeV$ and $|\eta_{\text{cluster}}| < 2.47$, where
$E_{\text{cluster}}$ and $\eta_{\text{cluster}}$ denote the energy and
the pseudorapidity of the cluster of energy deposits in the EM
calorimeter, and $\theta_{\text{track}}$ denotes the polar angle of the ID track associated with this cluster. 
Clusters in the EM calorimeter barrel--endcap transition region,
corresponding to $1.37 < |\eta_{\text{cluster}}| < 1.52$, are
excluded. Muon candidates are reconstructed using combined information
from the ID tracks and the MS~\cite{Aad:2014rra}. They are required to
have $\pT > 25~\GeV$ and $|\eta| < 2.5$. The electron and muon
candidates must fulfil additional isolation requirements, as described
in Ref.~\cite{ATLAS:2014ffa}, in order to reduce contributions from
misidentified jets, non-prompt leptons from the decay of heavy-flavour
quarks and non-prompt electrons from photon conversions.

Jets are reconstructed using the anti-$k_{t}$ algorithm~\cite{Cacciari:2008gp,Cacciari:2011ma} with a radius parameter of 0.4, using topological clusters of calorimeter energy deposits~\cite{Aad:2016upy} as inputs to the jet finding. The clusters are calibrated with a local cluster weighting method~\cite{Aad:2016upy}. The jet energy is further corrected for the effect of multiple $pp$ interactions. Jets are calibrated using an energy- and $\eta$-dependent simulation-based scheme, with in situ corrections based on data~\cite{Aad:2014bia}. To reject jets from pile-up events, a so-called jet-vertex-fraction (JVF) criterion~\cite{Aad:2015ina} is applied to the jets with $\pT < 50~\GeV$ and $|\eta| < 2.4$: at least 50\% of the scalar sum of the \pT\ of the tracks associated with a jet is required to be from tracks compatible with the primary vertex.\footnote{A primary-vertex candidate is defined as a reconstructed vertex with at least five associated tracks with $\pT > 400~\MeV$. The primary vertex associated with the hard-scattering collision is the candidate with the largest sum of the squared \pT\ of the associated tracks.} Only events containing reconstructed jets with $\pT > 30~\GeV$ and $|\eta| < 4.5$ are considered. The \pT\ threshold is raised to 35~\GeV\ for the jets in the calorimeter endcap--forward transition region, corresponding to $2.7 < |\eta| < 3.5$~\cite{Aad:2014fwa}.  Jets identified as likely to contain $b$-hadrons are tagged as $b$-jets. The $b$-tagging is performed using a neural network (NN) which combines three different algorithms exploiting the properties of a $b$-hadron decay in a jet~\cite{Aad:2015ydr}. The $b$-tagging algorithm, only applied to jets within the coverage of the ID (i.e. $|\eta| < 2.5$), is optimised to improve the rejection of $c$-quark jets, since $W$ boson production in association with $c$-quarks is a major background for the selected final state. The requirement applied to the NN discriminant corresponds to a $b$-tagging efficiency of 50\%, with mis-tagging rates of 3.9\% and 0.07\% for $c$-quark jets and light-flavour jets ($u$-, $d$-, $s$-quark or gluon $g$), respectively, as predicted in simulated \ttbar\ events and calibrated with data~\cite{ATLAS:2014pla, ATLAS:2014jfa}.

The missing transverse momentum, with magnitude \MET, is reconstructed from the vector sum of energy deposits in the calorimeter projected onto the transverse plane~\cite{TheATLAScollaboration:2013oia}. The energies of all clusters are corrected using the local cluster weighting method. Clusters associated with high-\pT\ jets and electrons are further calibrated using their respective energy corrections. In addition, contributions from the \pT\ of the selected muons are also included in the calculation. The \MET\ is taken as a measurement of the undetectable particles, and is affected by energy losses due to detector inefficiencies and acceptance, and by energy resolution.

\section{Event selection in the signal, control, and validation regions}
\label{sec:event_selection}

The signal event candidates are selected by requiring a single prompt isolated lepton,\footnote{This analysis considers only $W$ boson decay modes to an electron or a muon. Events in which the $W$ boson decays to a $\tau$ lepton are included if the $\tau$ subsequently decays to an electron or a muon.} significant \MET, and exactly two jets. All these objects must satisfy the criteria described in \Section{sec:object_selection}, and the \MET\ is required to be larger than 30~\GeV.
One of the jets must be identified as a $b$-tagged jet with $|\eta| < 2.5$ while the second jet, also called the spectator jet, is required to be untagged and produced in the forward direction. Events containing additional jets are vetoed to suppress background from \ttbar\ production. The spectator $\bar{b}$-quark originating from the gluon splitting (4FS), as shown in \Figure{fig:tchannel}(b), can result in an additional $b$-tagged jet. This jet is expected to have a softer $\pT$ spectrum and a broader $\eta$ distribution than the $b$-tagged jet produced in the top-quark decay. It is generally not detected in the experiment and these events pass the event selection.
Events are required to contain at least one good primary vertex candidate, and no jets failing to satisfy reconstruction quality criteria. 
In addition, the transverse mass of the lepton--\MET\ system,
\begin{displaymath}
  \mtw=\sqrt{2\pT(\ell)\cdot\MET\left[1-\cos\left( \Delta\phi(\ell,\MET)\right)\right]} \,,
\end{displaymath}
where $\Delta\phi(\ell,\MET)$ is the difference in azimuthal angle between the lepton momentum and the \MET\ direction, is required to be larger than 50~\GeV\ in order to reduce the \qcd\ background contribution.
Further reduction of this background is achieved by imposing a requirement on the lepton \pT\ to events in which the lepton and leading jet ($j_1$) are back-to-back~\cite{Aad:2014fwa,Aad:2015yem,Aaboud:2017pdi},
\begin{displaymath}
  \pT(\ell) > 40 \left( \frac{|\Delta \phi(j_1,\mbox{$\ell$})| - 1}{\pi-1} \right)~\mbox{\GeV} \,,
\end{displaymath}
where $\Delta \phi(j_1,\ell)$ is the difference in azimuthal angle between the lepton momentum and the leading jet.
To reduce the dilepton backgrounds, events containing an additional lepton, identified with less stringent criteria (referred to as a loose lepton) and with a \pT\ threshold lowered to 10~\GeV, are rejected. Finally, two additional requirements are applied in order to remove a mis-modelling between data and prediction seen in the $W$+jets control and validation regions, in the $|\eta|$ distribution of the non-$b$-jet and in the $|\Delta\eta|$ distribution between the two required jets: $|\eta(\text{non-$b$-jet})| < 3.6$ and $|\Delta\eta(\text{non-$b$-jet, $b$-jet})| < 4.5$.

The $W$ boson originating from the decay of the top quark is reconstructed from the momenta of the lepton and the neutrino by imposing four-momentum conservation. Since the neutrino escapes undetected, the $x$ and $y$ components of the reconstructed \MET are assumed to correspond to the \pT\ of the neutrino. The unmeasured longitudinal component of the neutrino momentum, $p_{\nu}^z$, is computed by imposing a $W$ boson mass constraint on the lepton--neutrino system. A quadratic expression is found for $p_{\nu}^z$. If there are two real solutions, the solution closer to zero is taken. If the solutions are complex, the assumption of the neutrino being the only contributor to the \MET\ is not valid.\footnote{Although it is true that at LO the neutrino is the main contributor to the \MET, there may be other contributors, such as extra neutrinos (from $b$-hadron and $\tau$ decays), additional \pT\ contributions (initial/final-state radiation effects), miscalibration of \MET, fake \MET\ due to the detector energy resolution and acceptance.} Therefore, the reconstructed \MET\ is rescaled, preserving its direction, in order to have physical (real) solutions for $p_{\nu}^z$. This generally results in two solutions for the rescaled \MET. If just one solution of the rescaled \MET\ is positive, this is chosen. If both are positive, the one closer to the initial \MET\ is chosen. 
The top-quark candidate is then reconstructed by combining the four-momenta of the reconstructed $W$ boson and the selected $b$-tagged jet. Finally, the momenta of the $W$ boson and spectator jet are boosted into the top-quark rest frame to obtain \Wmom\ and \Smom, used to define the coordinate system in \Figure{fig:axes}, and the lepton is boosted into the $W$ boson rest frame to obtain \Lmom.

In addition to this basic event selection, which defines the preselected region, further discrimination between the \tch\ signal events and background events is achieved by applying additional criteria:

\begin{itemize}
\item The pseudorapidity of the non-$b$-tagged jet must satisfy $|\eta(\text{non-$b$-jet})| > 2.0$, since the spectator jet tends to be produced in the forward region in the \tch\ signature.
\item The scalar sum of the \pT\ of all final-state objects (lepton, jets and \MET), $\HT$, must be larger than 195~\GeV, since the $\HT$ distributions of the backgrounds peak at lower values (in particular for the $W$+jets contribution) than the \tch\ signature.
\item The mass of the top quark reconstructed from its decay products, \mtop,  is required to be within 130--200~\GeV, to reject background events from processes not involving top quarks. 
\item The absolute difference in $\eta$ between the non-$b$-tagged jet and the $b$-jet, $|\Delta\eta(\text{non-$b$-jet, $b$-jet})|$, must be larger than 1.5,  to further reduce \ttbar\ contributions.
\end{itemize}

These criteria are based on the selection requirements used in Ref.~\cite{Aad:2015yem}, re-optimised using MC simulation at \cmenergy~\cite{Aaboud:2017aqp}. Thus, these criteria together with the signal preselection define the signal region of this analysis.

\begin{figure}[!htb]
  \centerline{\mbox{
      \subfloat[]  {\includegraphics[width=0.48\textwidth]{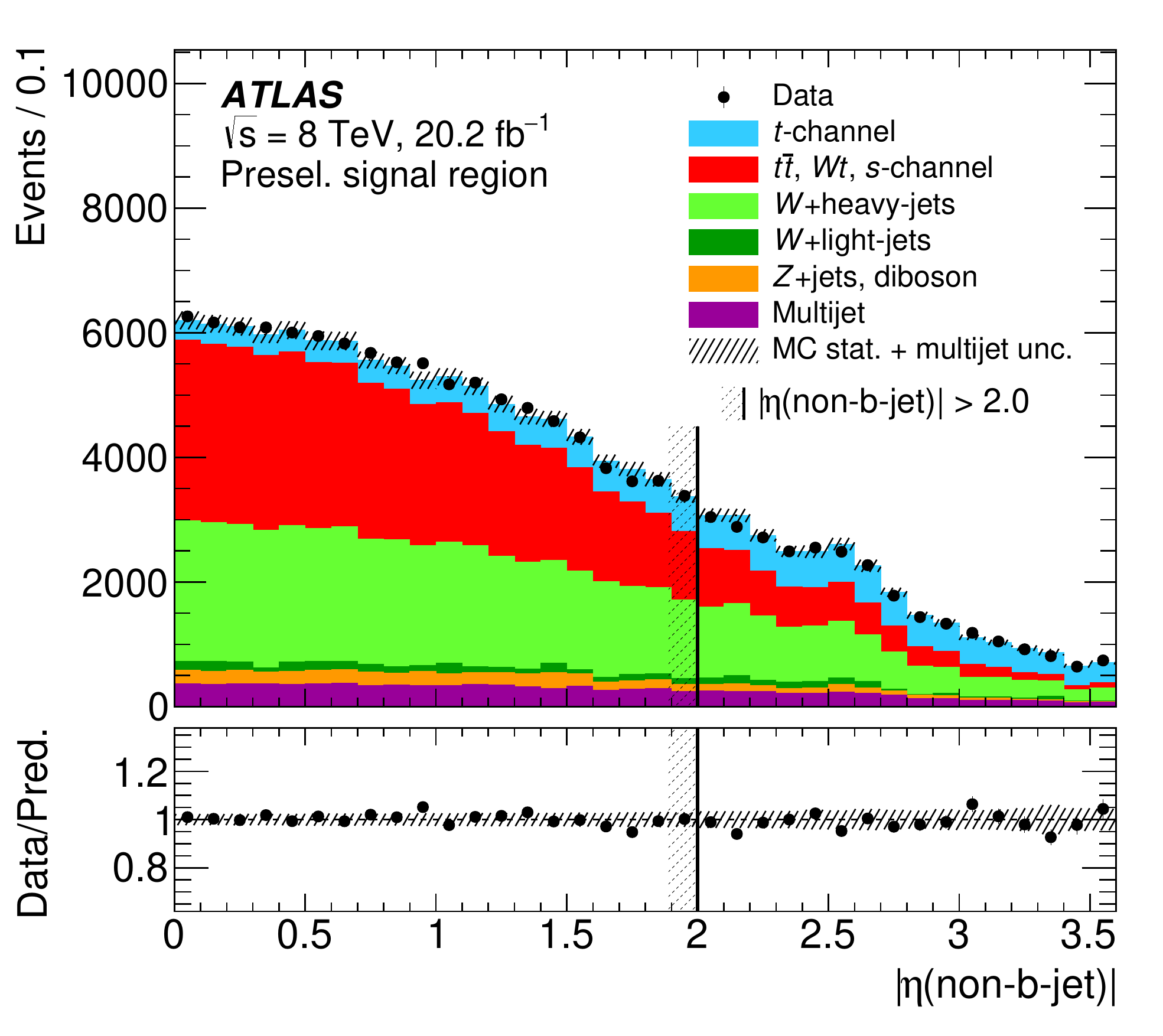}}
      \subfloat[]  {\includegraphics[width=0.48\textwidth]{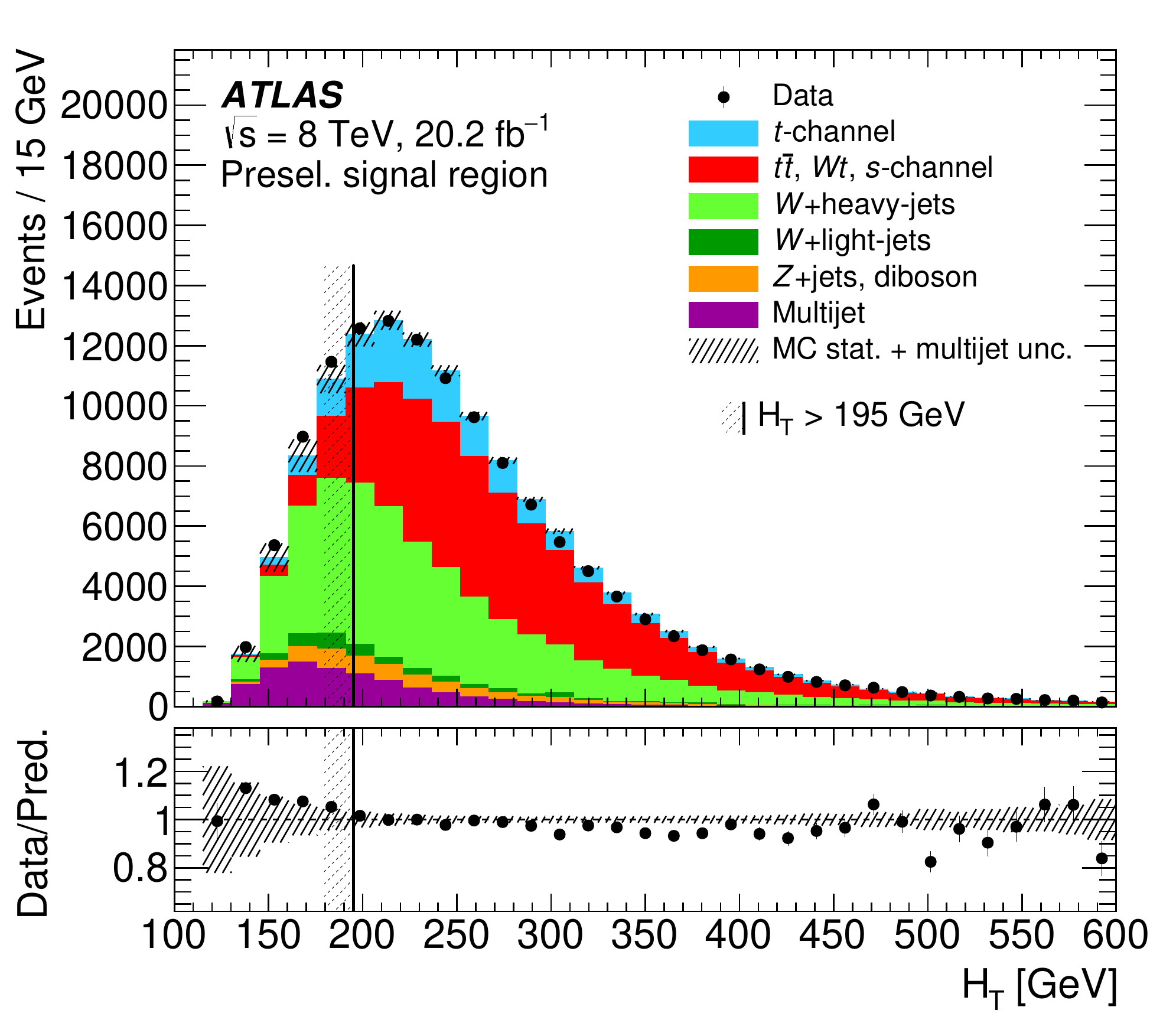}}
    }}
  \centerline{\mbox{
      \subfloat[]  {\includegraphics[width=0.48\textwidth]{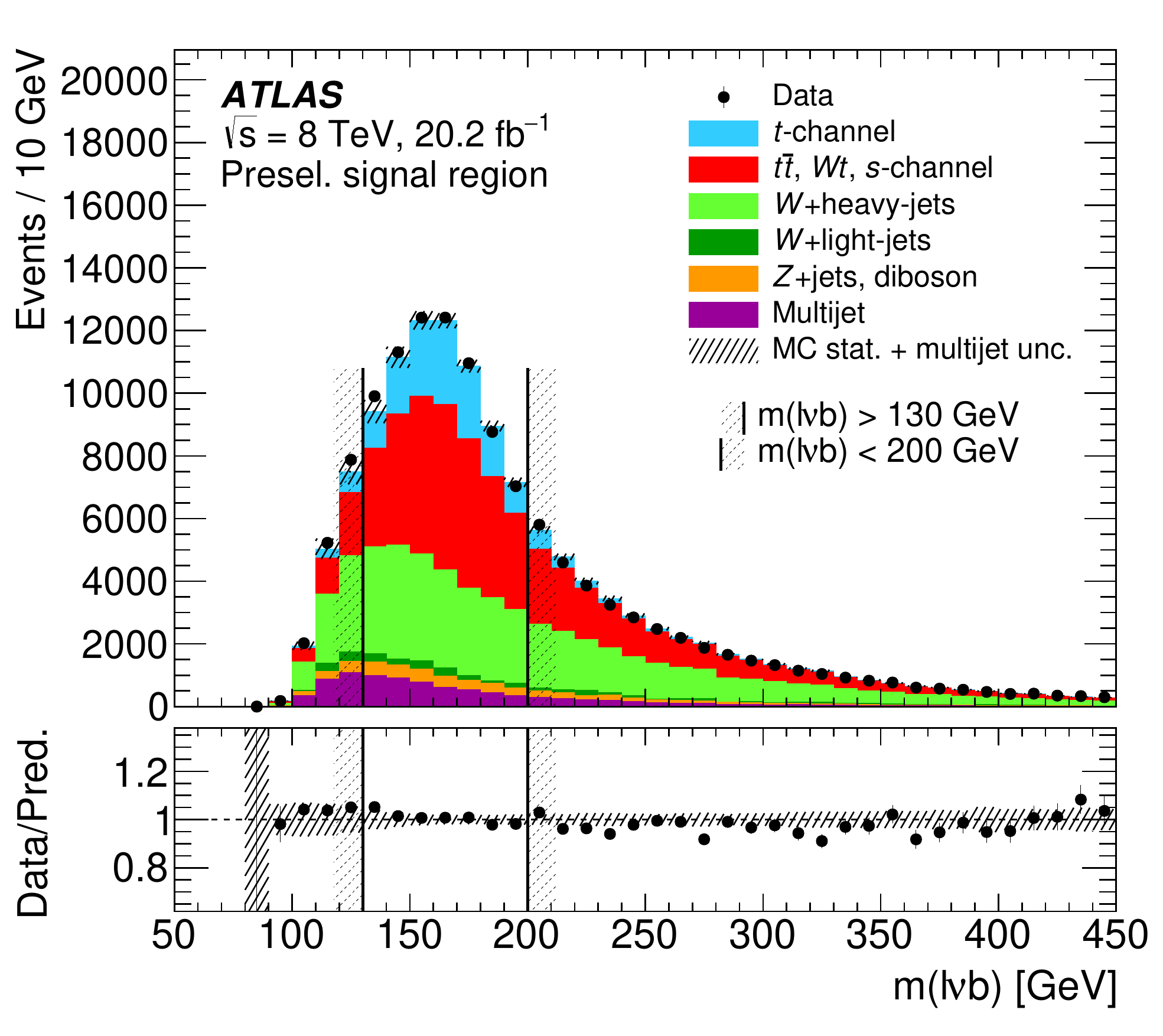}}
      \subfloat[]  {\includegraphics[width=0.48\textwidth]{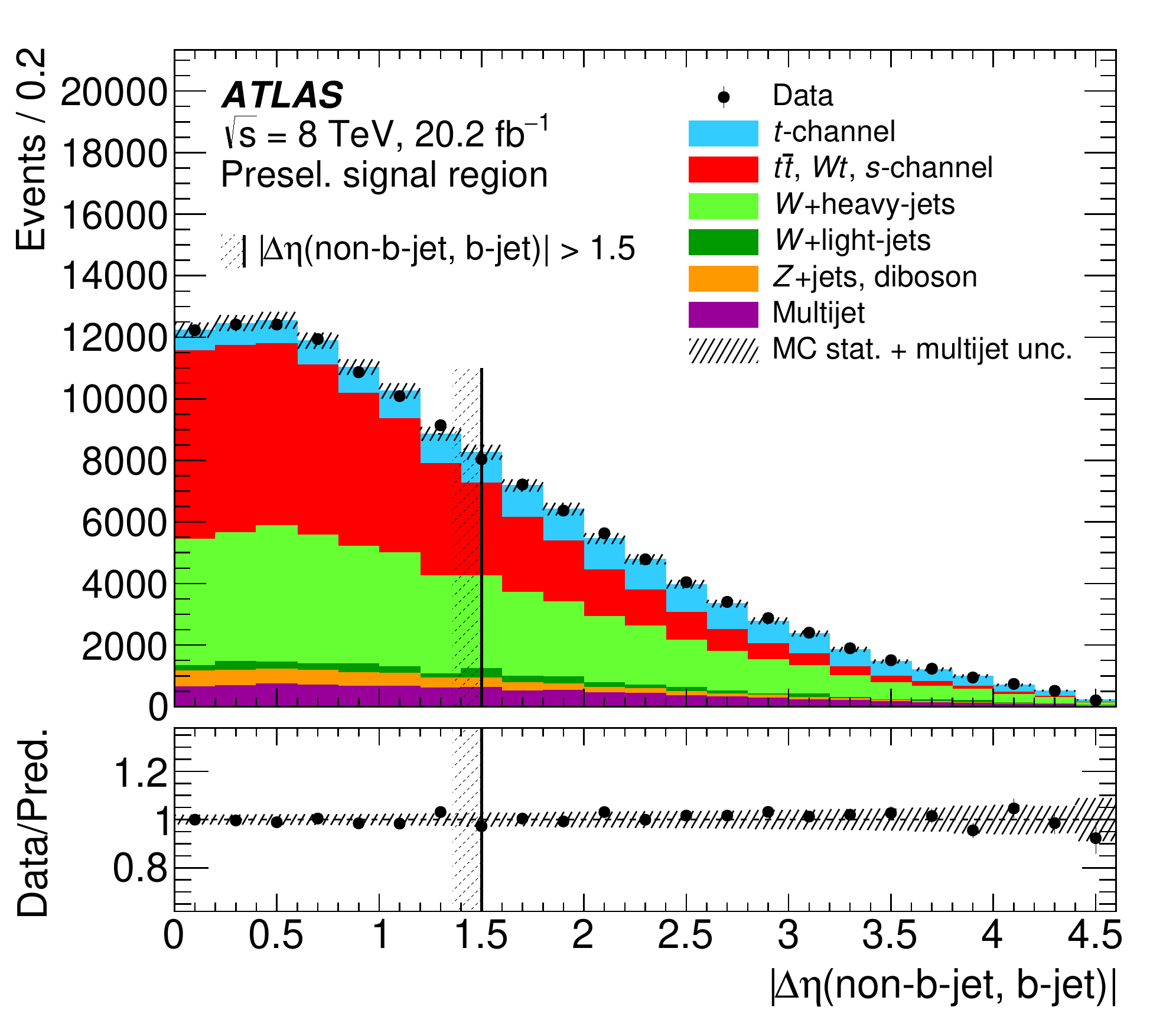}}
    }}
  \caption{Distributions of (a) $|\eta(\text{non-$b$-jet})|$, (b) the scalar sum of the \pT\ of all final-state objects, $\HT$,
   (c) reconstructed top-quark mass, $\mtop$, and (d) $|\Delta\eta(\text{non-$b$-jet, $b$-jet})|$ in the
    signal preselected region for the electron and muon channels merged.
    The prediction is compared to data, shown as the black points with statistical uncertainties.
    The \qcd\ background is estimated using data-driven techniques, while contributions from
    simulated $W$+jets, top-quark backgrounds and \tch\ event samples are normalised to the results of a
    maximum-likelihood fit to event yields in the signal and control
    regions. The uncertainty bands correspond to
    the uncertainties due to the size of the simulated event samples added
    in quadrature with the data-driven normalisation uncertainty of 70\% estimated for the \qcd\
    contribution. The lower plots show the
    ratio of data to prediction in each bin. The regions excluded by the selection criteria are shown by vertical black lines and dashed areas.}
  \label{fig:SelectionCuts}
\end{figure}

The distributions of the four variables used to define the signal region are shown in \Figure{fig:SelectionCuts} at the preselection stage.
The simulated signal and background distributions are scaled to their theoretical predictions except the \qcd\ background, which is estimated using data-driven techniques described in \Section{sec:background_estimation}. The $W$+jets, top-quark backgrounds and \tch\ distributions are normalised to the results of the maximum-likelihood fit, also described in \Section{sec:background_estimation}. In \Figure{fig:SelectionCuts}(a), the well-modelled bump around $|\eta| = 2.5$ is due to a combination of the JVF requirement, which is applied to jets with $\pT < 50~\GeV$ and $|\eta| < 2.4$, and the increased \pT\ requirement on jets in the calorimeter endcap--forward transition region ($2.7 < |\eta| < 3.5$). These two requirements are described in \Section{sec:object_selection}.

To estimate the rates and validate the modelling of the dominant background contributions, the simulated events are compared to the data in three dedicated background-enriched regions:

\begin{itemize}
\item A control region dominated by \ttbar\ events is defined by considering preselected events containing two additional non-$b$-tagged jets (i.e. four jets are required since just one of them is required to be $b$-tagged).
\item A control region enriched in $W$+jets events, and dominated by $W$+heavy-jets, is defined in order to control the modelling of the background. The events selected in this control region are the ones satisfying the preselection criteria and failing to satisfy any of the four requirements in the selection criteria.
The flavour composition of this control region is similar to that of the signal region.
\item A third region is defined as a validation region dominated by $W$+jets events to further control the modelling of the shapes of the $W$+jets background. Events in this validation region are selected by considering the preselection criteria with a relaxed $b$-tagging efficiency requirement of 80\%. In addition, all events satisfying the tighter signal $b$-tagging efficiency requirement of 50\% are excluded. This region has much larger enrichment in $W$+jets events although the flavour composition differs from that of the signal region.
\end{itemize}

The two control regions are used to extract the normalisation of
\ttbar\ and $W$+jets as described in \Section{sec:background_estimation}.

\section{Background estimation and normalisation}
\label{sec:background_estimation}

The largest background contributions to single-top-quark \tch\ production arise from \ttbar\ and $W$+jets production. The former is difficult to distinguish from the signal since \ttbar\ events contain real top quarks in the final state. The $W$+jets production contributes to the background if there is a $b$-quark in the final state or due to mis-tagging of jets containing other quark flavours. \QCD\ production via the strong interaction can contribute as well if, in addition to two reconstructed jets, an extra jet is misidentified as an isolated lepton, or if a non-prompt lepton appears to be isolated (both referred to as fake leptons). Other minor backgrounds originate from single-top-quark \Wt-channel and \sch, $Z$+jets and diboson production.

For all background processes, except \qcd\ production, the normalisation is initially estimated by using the MC simulation scaled with the theoretical cross-section prediction, and the event distribution modelling is taken from simulation.

The \ttbar\ events are normalised with the \ttbar\ production cross-section calculated at NNLO in QCD including resummation of NNLL soft gluon terms with Top++2.0~\cite{Baernreuther:2012ws,Czakon:2011xx,Czakon:2012zr,Czakon:2012pz,Czakon:2013goa,Cacciari:2011hy}. Its predicted value is $253^{+13}_{-15}$~pb calculated according to Ref.~\cite{Cacciari:2011hy}. The quoted uncertainty, evaluated according to the PDF4LHC prescription~\cite{Botje:2011sn}, corresponds to the sum in quadrature of the $\alphas$ uncertainty and the PDF uncertainty, calculated from the envelope of the uncertainties at 68\%~CL of the MSTW2008 NNLO, CT10 NNLO~\cite{Gao:2013xoa} and NNPDF2.3 5f FFN~\cite{Ball:2012cx} PDF sets. The associated \Wt-channel events are normalised with the predicted NNLO production cross-section of $22.4\pm1.5$~pb~\cite{Kidonakis:2010ux} and the \sch\ production to the predicted NNLO cross-section of $5.61\pm0.22$~pb~\cite{Kidonakis:2010tc}.
The uncertainties correspond to the sum in quadrature of the uncertainty derived from the MSTW2008 NNLO PDF set at 90\%~CL and the scale uncertainties.

The inclusive cross-sections of vector-boson production are calculated to NNLO with the FEWZ program~\cite{Anastasiou:2003ds} and the MSTW2008 NNLO PDF set, with a theoretical uncertainty of 4\% and 5\% for $W$+jets and $Z$+jets, respectively. The cross-sections of diboson processes are calculated at NLO using the MCFM program~\cite{Campbell:2011bn}, with a theoretical uncertainty of 5\%. For these three background processes the normalisation uncertainty is 34\% each. This is the result of adding in quadrature their theory uncertainty and 24\% per additional jet, accordingly to the
Berends--Giele scaling~\cite{Berends:1990ax}.

The normalisation as well as the event modelling of the \qcd\ background is estimated from data using a matrix method~\cite{ATLAS:2014ffa, Aad:2012qf}. This method allows the derivation of the true composition of the data sample in terms of prompt (real) and fake leptons from its observed composition in terms of tight (signal selection) and loose leptons. An alternative normalisation and modelling based on the mixed data--simulation jet-electron method~\cite{ATLAS:2014ffa, Aad:2012ux, Aad:2014fwa} and the purely data-driven anti-muon selection~\cite{ATLAS:2014ffa} are also considered. From the comparison of these two models with the results obtained using the matrix method, an overall normalisation uncertainty of 70\% is assigned to the \qcd\ contribution, irrespective of lepton flavour, as done in Ref.~\cite{Aaboud:2017aqp}.

The final \tch, $W$+jets and top-quark background (\ttbar,
associated \Wt\ and \sch) normalisations are estimated through a
simultaneous maximum-likelihood fit to the numbers of data events
observed in the signal region and the \ttbar\ and $W$+jets control regions, described
in \Section{sec:event_selection}. The likelihood function~\cite{Aad:2012ux} is given by the product of Poisson
probability terms associated with the fitted regions, combined with
the product of Gaussian priors to constrain the background rates to
their predictions within the associated uncertainties. 
In the fit, the \tch\ contribution, estimated using \powheg, is treated as
unconstrained. The top-quark background contributions are merged with their relative
fractions taken from simulation, and the applied constraint, 6\%, is derived
from the combination in quadrature of their cross-section uncertainties. The $W$+jets contribution is
constrained to the normalisation uncertainty of 34\% and its flavour
composition is taken from simulation. In these three fitted regions the
production of a $W$ boson in association with heavy-flavour jets is
the dominant contribution to the $W$+jets background, predicted to be
around 95\% in each region. The $Z$+jets and diboson
contributions, which are very low in the signal region (2\% of the
expected total), are merged and fixed to the predictions. The \qcd\
contribution is kept fixed to its data-driven estimate.
The overall normalisation scale factors obtained from the maximum-likelihood fit together with the
statistical post-fit uncertainties are found to be $1.010 \pm 0.005$ and $1.128 \pm 0.013$
for the top-quark and $W$+jets background contributions, respectively, and $0.909
\pm 0.022$ for the \tch\ signal. 
The impact on the analysis due to the deviation of these scale factors from unity is negligible and it is taken into account through the $W$+jets normalisation
uncertainty as discussed in \Section{sec:systematics}. In the case of the $W$+jets
validation region, used to validate the shapes of the predicted
templates, just an overall scale factor for the $W$+jets component is
estimated. It is extracted by matching the total predicted event yields to the number of events 
observed in this validation region. 
The results are found to be stable when the prior constraints
on the top-quark and $W$+jets backgrounds are relaxed to 100\% of their predicted cross-section in the signal and control regions.

The overall normalisation scale factors are used to control the
modelling of the kinematic and angular variable distributions in the
signal, control, and validation regions. In the subsequent steps of the analysis, the overall scaling of the \tch\ prediction is not relevant, since it is taken from background-subtracted data, while the $W$+jets and top-quark backgrounds are normalised using these overall scale factors.

\section{Event yields and kinematic distributions}
\label{sec:eventyields_and_controlplots}

\Table{tab:evtyield_summay_all} provides the
predicted signal and background event yields for the electron and muon channels
merged together in the signal, control, and validation regions after scaling
to the results of the maximum-likelihood fit to the data. Observed
data yields are also shown. The signal-to-background (S/B) ratio
is 0.97 in the signal region while $\lesssim 0.1$ in the control and
validation regions.

\begin{table}[!htb]
  \centering
  \begin{adjustbox}{max width=0.98\textwidth}
    \input{tables/EventYield_Merged_2-jetbin_CustomGroupList_combined_nominal_AllRegions_rounded.tex}
  \end{adjustbox}
  \caption{Predicted and observed data event yields are shown for the merged electron and muon channels in the
    signal, \ttbar\ and $W$+jets control and validation regions.
    The \qcd\ background is estimated using data-driven techniques, while contributions from
    simulated $W$+jets, top-quark backgrounds and \tch\ event samples are normalised to the results of a
    maximum-likelihood fit to event yields in the signal and control
    regions. The uncertainties shown are statistical only. Individual predictions are rounded to two significant digits of the uncertainty
    while ``Total expected'' corresponds to the rounding of the sum of
    full-precision individual predictions. The expected S/B ratios are also given.}
  \label{tab:evtyield_summay_all}
\end{table}

\begin{figure}[!htb]
  \centerline{\makebox{
      \subfloat[]  {\includegraphics[width=0.48\textwidth]{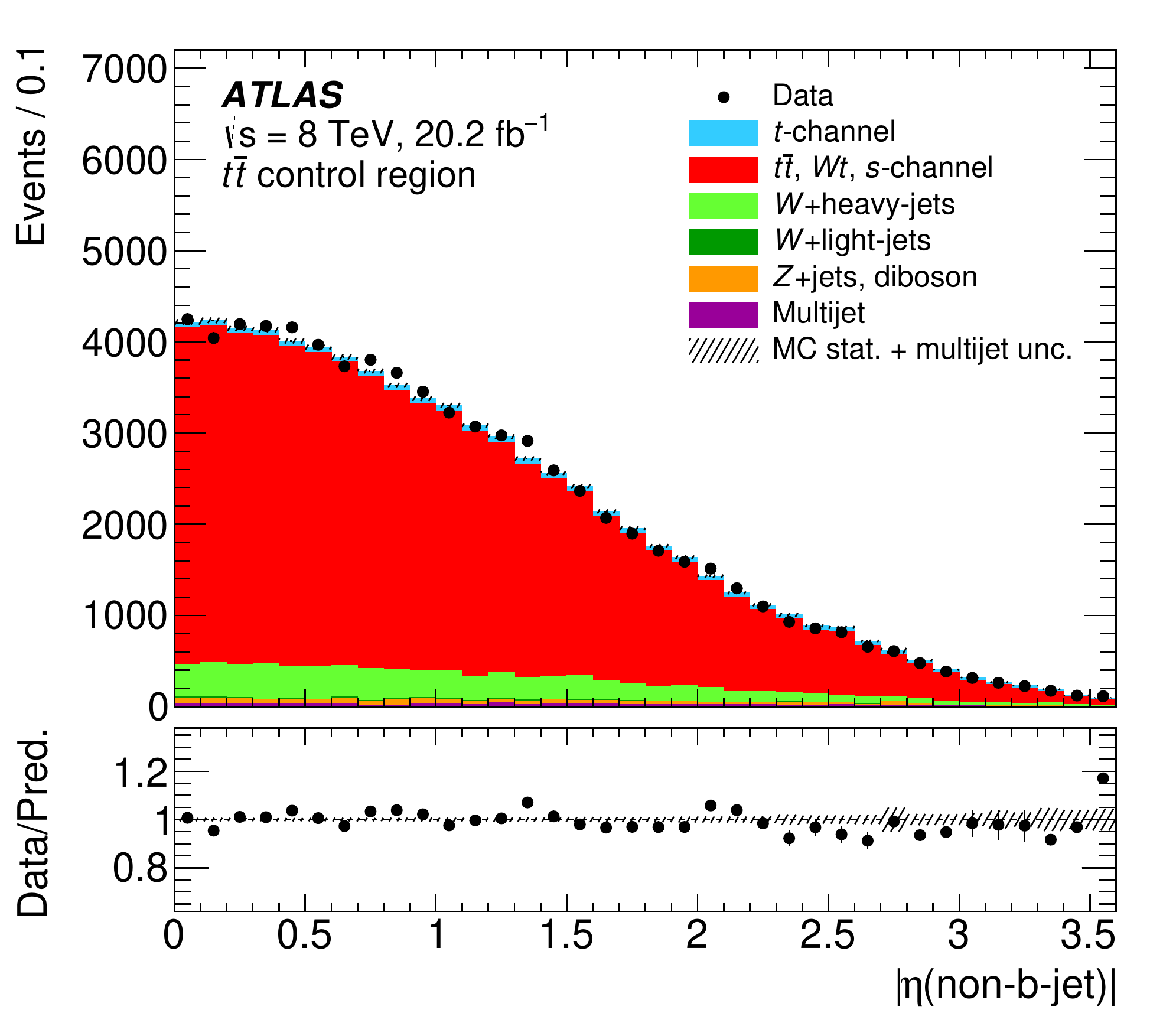}}
      \subfloat[]  {\includegraphics[width=0.48\textwidth]{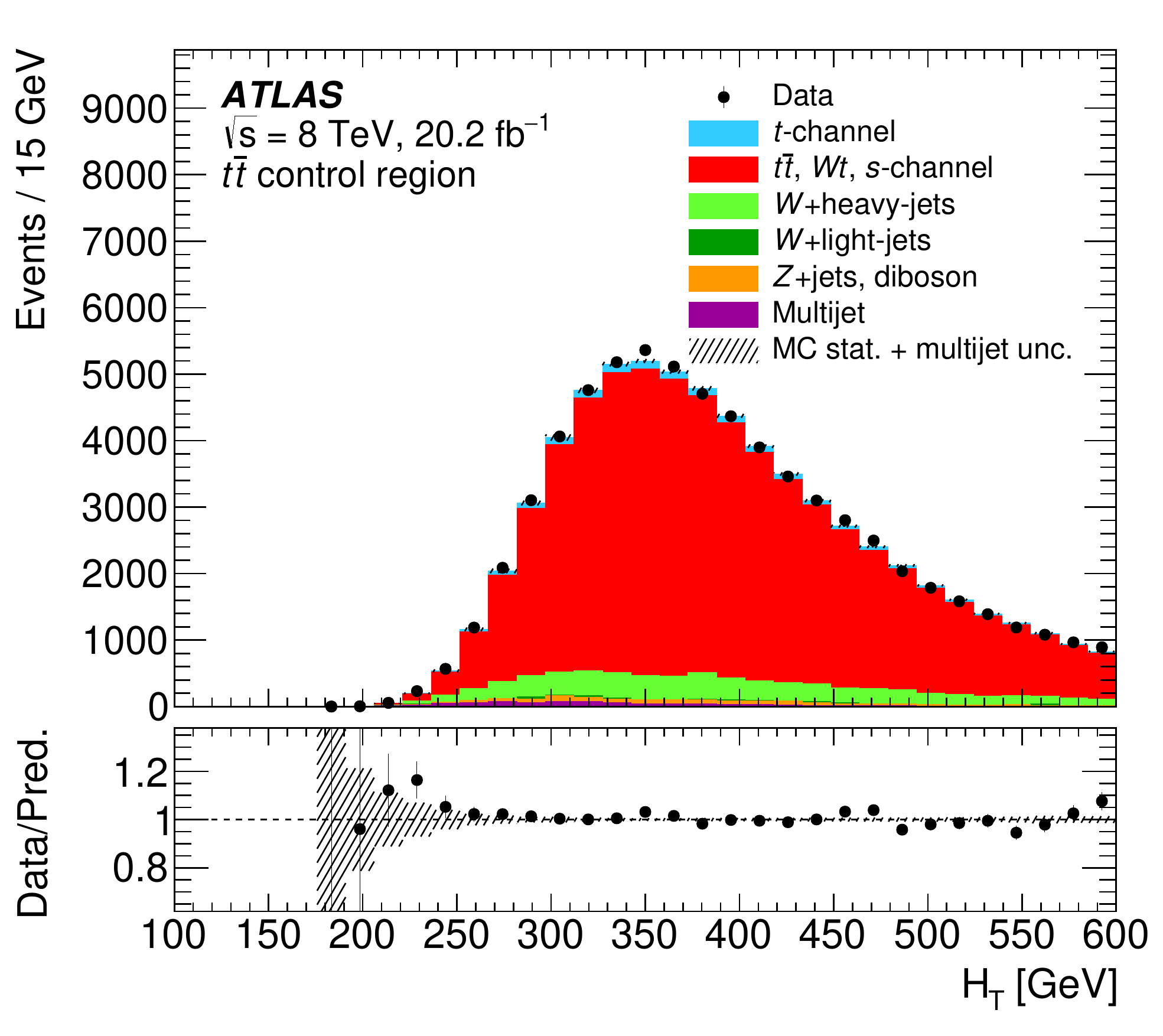}}
    }}
  \centerline{\makebox{
      \subfloat[]  {\includegraphics[width=0.48\textwidth]{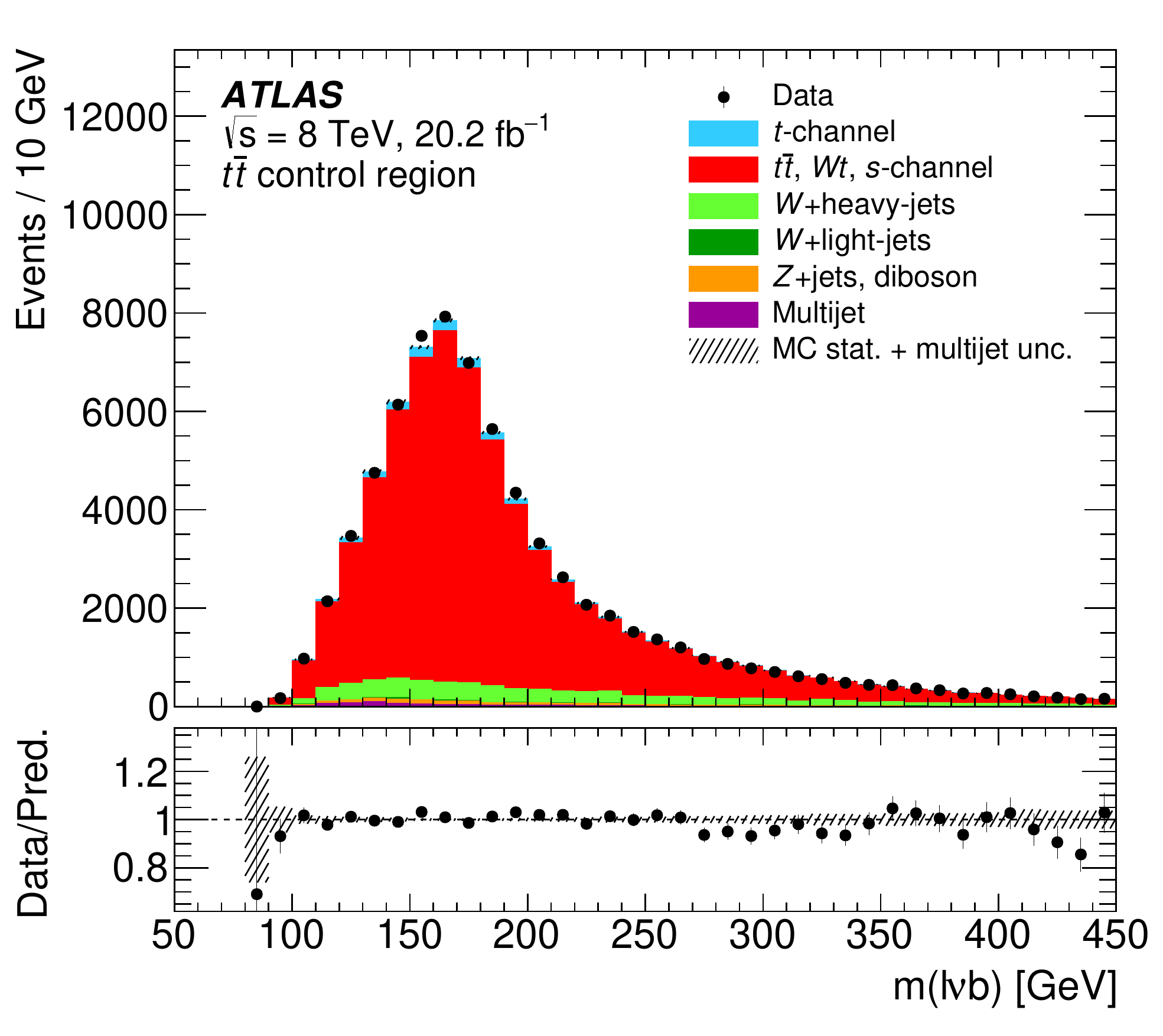}}
      \subfloat[]  {\includegraphics[width=0.48\textwidth]{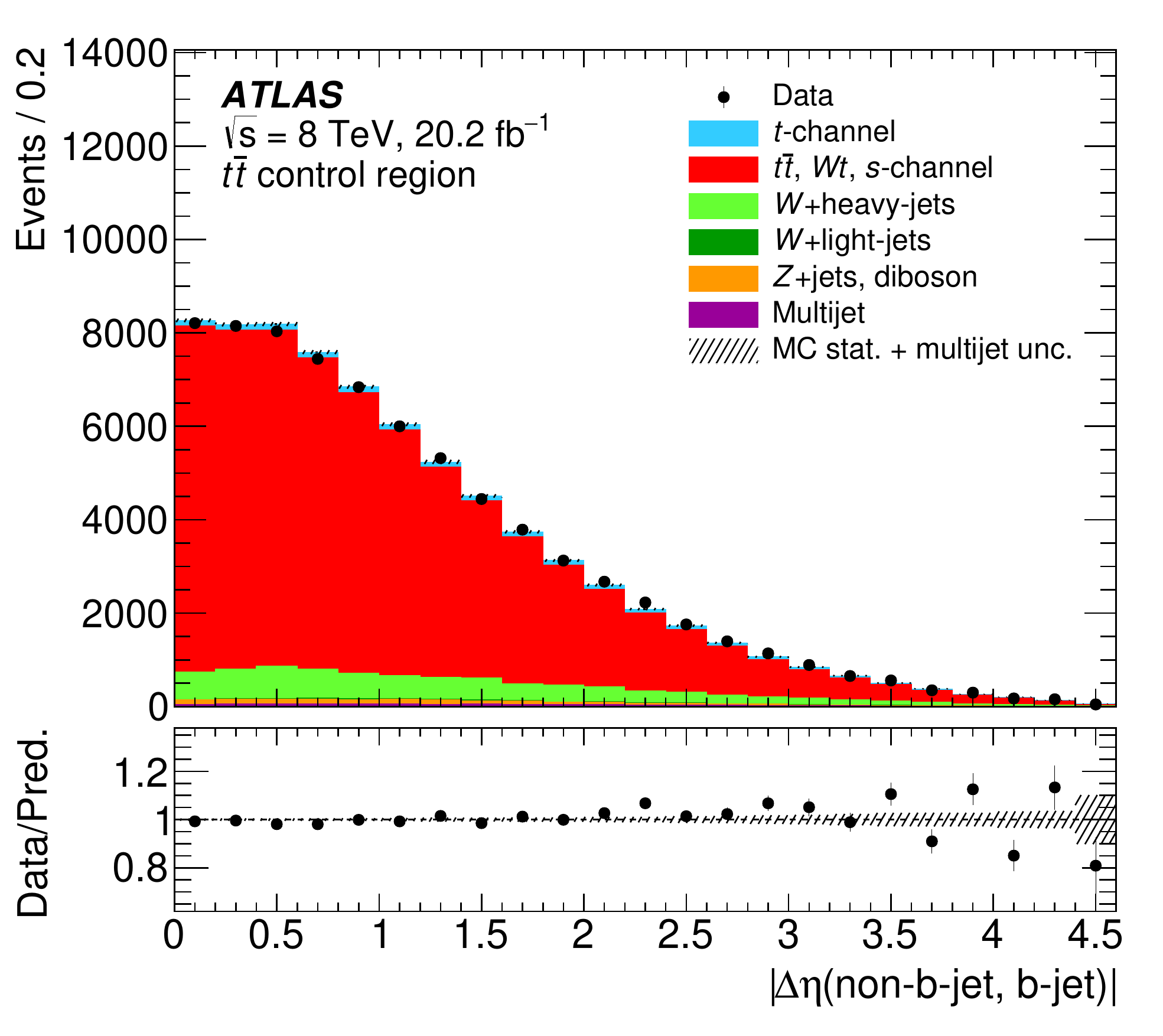}}
    }}
  \caption{Distributions of (a) $|\eta(\text{non $b$-jet})|$, (b) the scalar sum of the \pT\ of all final-state objects, $\HT$, (c)
    reconstructed top-quark mass, $\mtop$, and (d) $|\Delta\eta(\text{non $b$-jet}, b\text{-jet})|$ in the
    \ttbar\ control region for the merged electron and muon channels.
    The \qcd\ background is estimated using data-driven techniques, while contributions from
    simulated $W$+jets, top-quark backgrounds and \tch\ event samples are normalised to the results of a
    maximum-likelihood fit to event yields in the signal and control
    regions. The uncertainty bands correspond to
    the uncertainties due to the  size of the simulated event samples added
    in quadrature with the data-driven normalisation uncertainty of 70\% estimated for the \qcd\
    contribution. The lower plots show the ratio of data to prediction in each bin.}
  \label{fig:ttbarCR_controlPlots_merged}
\end{figure}

\begin{figure}[!htb]
  \centerline{\makebox{
      \subfloat[]  {\includegraphics[width=0.48\textwidth]{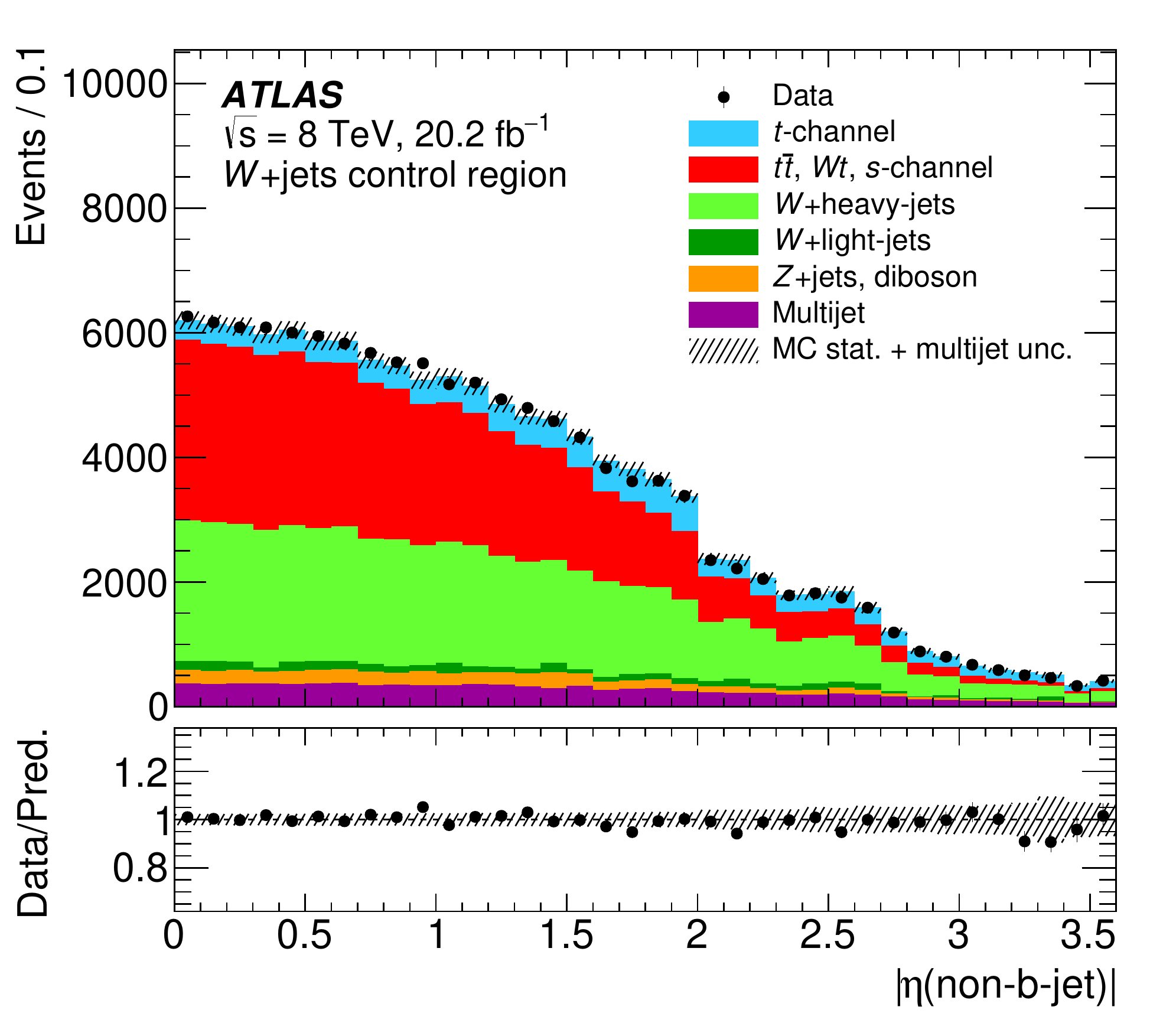}}
      \subfloat[]  {\includegraphics[width=0.48\textwidth]{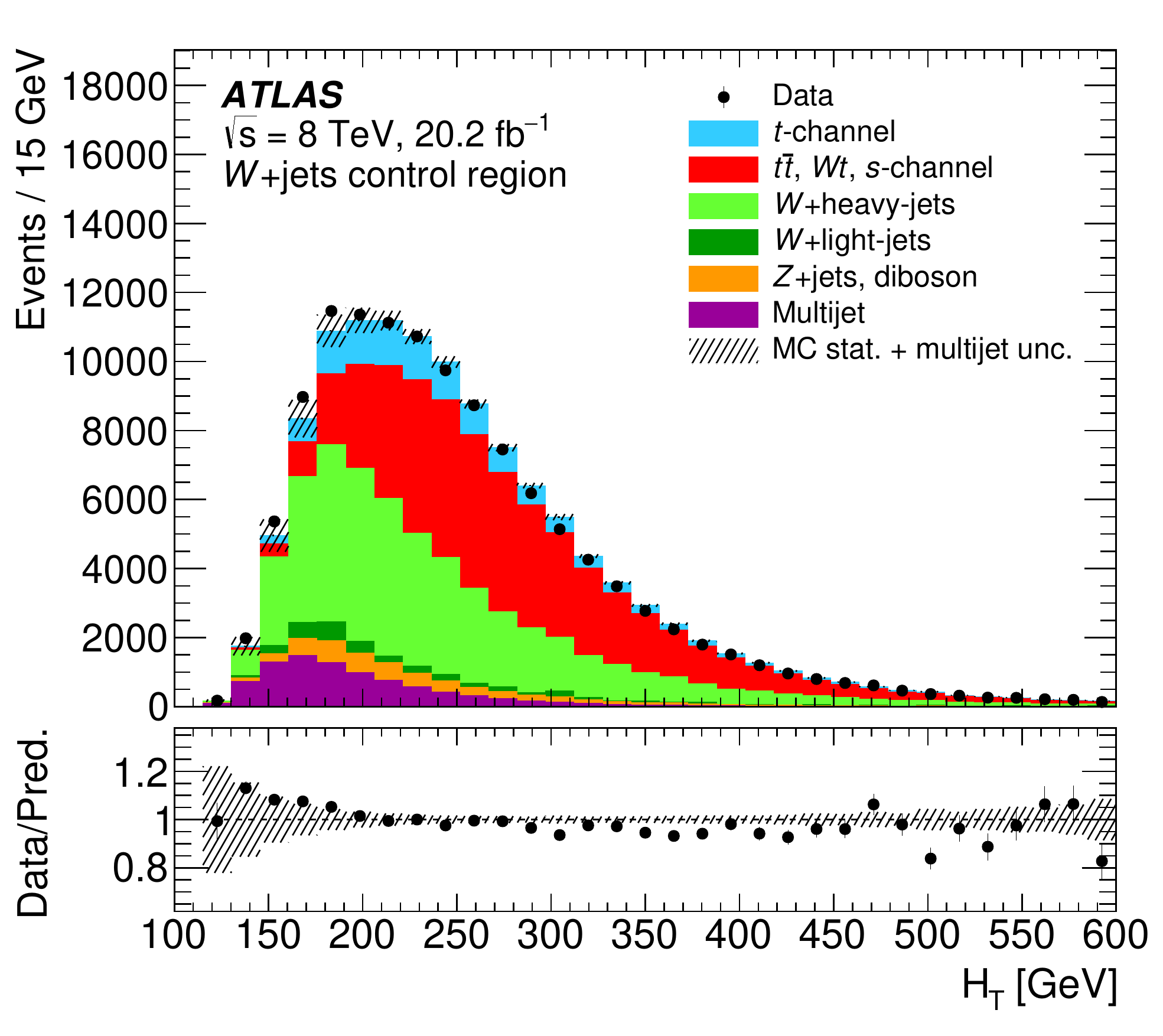}}
    }}
  \centerline{\makebox{
      \subfloat[]  {\includegraphics[width=0.48\textwidth]{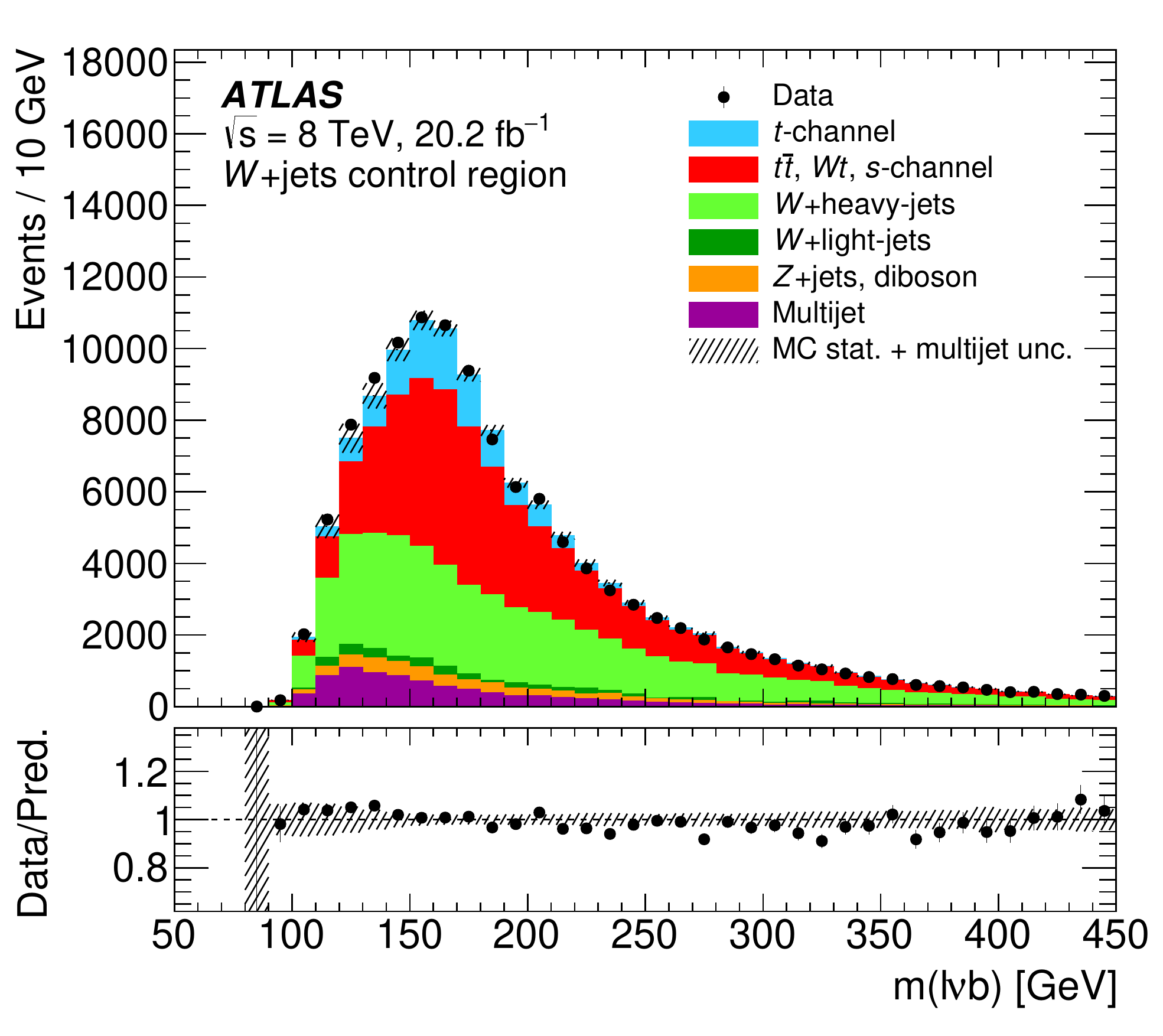}}
      \subfloat[]  {\includegraphics[width=0.48\textwidth]{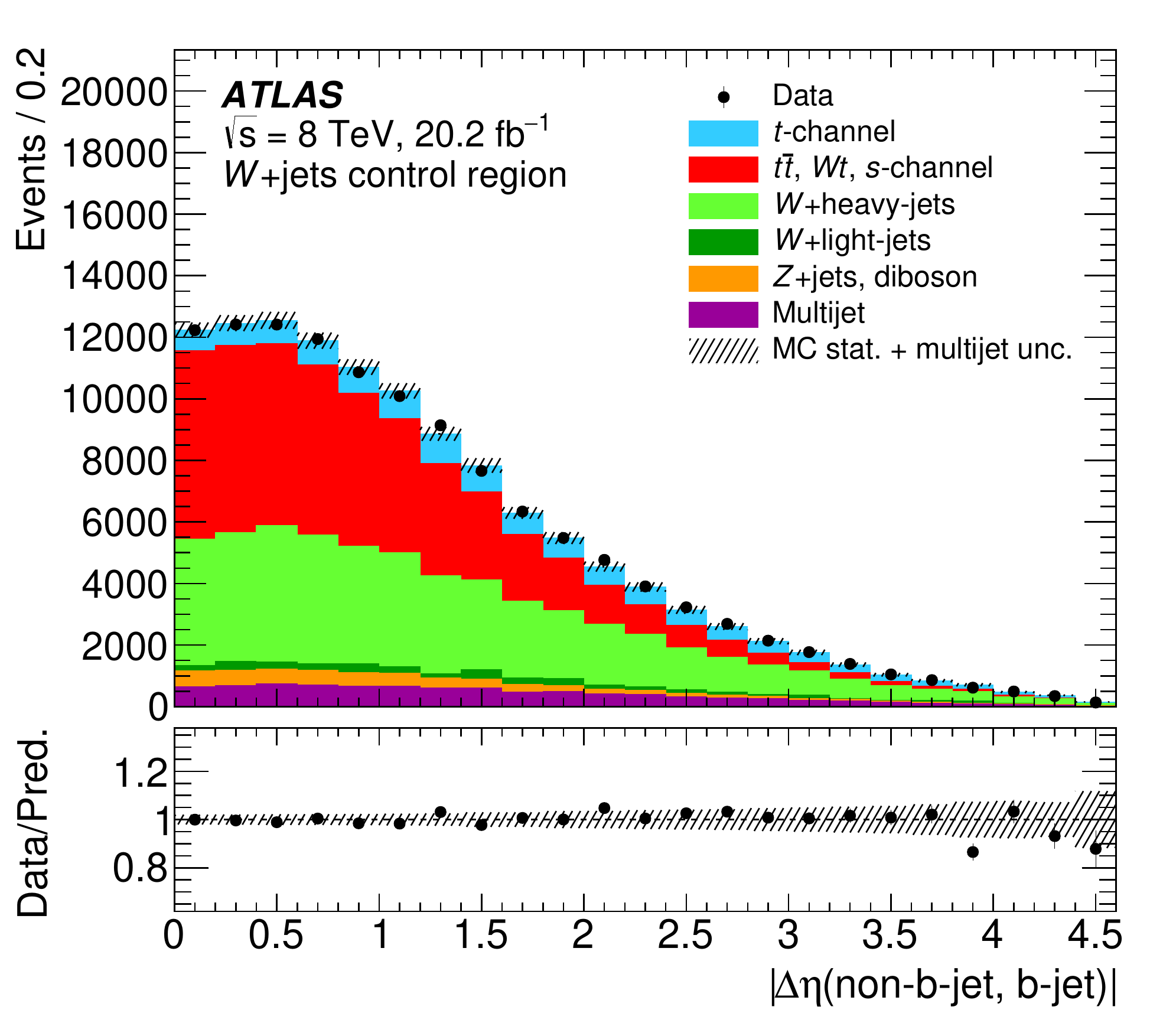}}
    }}
  \caption{Distributions of (a) $|\eta(\text{non $b$-jet})|$, (b) the scalar sum of the \pT\ of all final-state objects, $\HT$, (c)
    reconstructed top-quark mass, $\mtop$, and (d) $|\Delta\eta(\text{non $b$-jet}, b\text{-jet})|$ in the
    $W$+jets control region for the merged electron and muon channels.
    The \qcd\ background is estimated using data-driven techniques, while contributions from
    simulated $W$+jets, top-quark backgrounds and \tch\ event samples are normalised to the results of a
    maximum-likelihood fit to event yields in the signal and control
    regions. The uncertainty bands correspond to
    the uncertainties due to the size of the simulated event samples added
    in quadrature with the data-driven normalisation uncertainty of 70\% estimated for the \qcd\
    contribution. The lower plots show the ratio of data to prediction in each bin.}
  \label{fig:WjetsCR_controlPlots_merged}
\end{figure}

\begin{figure}[!htb]
  \centerline{\makebox{
      \subfloat[]  {\includegraphics[width=0.48\textwidth]{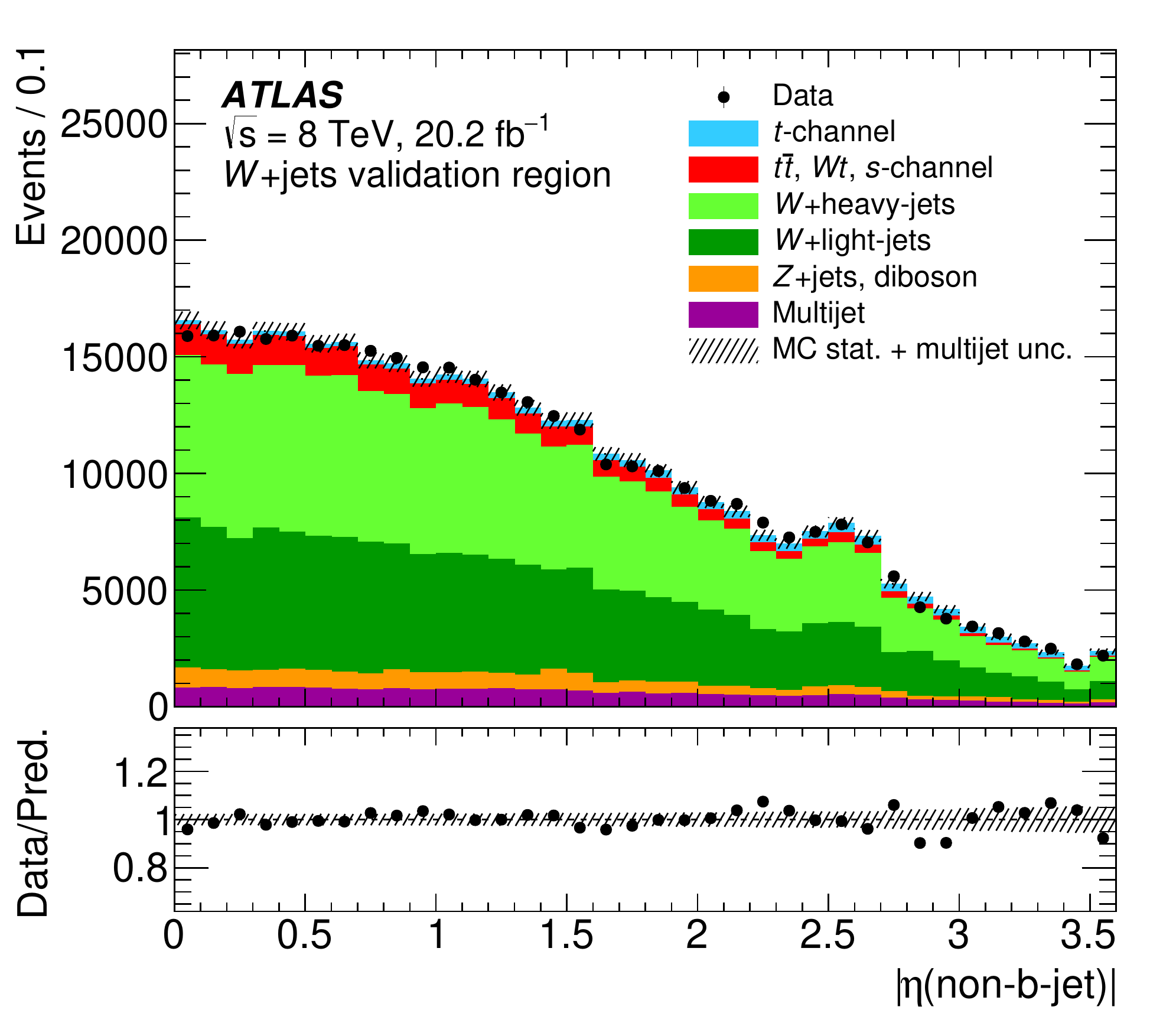}}
      \subfloat[]  {\includegraphics[width=0.48\textwidth]{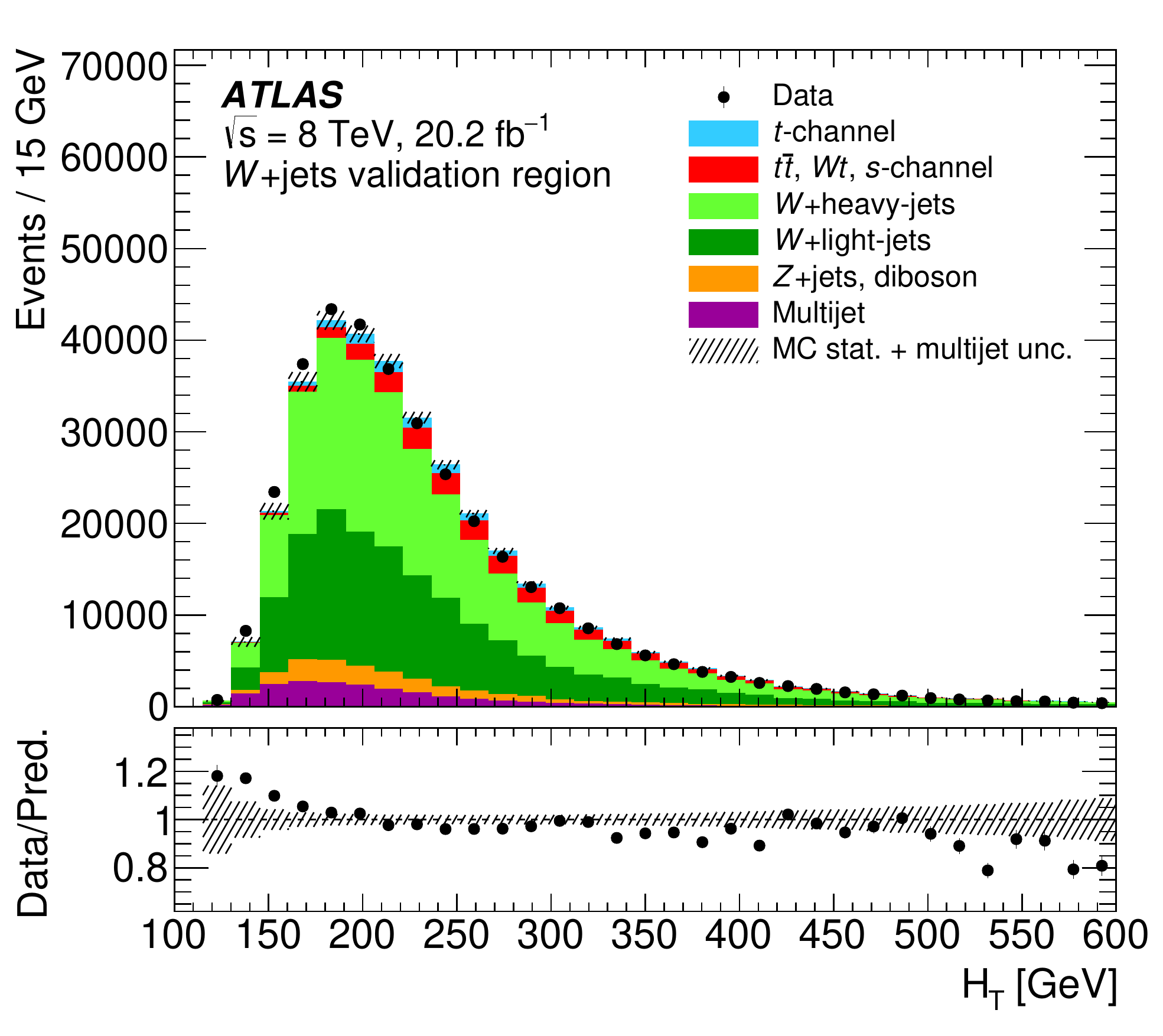}}
    }}
  \centerline{\makebox{
      \subfloat[]  {\includegraphics[width=0.48\textwidth]{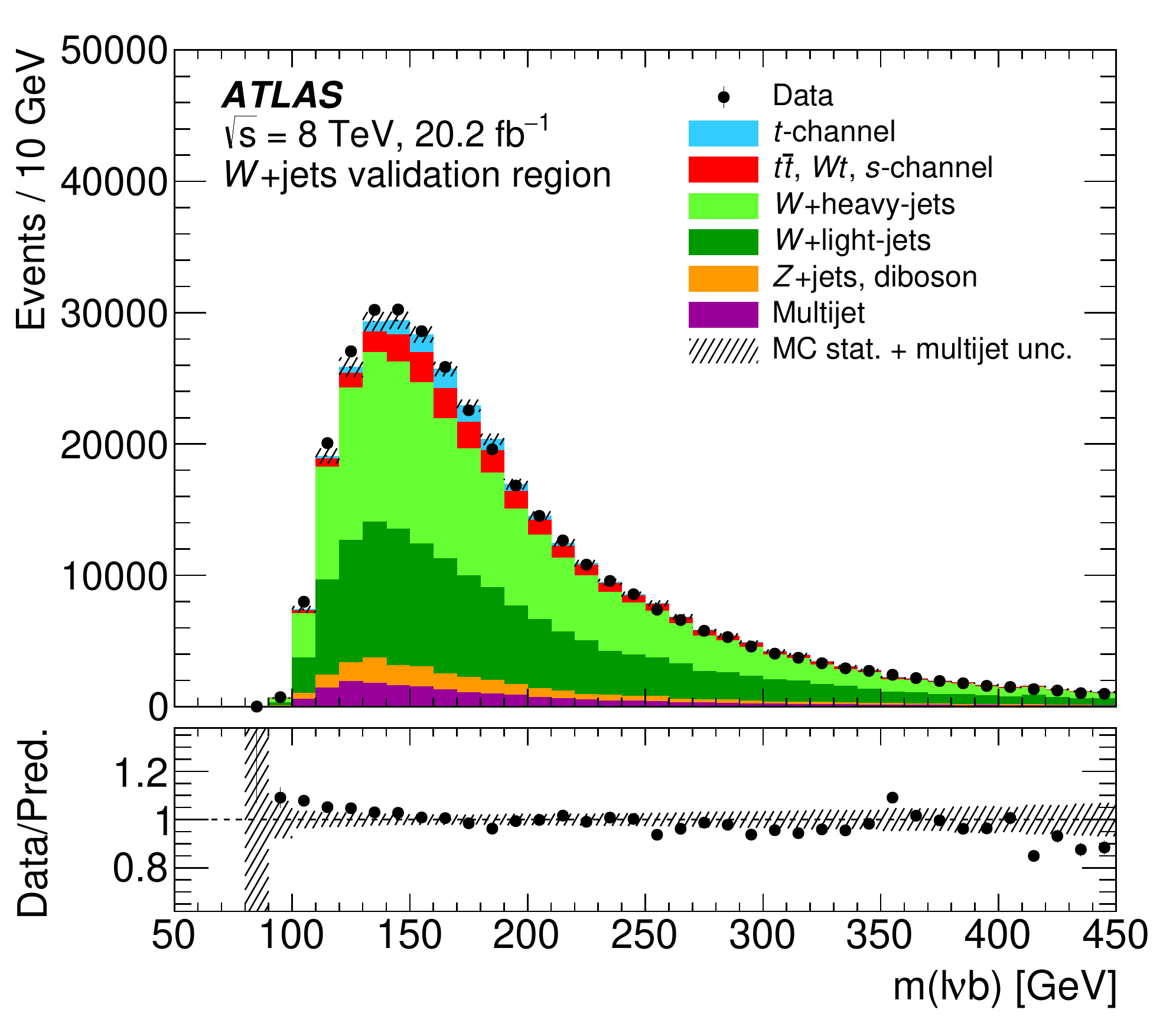}}
      \subfloat[]  {\includegraphics[width=0.48\textwidth]{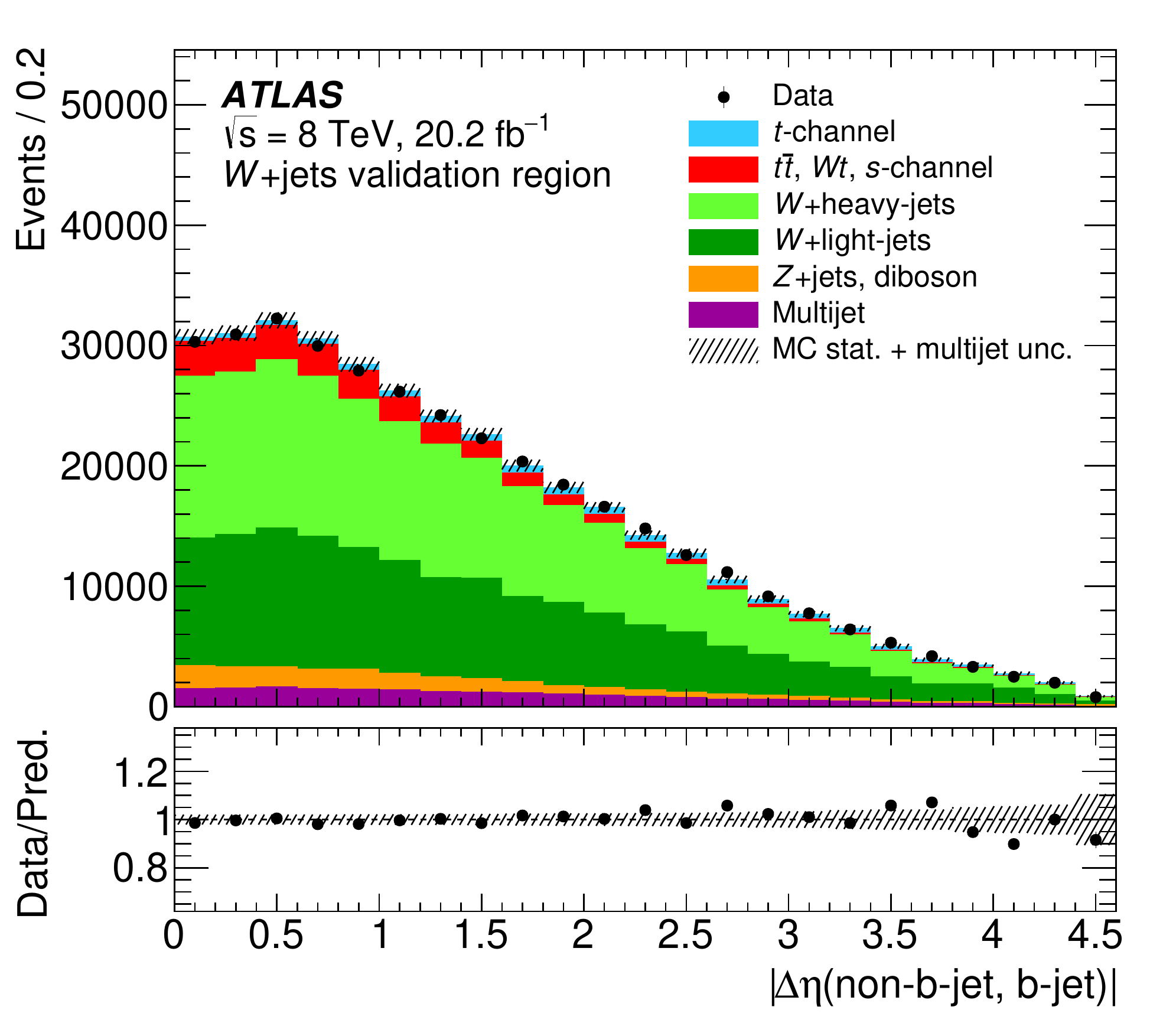}}
    }}
  \caption{Distributions of (a) $|\eta(\text{non $b$-jet})|$, (b) the scalar sum of the \pT\ of all final-state objects, $\HT$, (c)
    reconstructed top-quark mass, $\mtop$, and (d) $|\Delta\eta(\text{non $b$-jet}, b\text{-jet})|$ in the
    $W$+jets validation region for the merged electron and muon channels.
    The \qcd\ background is estimated using data-driven techniques, while contributions from
    simulated $W$+jets, top-quark backgrounds and \tch\ event samples are normalised to the results of a
    maximum-likelihood fit to event yields in the signal and control
    regions. The uncertainty bands correspond to
    the uncertainties due to the size of the simulated event samples added
    in quadrature with the data-driven normalisation uncertainty of 70\% estimated for the \qcd\
    contribution. The lower plots show the ratio of data to prediction in each bin.}
  \label{fig:WjetsVR_controlPlots_merged}
\end{figure}

Figures \ref{fig:ttbarCR_controlPlots_merged} and \ref{fig:WjetsCR_controlPlots_merged}
show the distributions of the relevant kinematic distributions used to
define the signal region in the \ttbar\ and
$W$+jets control regions while \Figure{fig:WjetsVR_controlPlots_merged} shows the same distributions
in the $W$+jets validation region.
Good overall data-to-prediction
agreement is found within the uncertainty band shown in these
distributions, which only includes the uncertainty due to the size of the simulation samples and the uncertainty in the
normalisation of the \qcd\ background, added in quadrature. 
Any data-to-prediction disagreement is covered by the  \ttbar\ and/or
$W$+jets normalisation and modelling uncertainties detailed
in \Section{sec:systematics}. In \Figure{fig:WjetsCR_controlPlots_merged}(a) and
\Figure{fig:WjetsVR_controlPlots_merged}(a), the origin of the well-modelled
bumps around $|\eta| = 2.5$ is the same as for
\Figure{fig:SelectionCuts}(a). In addition, the well-modelled decrease at $|\eta| =
2$ shown in \Figure{fig:WjetsCR_controlPlots_merged}(a) is due to the
rejected events in the $W$+jets control region, which satisfy the signal selection requirement of $|\eta(\text{non $b$-jet})| > 2.0$.

\section{Analysis of angular distributions}
\label{sec:analysis}

\begin{figure}[!htb]
  \centerline{\makebox{
      \subfloat[]  {\includegraphics[width=0.48\textwidth]{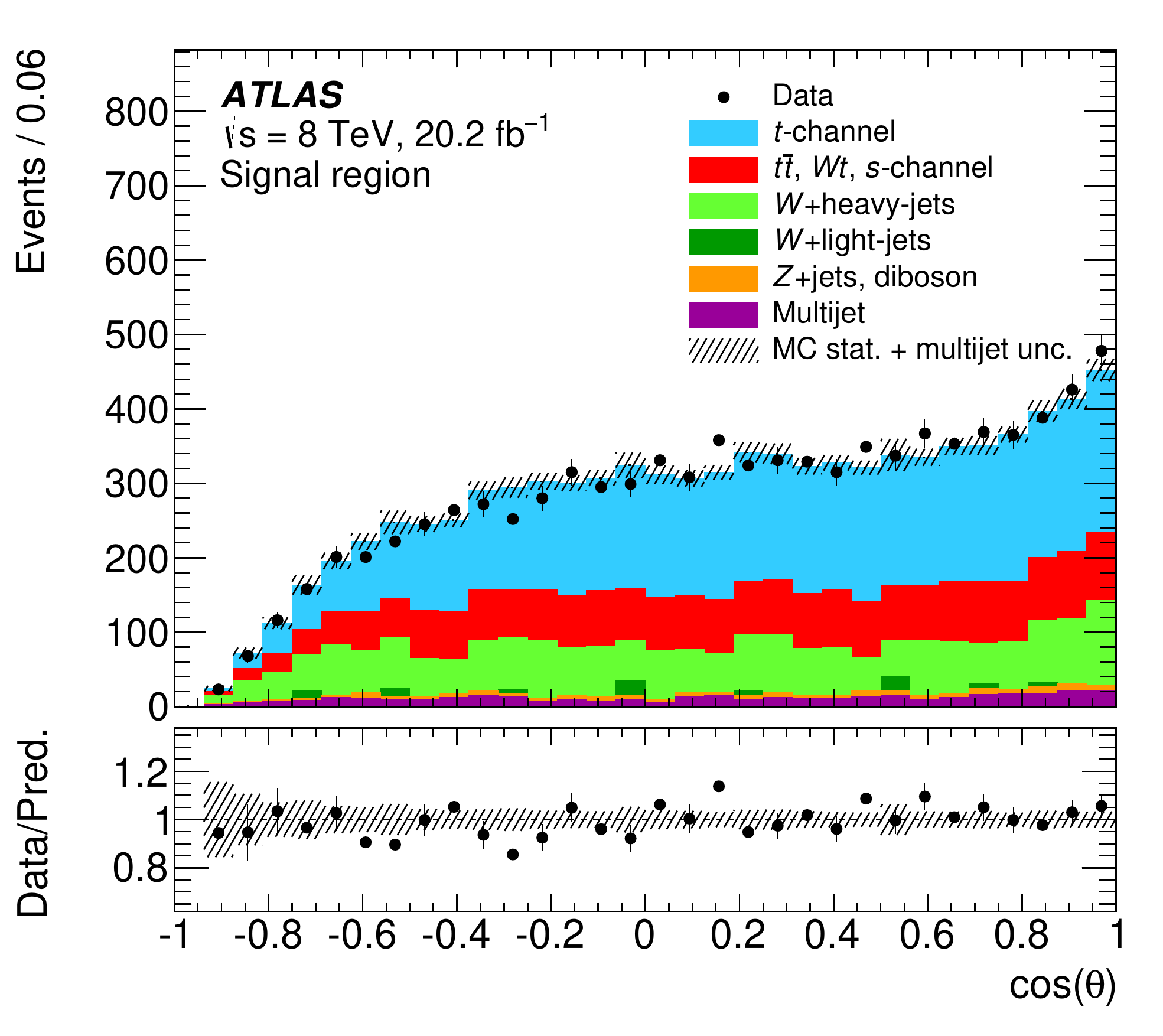}}
      \subfloat[]  {\includegraphics[width=0.48\textwidth]{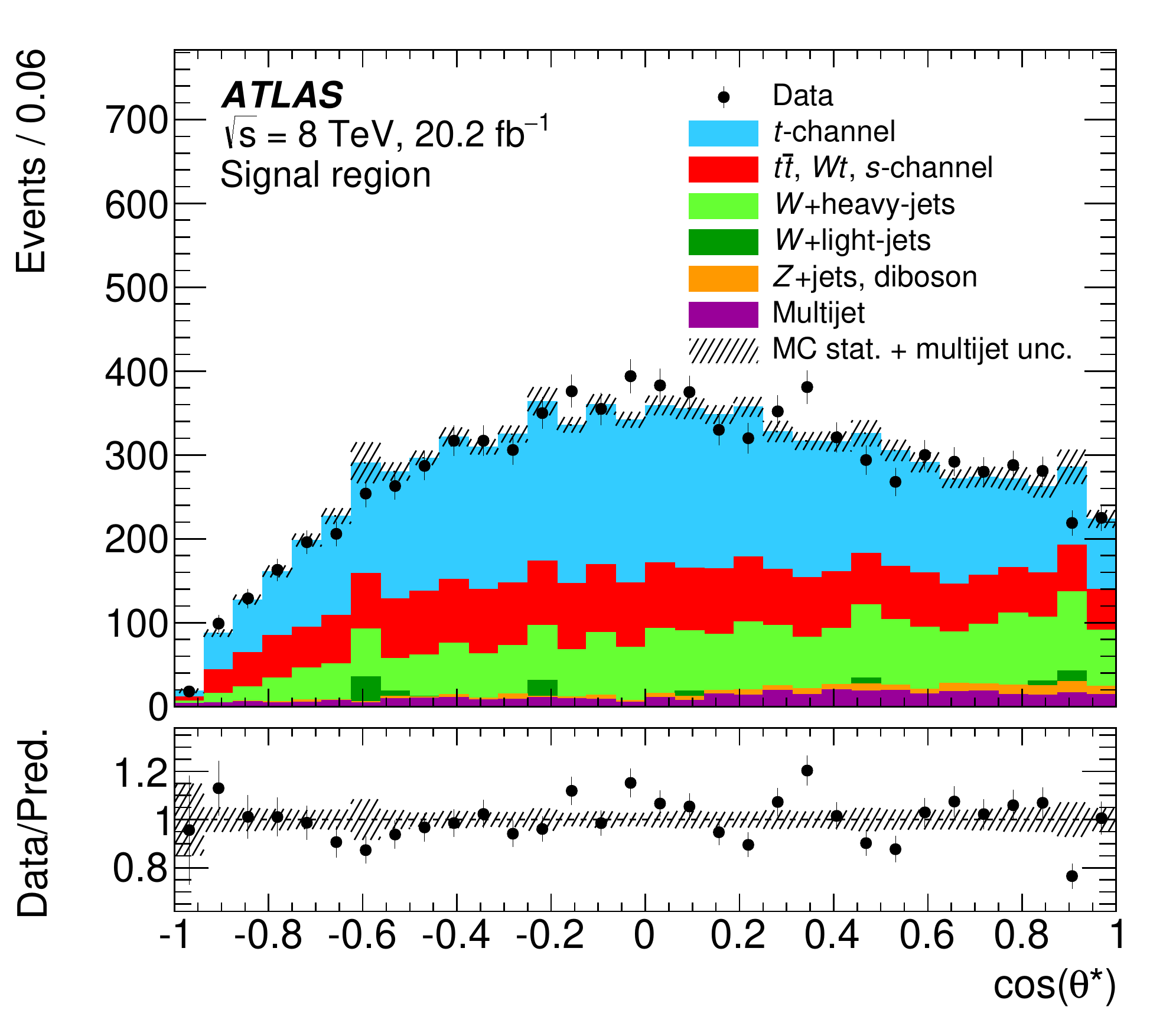}}
    }}
  \centerline{\makebox{
      \subfloat[]  {\includegraphics[width=0.48\textwidth]{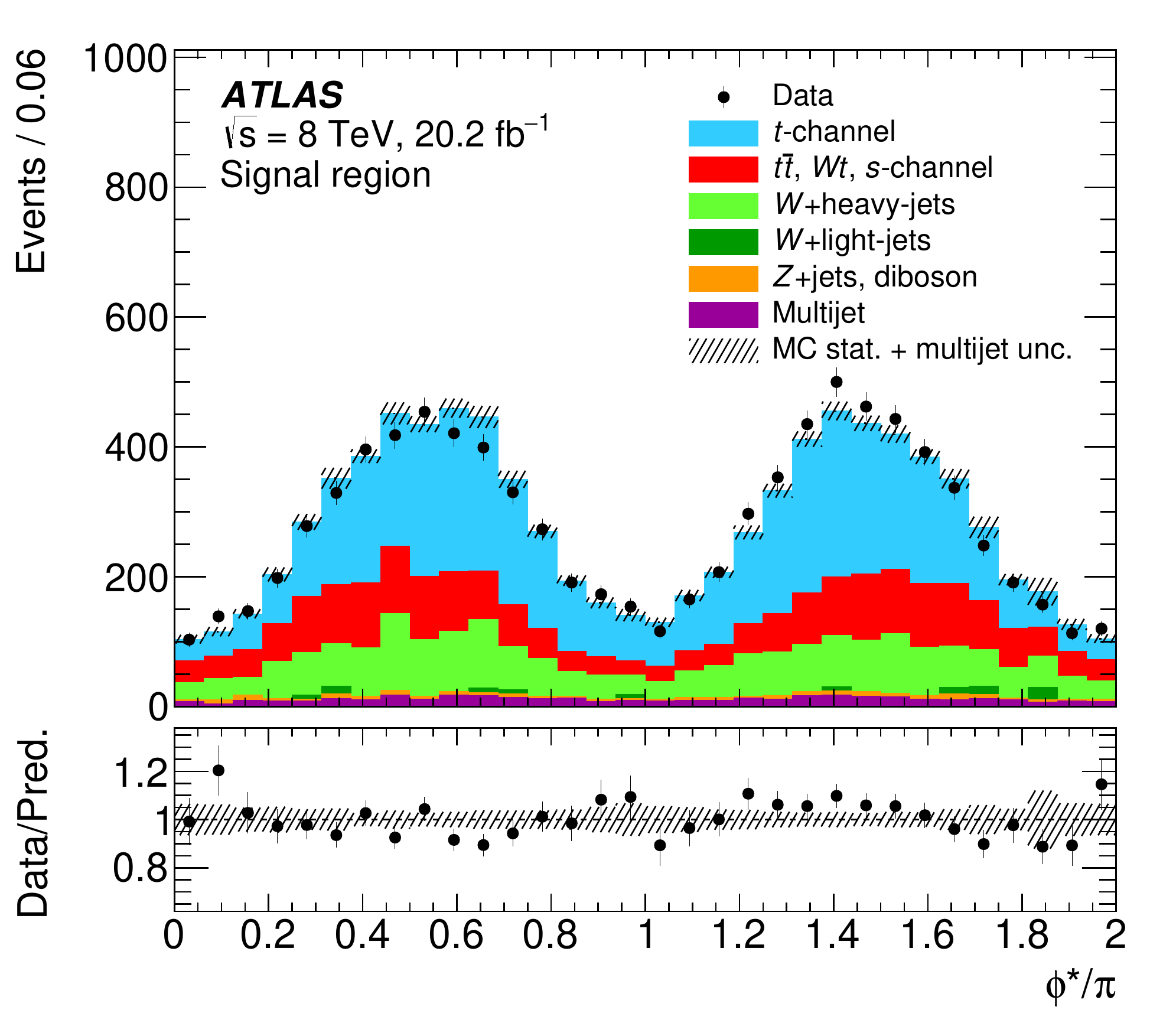}}
    }}
  \caption{Angular distributions of (a) $\cosTheta$, (b) $\cosThetaS$
    and (c) $\phiS$ in the signal region for the electron and muon
    channels merged, comparing observed data,
    shown as the black points with statistical uncertainties, to SM signal
    and background predictions. The \qcd\ background is estimated using data-driven techniques, while contributions from
    simulated $W$+jets, top-quark backgrounds and \tch\ event samples are normalised to the results of a
    maximum-likelihood fit to event yields in the signal and control
    regions. The uncertainty bands correspond to
    the uncertainties due to the size of the simulated event samples added
    in quadrature with the data-driven normalisation uncertainty of 70\% estimated for the \qcd\
    contribution. The lower plots show the ratio of data to prediction in each bin.}
  \label{fig:SR_angular_plots_theta_thetaS_phiS_merged}
\end{figure}

The model introduced in \Section{sec:triplediffrate} is based on
the angles $\theta$, $\thetaS$ and $\phiS$. The distributions of these
angular observables, for events satisfying the signal selection
criteria, are shown in \Figure{fig:SR_angular_plots_theta_thetaS_phiS_merged}.
Isolation requirements placed on the leptons influence the shape
of these angular distributions.
Thus from \Figure{fig:axes}
one can see that for $\cosTheta = -1$, the spectator jet overlaps with the $b$-tagged
jet. Similarly, for $\cosThetaS = -1$, the lepton overlaps with the
$b$-tagged jet. Therefore, in both cases, the acceptance is significantly reduced.
For $\cosTheta = +1$, the acceptance is maximal since the spectator jet and the $b$-tagged jet are back-to-back. 
For $\cosThetaS = +1$, although the lepton and the $b$-tagged jet are back-to-back, the
acceptance is not maximal since the lepton is in the same
plane as the spectator jet and therefore it may overlap with
this jet. For $\phiS = 0$, $\pi$ or $2\pi$, the lepton is in the
same plane as the spectator jet and therefore it may overlap with this
jet. This is disfavoured by the isolation criteria, so acceptance reduces
in these three regions. 
Acceptance is maximal for
$\phiS = \pm\pi\text{/2}$, since the lepton is in a plane perpendicular to the spectator.

Just as the angular distribution for the true signal can be expressed
in terms of the angular coefficients, $a_{k,l,m}$, of a finite series
of orthonormal functions, the reconstructed angular distribution can
be expressed as an infinite series of the same functions, similarly to
\Equation{eq:tripleDiff}:
\begin{equation}
    \varrho_\mathrm{r}(\theta, \thetaS,\phiS;\ \vec{\alpha}, P)  = \sum\limits_{\kappa,\lambda,\mu}\mathcal{A}_{\kappa,\lambda,\mu}(\params,P) M_{\kappa,\lambda}^{\mu}(\theta,\thetaS,\phiS) \,,
  \label{eq:reconstruction_model}
\end{equation}
where $|\mu|\le \mathrm{min}(\kappa,\lambda)$. Multiplying \Equation{eq:reconstruction_model} by $M_{\kappa,\lambda}^{\mu*}(\theta,\thetaS,\phiS)$, integrating, and applying the orthonormality of the $M$-functions, one projects out the angular coefficients, obtaining
\begin{equation*}
\mathcal{A}_{\kappa,\lambda,\mu}= \int  \varrho_\mathrm{r}(\theta, \thetaS,\phiS;\ \vec{\alpha}, P) M_{\kappa,\lambda}^{\mu*}(\theta,\thetaS,\phiS) \, \mathrm{d}(\cos\theta) \mathrm{d}\Omega^* \,. 
\end{equation*}
For a discrete set of data that follows $\varrho_\mathrm{r}$, the angular coefficients can be estimated as the average value of the function over the data:
\begin{equation*}
\mathcal{A}_{\kappa,\lambda,\mu} =  \langle M_{\kappa,\lambda}^{\mu*}(\theta,\thetaS,\phiS)\rangle \,,
\end{equation*}
similar to a MC estimation of an integral.
Experimental values of these coefficients can thus be obtained by taking
this average over a set of discrete data for terms up to a maximum $\kappa$ and $\lambda$, determined by the precision of the data.
A similar approach to sequential decays is suggested in Ref.~\cite{Bialas:1992ny}.
This technique, called orthogonal series density estimation (OSDE)~\cite{OSDE}, is essentially a Fourier technique to determine moments of the angular distribution.
Since $\mathcal{A}_{\kappa,\lambda,\mu} = \mathcal{A}_{\kappa,\lambda,-\mu}^*$, the coefficients with $\mu=0$ are purely real, while those with $\mu\ne 0$ can be represented by the real and imaginary components of $\mathcal{A}_{\kappa,\lambda,|\mu |} $.
These sets of reconstructed and true angular coefficients, $\mathcal{A}_{\kappa,\lambda,\mu}$ and $a_{k,l,m}$, can be represented by two vectors of coefficients, $\vec{\cal A}$ and $\vec{a}$.
A covariance matrix, ${\mathbf C}={\mathrm{Cov}}(\vec{\cal A})$, is also determined using OSDE, in the standard way by averaging products of two $M$-functions.

The background's shape and its covariance matrix are determined through an OSDE analysis of a hybrid sample consisting of background events from simulation samples, and selected data events from samples enriched in \qcd\ events as reported in \Section{sec:background_estimation}.
The vector of reconstructed and background-subtracted coefficients, $\vec{\cal A^\prime}$, is
\begin{equation*}
  \vec{\cal A^\prime}= \frac{1}{f_{\text{s}}}\vec{\cal A} - \left( \frac{1}{f_{\text{s}}} -1 \right) \vec{\cal A}_b \,,
\label{eq:backgroundCoefficientSubtract}
\end{equation*}
where $\vec{\cal A}_b$ is the vector of coefficients for the background and $f_{\text{s}}$ is the signal fraction. 
On the other hand, the covariance matrix ${\mathbf C}$ is modified to include the contribution from the background, 
\begin{equation}
{\mathbf C^\prime} = \left( \frac{1}{f_{\text{s}}} \right)^2 {\mathbf C} + \left( \frac{1}{f_{\text{s}}} -1\right)^2 {\mathbf C}_{\text{b}} \,,
\label{eq:Cprime}
\end{equation}
where ${\mathbf C^\prime}$ and ${\mathbf C}_{\text{b}}$ are the
covariance matrices of the background-subtracted coefficients and the
background coefficients alone, respectively. 
The second term in \Equation{eq:Cprime} represents a systematic uncertainty in ${\mathbf C^\prime}$ due to statistical
uncertainties in the background estimate.

Detector effects, both efficiency and resolution, are incorporated through a migration matrix that relates true coefficients, $\vec{a}$, to reconstructed and background-subtracted coefficients, $\vec{\cal A^\prime}$.  
This matrix, denoted by ${\mathbf G}$, translates all of the nine true
coefficients (not counting $a_{0,0,0}$) to the reconstructed coefficients.  
It is determined from MC samples produced with the \protos\ generator
using a Fourier analysis of the joint probability density function of true and reconstructed angles, followed
by a transformation to coefficients of a conditional probability density function. The procedure is described in more detail in Refs.~\cite{Boudreau:2013yna, Boudreau:2016pdi}. 
In terms of $\mathbf{G}$, 
\begin{equation}
\vec{\cal A^\prime} = {\mathbf G}\cdot\vec{a} \,.
\label{eq:MME}
\end{equation}

Equation~(\ref{eq:MME}) cannot be inverted in practice because the matrix
${\mathbf G}$ has more rows than columns, indicating a situation with more equations than unknown variables.  
Owing to statistical fluctuations or systematic shifts in the measured quantities, 
it is possible that they cannot all be satisfied simultaneously. 
The number of rows can be reduced by considering fewer equations. 
The higher-order terms in 
$\vec{\cal A}$ and $\vec{\cal A}_b$, of which there are an infinite number, are truncated since they represent high-frequency components bringing little information about
the true coefficients. 
In what follows, a truncation is done at
$\lambda_{\mathrm{max}}=\kappa_{\mathrm{max}}=2$
(subscript ``max'' is the maximum index value of a given series).  
The maximum values of $k$ and $l$ are chosen to obtain the optimal
statistical uncertainty in physics parameters. With this truncation
the number of background-subtracted coefficients is 18.

\begin{figure}[!htb]
  \centerline{\makebox{
      \includegraphics[width=4.1in]{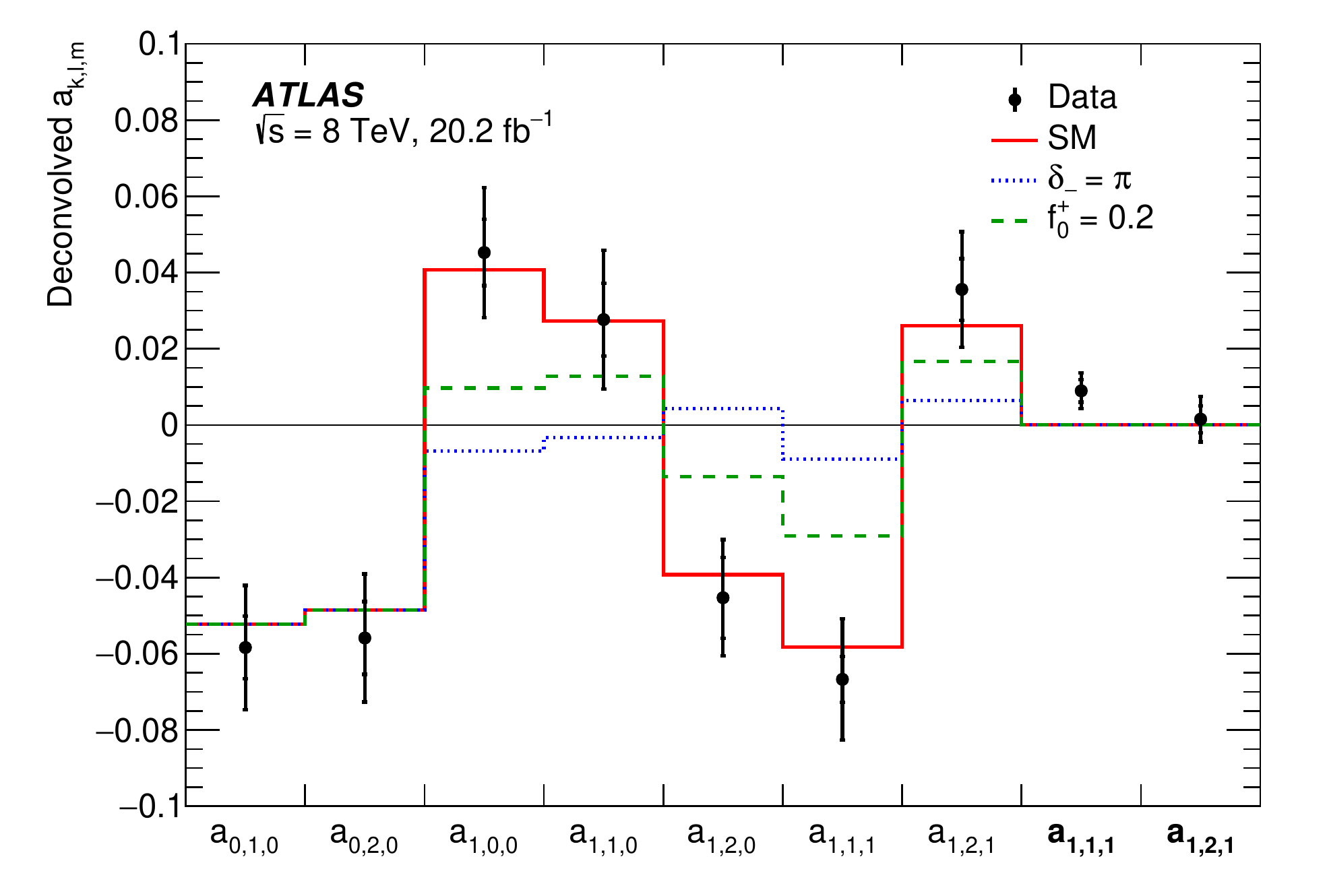}
    }}
  \caption{Deconvolved angular coefficients from data using the migration
    matrix from the SM simulation. Data are shown as black points
    with statistical uncertainties (inner error bar) and statistical and systematic
    uncertainties added in quadrature (outer error bar), while SM prediction
    is shown as a red line. In addition, two new physics scenarios,
    one with $\delm = \pi$ and another one with $\fzp = 0.2$, are also
    shown as a dotted blue line and dashed green line,
    respectively.
    The $x$-axis shows the real and imaginary parts of the angular
    coefficients, where the latter appears in boldface.}
  \label{fig:coefficientsFromData_NP}
\end{figure}

Since a covariance
matrix, ${\mathbf C^\prime}={\mathrm{Cov}}(\vec{\cal A^\prime})$, is available, one can minimise the function
\begin{equation*}
  \chi^2(\vec{a}) = \left( \vec{\cal A^\prime}-{\mathbf G}\cdot\vec{a}
  \right)^\mathrm{T}\cdot ({\mathbf C}^\prime)^{-1} \cdot \left( \vec{\cal
      A^\prime}-{\mathbf G}\cdot\vec{a} \right) \,,
  \label{eq:decoChi2}
\end{equation*}
over the vector $\vec{a}$.  This can be done analytically, and yields the solution 
\begin{equation}
  \vec{a} = {\mathbf V} \cdot {\mathbf G}^\mathrm{T}  \cdot ({\mathbf C}^\prime)^{-1} \cdot \vec{\cal A^\prime} \,,
  \label{eq:unfoldeda}
\end{equation}
with
\begin{equation}
{\mathbf V} = {\mathrm{Cov}}(\vec{a})=  \left( {\mathbf G}^\mathrm{T} \cdot
  ({\mathbf C}^\prime)^{-1} \cdot {\mathbf G} \right)^{-1} \,.
\label{eq:decoProc}
\end{equation}
The deconvolved coefficients, using a migration matrix derived from
simulated SM event samples produced with the \protos\ generator, are
shown in \Figure{fig:coefficientsFromData_NP}. Correlations between the
different coefficients range from nearly zero to almost 70\%. Also shown are the SM
predictions, obtained from \Equation{eq:physicsCoefficients}, using SM values for $\vec{\alpha}$, and a \protos\ simulation for the polarisation.
Moreover, two new physics scenarios, obtained from \protos\
simulations, are also shown. The scenario with $\delta_- = \pi$ corresponds to a region where
$\Re{\gr/\vl} \approx 0.77$, allowed by the fit in measurements of $W$
boson helicity fractions in top-quark decays~\cite{Aad:2012ky,Chatrchyan:2013jna, Aaboud:2016hsq, Khachatryan:2016fky,Khachatryan:2014vma}.
The scenario with $\fzp=0.2$ corresponds to a set of couplings
($|\vr/\vl| \approx 0.65$, and $|\gl/\vl|\approx 0.27$) that are also
consistent with measurements of $W$ boson helicity fractions, but where $20\%$ of the longitudinal $W$ bosons are due to right-handed couplings.

The derivation of the migration matrix, $\mathbf{G}$, and background model, $\vec{\mathcal{A}}_b$, described above, is based on the form of 
these distributions in MC simulation.
For the background model, constructed from the sum of all predicted backgrounds with an appreciable effect on the 
distribution, this includes events containing top quarks, primarily from \ttbar\ production, the distribution of which is affected by changing the values of the 
anomalous couplings. The efficiency and resolution models are averages over all unmeasured distributions in the signal. Variations in the values of anomalous couplings
alter those unmeasured distributions, which could lead to a dependence on these couplings for the efficiency and resolution models. For instance, \tch\ single-top-quark production depends on anomalous couplings in both the top-quark production and decay vertices, so varying the couplings alters production-side 
distributions, such as the \pT\ and $\eta$ distributions of the top or spectator quark. 
Therefore $\mathbf{G}$ and $\vec{\mathcal{A}}_b$ both depend upon $\vec{\alpha}$. 
When evaluating $\vec{a}$ for different possible values of $\vec{\alpha}$, the appropriate values of
$\mathbf{G}(\vec{\alpha})$ and $\vec{\mathcal{A}}_b(\vec{\alpha})$  must be used.
Consequently, $\vec{a}$ also depends on $\vec{\alpha}$.

To interpret the measurement of the coefficients $\vec{a}(\vec{\alpha})$ as a measurement
of the parameters $\vec{\alpha}$, the real and imaginary parts of 
the predicted coefficients $a_{k,l,m}$ obtained from \Equation{eq:physicsCoefficients} are packed into a vector
$\vec{a}_{\mathrm{th}}$.
The coefficient 
$a_{0,0,0}$ is omitted in this procedure because it is constrained by normalisation.  
Since the number of parameters used to describe the complex coefficients 
$\text{dim}(\vec{a})=9$ exceeds $\text{dim}(\vec{\alpha})=6$, an 
over-constrained system is found.
Using $\vec{a}(\vec{\alpha})$ from \Equation{eq:unfoldeda} and $\mathbf{V}$ from \Equation{eq:decoProc}, an additional $\chi^2$ contribution is defined as 
\begin{equation}
\chi^2 (\vec{\alpha})= \left( \vec{a}_{\mathrm{th}}(\vec{\alpha}) - \vec{a}(\vec{\alpha})\right)^\mathrm{T} \cdot \mathbf{V}^{-1} \cdot \left( \vec{a}_{\mathrm{th}}(\vec{\alpha}) - \vec{a}(\vec{\alpha}) \right) \,.
\label{eq:propagationToGHF}
\end{equation}
The final fit uses the combined likelihood
\begin{equation}
-2\ln\mathcal{L} = \chi^2 (\vec{\alpha}) + \chi^2(\vec{a}) \,.
\label{eq:finallikelihood}
\end{equation}
Likelihood profiles over the parameters $\vec{\alpha}$ are computed using a Markov chain MC method~\cite{brooks2011handbook}.
In order to correct for the dependence of ${\mathbf G}$ on $\vec{\alpha}$, the migration matrix is computed on a four-dimensional grid
in $\fu$, $\fup$, $\fzp$, and $\delm$ using Lagrange interpolation
between the grid points.  Two points are used in  $\fup$, $\fzp$,
while four are used in $\fu$ and $\delm$. The range of interpolation
is $\fu \in [0.24, 0.36]$, $\fup \in [0.0, 0.25]$,  $\fzp \in [0.0,
0.25]$, and $\delm \in [-0.5, 0.5]$. The background coefficients
$\vec{\mathcal{A}}_b$ are also corrected for the dependence of the
\ttbar\ background on $\vec{\alpha}$ in the same manner.

The procedure for deconvolving detector effects has been validated
with closure tests, performed using simulation samples produced with
the \protos\ and \acermc\ generators. The model independence of this
procedure has been validated using the various simulation samples with anomalous couplings
enabled in both the production and the decay vertices, as mentioned in \Section{sec:data_mc_samples}.

\section{Sources of systematic uncertainty}
\label{sec:systematics}

Systematic uncertainties are estimated for the angular coefficients
$a_{k,l,m}$. The systematic uncertainties are better behaved in these angular coefficients than in the parameters $\vec{\alpha}$, where they might be close to physical boundaries, e.g. $\fup=0$ or $\fzp=0$.
These systematic uncertainties are used to construct a $9\times 9$ covariance matrix including all correlations between different angular coefficients for each uncertainty considered.
The full systematic covariance matrix, ${\mathbf V}_{\text{syst}}$, is then formed by summing the individual matrices.
For evaluating the likelihood including the total uncertainty, ${\mathbf V}_{\text{syst}}$ is added to the covariance matrix determined from \Equation{eq:decoProc} before evaluating \Equation{eq:propagationToGHF}.

Unless addressed specifically, the efficiency and resolution models
(i.e. migration matrix) in \tch\ events used to estimate
the impact of the various sources
of uncertainty on the deconvolved measurements are those extracted from the nominal
simulation sample produced with the \protos\
generator and SM couplings. The nominal acceptance and template shape
of the \tch\ signal is predicted using the \powheg\ generator.
Various signal and background models are determined
from MC simulation samples with either alternative generators or
parameters varied by their uncertainty in order to estimate systematic uncertainty from different sources. For each source, a likelihood is constructed from the
resulting background-subtracted-data model, using events generated
with varied parameters. 
The difference is calculated between the central values estimated
at the nominal value of a parameter and at the value varied by its
uncertainty, or half the difference between central values estimated
with the parameter varied up and down by its uncertainty.
These differences are used to
construct a covariance matrix for each source of systematic
uncertainty. The total covariance matrix for the systematic
uncertainties and its correlation matrix are found from the sum of the
covariance matrices determined for individual uncertainties.

When estimating the impact of the various sources
of uncertainty, the variations are propagated in a correlated way to
the rates and to the shapes. The variations due to the systematic
uncertainties are also propagated in a correlated way to the signal
region and to the two control regions used to constrain the top-quark
and $W$+jets background contributions. For the statistical
uncertainties, the variations in the signal and control regions are
considered as independent. A set of overall scale factors associated
with the top-quark and $W$+jets backgrounds and with the signal events
are extracted for each source of systematic or statistical variation,
through the procedure explained
in \Section{sec:background_estimation}. 
The background normalisation is obtained for each systematic
uncertainty shift before being subtracted from the observed data. Then the systematic and
statistical uncertainties in the fitted normalisation factors are
propagated to the measurement.

The sources of systematic uncertainty are split into the following
categories:

\textbf{Detector modelling:} The systematic uncertainties in the reconstruction, and
energy calibration of electrons and jets and momentum calibration of muons are propagated in
the analysis through variations in the modelling of the detector
response. Uncertainties related to leptons come from trigger, identification and
isolation efficiencies, as well as from the energy or momentum scale and
resolution~\cite{Aad:2014fxa, Aad:2014rra}. For jets, the main
source of uncertainty is the jet energy
scale (JES), evaluated using a combination of in situ
techniques~\cite{Aad:2014bia}.
Other jet-related uncertainty sources
are the modelling of the energy resolution~\cite{Aad:2012ag} and
reconstruction efficiency~\cite{Aad:2014bia}, the JVF efficiency~\cite{Aad:2015ina}, and the modelling of the
tagging efficiencies of $b$-quark jets, $c$-quark jets and light-quark
jets~\cite{ATLAS:2014pla, ATLAS:2014jfa}. The uncertainties from the
energy or momentum scale and resolution corrections applied to leptons and jets
are propagated to the computation of the \MET. The scale and
resolution uncertainties due to soft jets and to contributions of
calorimeter energy deposits not associated with any reconstructed
objects are also considered independently. For all detector
modelling uncertainties, positive and negative uncertainties are
estimated separately from the corresponding shifts.

\textbf{Background normalisation:} The uncertainties in the normalisation of
the top-quark and $W$+jets background processes are determined from
the scale factor obtained from the maximum-likelihood fit to data. For the top-quark background processes, the
statistical post-fit uncertainty of 1\% in its overall scale factor is
considered. For the $W$+jets background process, the difference
between its nominal overall scale factor and the one estimated when
constraining the scale factor of the \tch\ contribution to 1.0 in the maximum-likelihood
fit (3\%) is considered. For the $Z$+jets and diboson
processes, a normalisation uncertainty of 34\% is applied to the
predictions. For the data-driven normalisation of
the \qcd\ background the uncertainty of 70\% estimated from the
comparison of the matrix method estimates with those given by the
jet-electron and anti-muon methods is used. The uncertainty in the
integrated luminosity is 1.9\%~\cite{Aaboud:2016hhf} and it is
propagated through the normalisation of the simulated background events.

\textbf{Signal and background modelling:} Systematic uncertainties
associated with the signal and background modelling are estimated by
comparing different generators and by varying parameters in the event
generation. The uncertainty in the predicted efficiency and resolution models for
the \tch\ single-top-quark process, used to deconvolve
reconstructed quantities (from \powheg\ interfaced to \pythia), is
estimated by comparing the nominal \protos\ with \acermc, both
interfaced to \pythia. This uncertainty also accounts for the
difference between models which consider the 4FS in \protos\
and the 5FS+4FS in \acermc. The uncertainty in the ME calculation in the simulation of
the \tch\ process is estimated in two ways; by comparing \protos\
with \powheg, both interfaced to \pythia, to account for the mis-modelling of
an NLO process by a LO generator, and by comparing \powheg\
with \amcatnlo\ (ver. 2.2.2)~\cite{Alwall:2014hca}, both interfaced to \herwig\
(ver. 6.5.20.2)~\cite{Corcella:2000bw} using ATLAS underlying event tune 2 (AUET2)~\cite{ATL-PHYS-PUB-2011-008}, to account for modelling differences
between NLO generators. For the \ttbar\ process, \powheg\ is compared
with \mcatnlo\ (version 4.06)~\cite{Frixione:2002ik}, both also interfaced to
\herwig\ using the AUET2 tune. The uncertainty in the PS and hadronisation is
estimated by
comparing \powheg\ interfaced with \pythia\ and \herwig\ for both the
\tch\ and \ttbar\ processes. The uncertainty in the amount of radiation is evaluated for the \tch\
and \ttbar\ processes by comparing the nominal samples with the \powheg\
samples generated with varied factorisation and renormalisation scales
(and different values of the $h_{\text{damp}}$ parameter in the case of the
\ttbar\ samples), interfaced to \pythia\ with different hadronisation
scales or configurations via alternative Perugia sets of tuned
parameters (P2012radHi, P2012radLo, P2012mpiHi and P2012loCR)~\cite{Skands:2010ak}. In this case, the uncertainty is
defined by the shift from the nominal measurement. All
these signal and background modelling uncertainties are treated as uncorrelated between \tch\ and \ttbar.

The impact of the flavour composition on the modelling of the $W$+jets
distributions is determined by propagating an uncertainty of 50\% in
the ratio of the $W$+$bb$ and $W$+$cc$ contributions. As reported in \Section{sec:eventyields_and_controlplots},
$W$+light-jets events give a small contribution in the signal region and
no associated modelling uncertainty is taken into account. An
additional shape modelling uncertainty is considered for the $W$+jets
contribution by applying an event-by-event shape reweighting
procedure. This reweighting is derived in the $W$+jets validation region from the matching to the data (after subtraction of all
processes other than $W$+jets) in the distribution of the \pT\ of the
$W$ boson.

Systematic uncertainties related to the PDF sets are evaluated for all
processes, except for the \qcd\ contribution, in a correlated way. The uncertainty is
estimated, following a procedure based on the PDF4LHC prescription~\cite{Botje:2011sn}, by calculating
a multidimensional envelope of the uncertainties at 68\% CL of the
CT10, MSTW2008 NLO and NNPDF2.3~\cite{Ball:2012cx} PDF sets. Additionally, an uncertainty due to possible non-linearities in the
polarisation, while not statistically significant, is propagated to
the final likelihood contours.

\textbf{The size of simulation samples:} The statistical uncertainty
due to the size of simulated background event samples enters through the background coefficients and is estimated during
the OSDE analysis of simulated background events. It is evaluated by subtracting, in quadrature, the covariance of the
deconvolved coefficients with and without the inclusion of the
statistical uncertainties from the background. The statistical uncertainty
due to the size of simulated signal event samples enters through the migration
matrix and is evaluated by subdividing the simulated signal event samples into
16 equally-sized subsamples. 
Migration matrices are computed for each subsample, each one being
used to deconvolve the full nominal simulation signal sample.
From the extracted values for $\vec{a}$, a covariance matrix is determined,
reflecting the size of the MC samples.

The expected statistical uncertainty due to the size of the data sample is
evaluated from pseudoexperiments. The covariance matrix is evaluated
for each experiment and the matrices are then averaged. The result is taken to be the expected covariance for the signal.  The square root of the diagonal elements are the predicted uncertainties in the coefficients. 

\Table{tab:bduncf_and_bduncg} shows the contribution of each source of systematic
uncertainty to the most sensitive helicity parameters and coupling ratios. The total systematic uncertainty is obtained by adding in
quadrature all the individual systematic uncertainties and the MC statistics uncertainties. 
Finally, the total statistical and systematic uncertainty is computed by adding all contributions in quadrature.

\begin{table}[!htb]
  \centering
  \input{tables/SystTable-PhysicsParams_and_CouplingRatios.tex}

  \caption{Statistical and systematic uncertainties in the
    measurement of helicity parameters $\fu$ and $\delm$, and of coupling ratios $\Re{\gr/\vl}$ and $\Im{\gr/\vl}$. Uncertainties from individual sources are estimated separately for
    shifts up and down, and symmetrised uncertainties $\sigma(\fu)$
    and $\sigma(\delm)$, and $\sigma(\Re{\gr/\vl})$
    and $\sigma(\Im{\gr/\vl})$ are given. 
    The statistical uncertainty is calculated by evaluating the likelihood including only the covariance matrix, ${\mathbf V}$, arising from the data statistics.
    The total uncertainty is calculated by including ${\mathbf V}_{\text{syst}}$ in the likelihood calculation as well as ${\mathbf V}$. 
    Finally, the total systematic uncertainty is computed by subtracting in
    quadrature the statistical uncertainty from the total
    uncertainty.
  }
  \label{tab:bduncf_and_bduncg}
\end{table}

The leading systematic uncertainties for \fu\ come
from the jet measurements and the generator modelling. For this
parameter, the size of the data sample is also an important source of uncertainty. In the
case of \delm, the leading systematic uncertainties are jet
measurements, the generator modelling and MC sample sizes. The
measurement of \delm\ is dominated by the statistical uncertainty in the data.
The leading systematic uncertainties for $\Re{\gr/\vl}$ and
$\Im{\gr/\vl}$ are the same as for \fu\ and \delm, respectively.

\section{Results}
\label{sec:results}

In this section, measurements, limits and distributions obtained from a
numerical calculation of the likelihood function
(\Equation{eq:finallikelihood}) are shown in the space of the generalised helicity fractions and phases $\params
\equiv \left \{ \fu, \fup, \fzp, \delp, \delm \right\}$ and $P$, or
alternatively of the anomalous couplings \vlr, \glr, and $P$. 
No external constraints or assumptions are imposed on couplings.
Values for parameters of interest can be obtained from likelihood profiles,
or joint likelihood contours which show the correlations between the extracted parameters.

\begin{figure}[!htb]
  \centering
  \centerline{\makebox{
      \subfloat[\fzp\ (stat. + syst.)]{\includegraphics[width=0.48\textwidth]{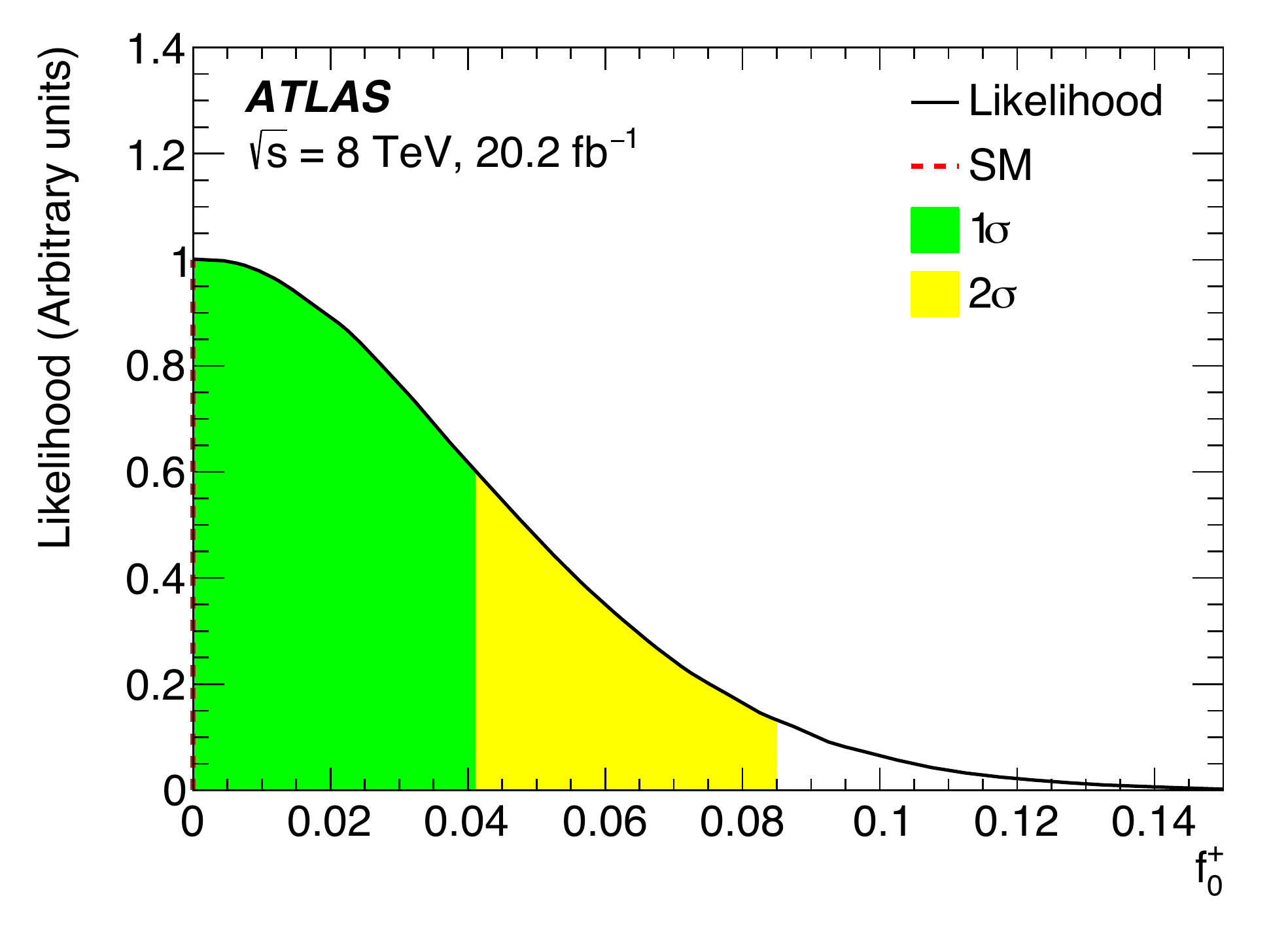}}
      \subfloat[\fup\ (stat. + syst.)]{\includegraphics[width=0.48\textwidth]{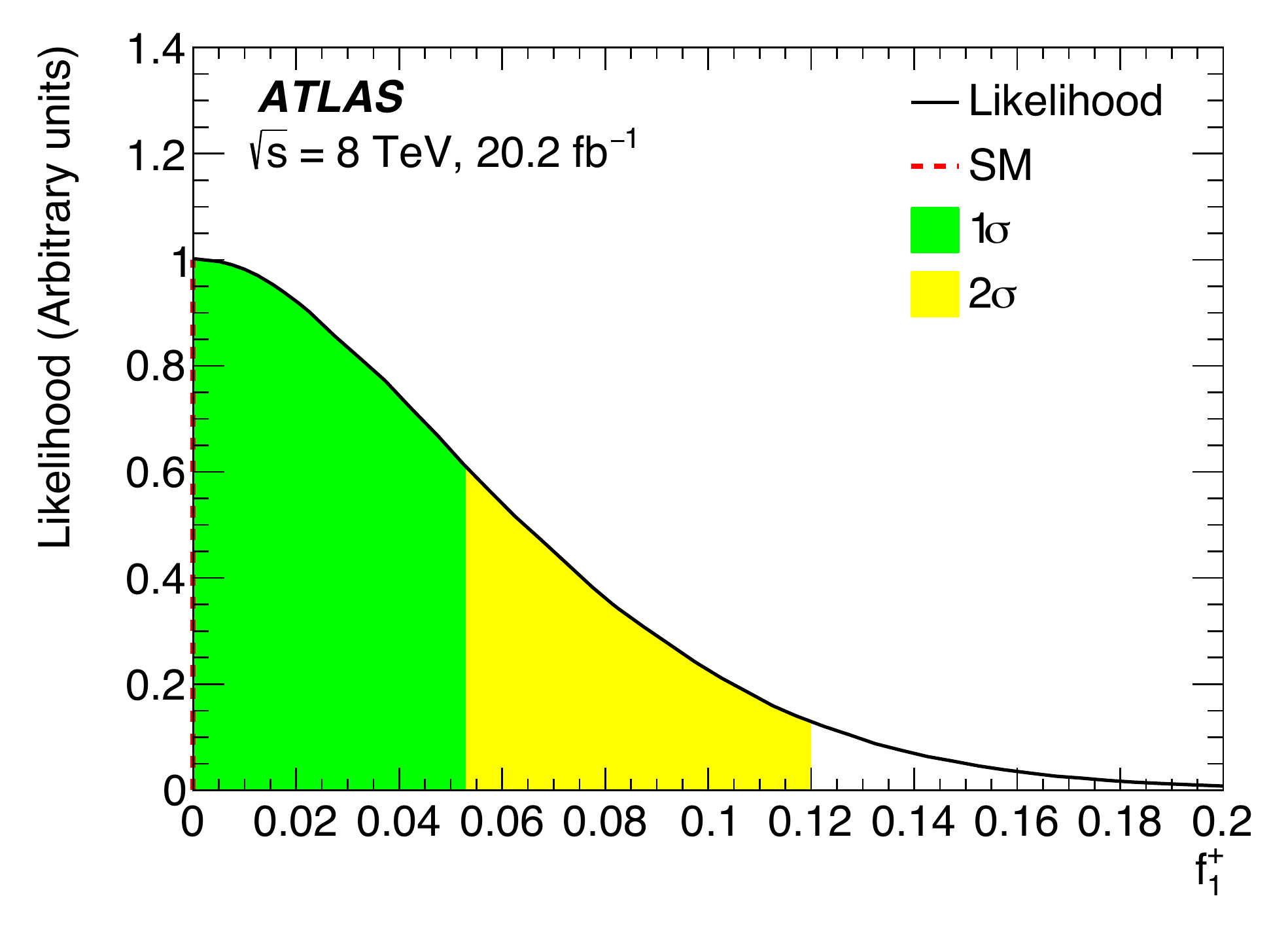}}
    }}
  \centerline{\makebox{
      \subfloat[\fzp\ vs. \fup\ (stat. + syst.)]{\includegraphics[width=0.48\textwidth]{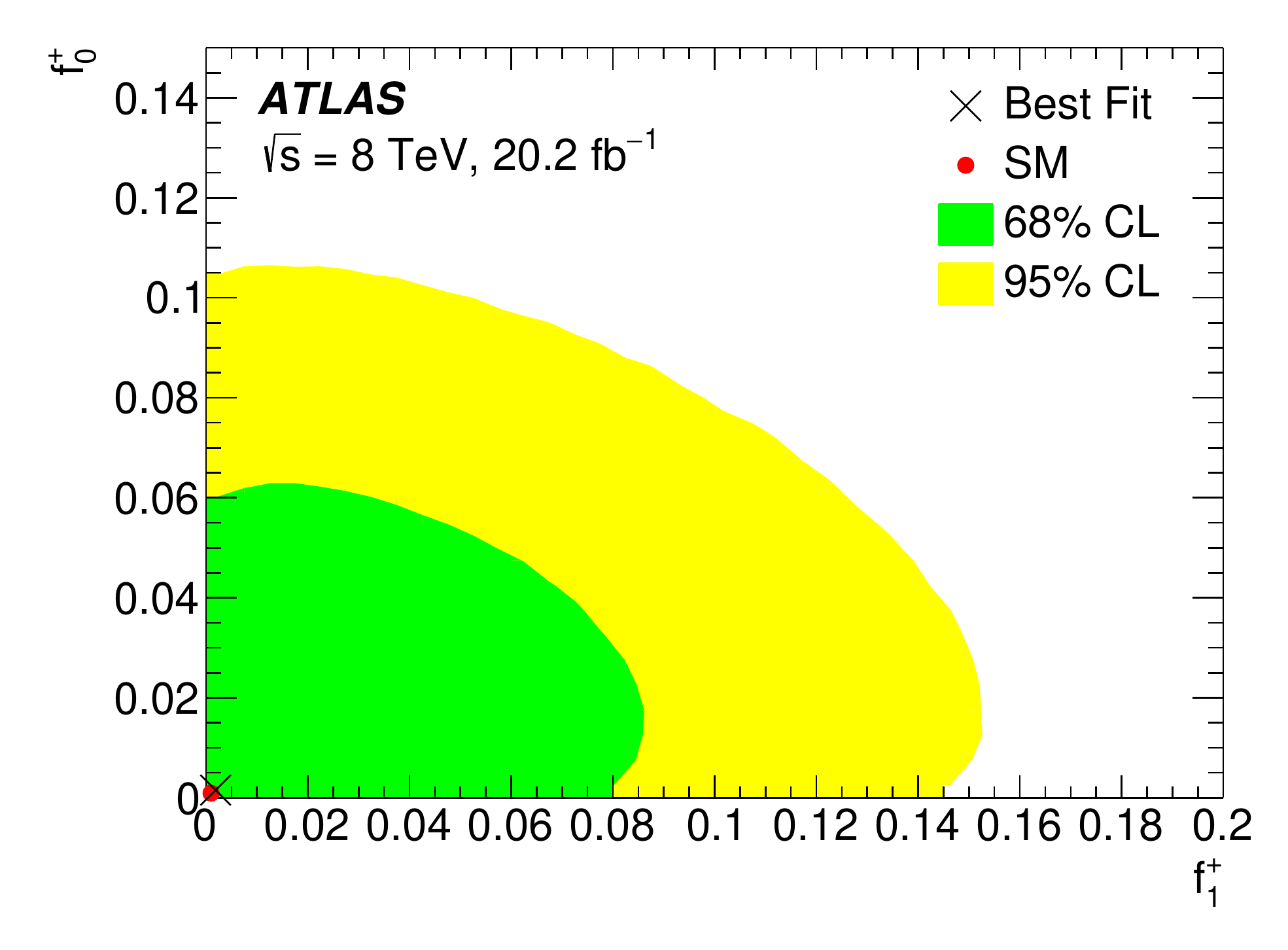}}
    }}
  \caption{The likelihood profiles for the parameters (a) \fzp\ and (b) \fup\
    are shown. The black line indicates the evaluated likelihood in each bin of the profiled variable. The
    red dashed line, which overlaps the $y$-axis, represents the SM
    expectation. Additionally (c), the joint likelihood profile of \fzp\ as a function of \fup\
    is shown. The red point represents the SM expectation while a black
    x mark indicates the observed value. Both points overlap with the
    origin of the $x$- and $y$-axis.  The 68\% and 95\%~\text{CL} regions are shown in green and yellow, respectively.}
  \label{fig:resultFZP-FUP}
\end{figure}

Likelihood profiles and a joint likelihood contour for the quantities $\fzp$ and $\fup$ are shown in
\Figure{fig:resultFZP-FUP}. The 68\% contours represent the total uncertainty in the measurement.

The limit for \fzp, i.e. for the fraction of $b$-quarks
that are right-handed in events with longitudinally polarised $W$
bosons, is
\begin{align*}
  \fzp &< \fzponesigma \qquad (68\%~\text{CL}) \,, \nonumber \\ 
  \fzp &< \fzptwosigma \qquad (95\%~\text{CL}) \,, \nonumber \\
\end{align*}
compared with the SM expectation of $\fzp = \fzpSM$. The 
limit for $\fup$, i.e. for the fraction of transversely polarised $W$
boson decays that are right-handed, is
\begin{align*}
  \fup &< \fuponesigma \qquad (68\%~\text{CL}) \,, \nonumber \\ 
  \fup &< \fuptwosigma \qquad (95\%~\text{CL}) \,, \nonumber \\ 
\end{align*}
compared with the SM expectation \fup = \fupSM.

The limits obtained for \fup\ in this analysis are comparable and complementary to those determined from $\Fp$~\cite{Aad:2012ky,
  Chatrchyan:2013jna, Aaboud:2016hsq,
  Khachatryan:2016fky,Khachatryan:2014vma}, since $\Fp=\fu\fup$.
However, the quantity \fzp\ is not accessible in measurements of the $W$ boson helicity fractions, as those analyses extract $\Fz$, 
which only measures the sum of the contributions of both longitudinal
amplitudes. The contributions can only be separated in an analysis with polarised top quarks.
Since \fup\ and \fzp\ are found to be very small, there is no sensitivity to the relative phase \delp.

\begin{figure}[!htb]
  \centering
  \centerline{\makebox{
      \subfloat[$P$ (stat. + syst.)]{\includegraphics[width=0.48\textwidth]{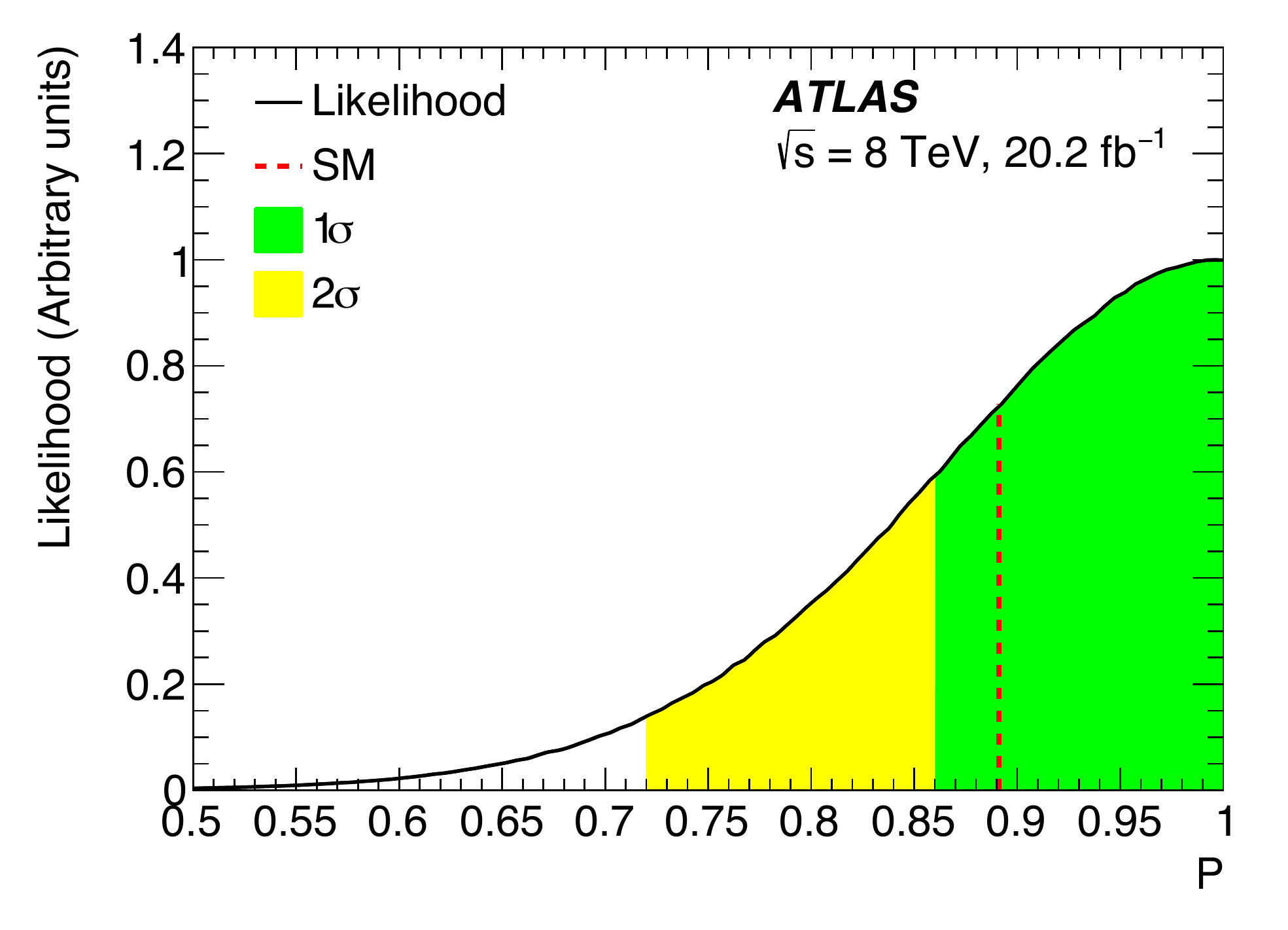}}
    }}
  \caption{The likelihood profile for the top-quark polarisation $P$ is
    shown. The black line indicates the evaluated likelihood in each bin of the profiled variable. The
    red dashed line represents the SM expectation. The 68\% and 95\%~\text{CL} regions are shown in green and yellow, respectively.}
  \label{fig:resultP}
\end{figure}

The likelihood profile for the top-quark polarisation $P$ is also obtained 
and it is shown in \Figure{fig:resultP}.
This leads to the following constraint on the top-quark polarisation:
\begin{align*}
  P &> \Ponesigma \qquad (68\%~\text{CL}) \,, \nonumber \\ 
  P &> \Ptwosigma \qquad (95\%~\text{CL}) \,. \nonumber \\\label{eq:polResult}
\end{align*}
This is compatible with the SM prediction of $P \approx \PSM$ at
\cmenergy\ as computed in
Refs.~\cite{Mahlon:1999gz,Schwienhorst:2010je,Jezabek:1994zv}, and with recent
measurements of the top-quark polarisation obtained from
asymmetries of angular distributions with additional inputs on the
values of the charged-lepton spin analysing
power~\cite{Khachatryan:2015dzz} and/or the $W$ boson helicity fractions~\cite{Aaboud:2017aqp}.

\begin{figure}[!htb]
  \centering
  \centerline{\makebox{
      \subfloat[\fu\ (stat. + syst.)]{\includegraphics[width=0.48\textwidth]{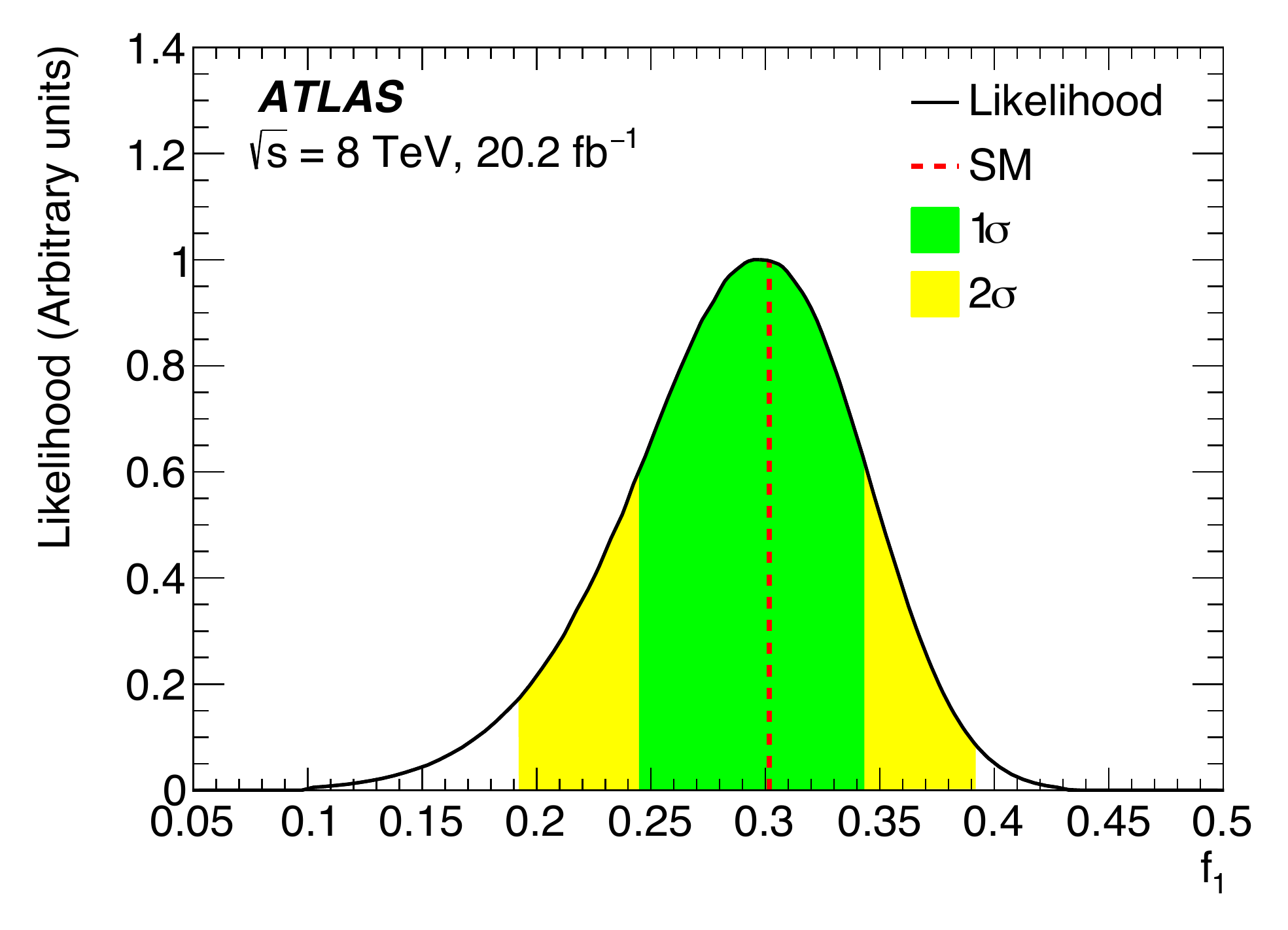}}
      \subfloat[$\delm$ (stat. + syst.)]{\includegraphics[width=0.48\textwidth]{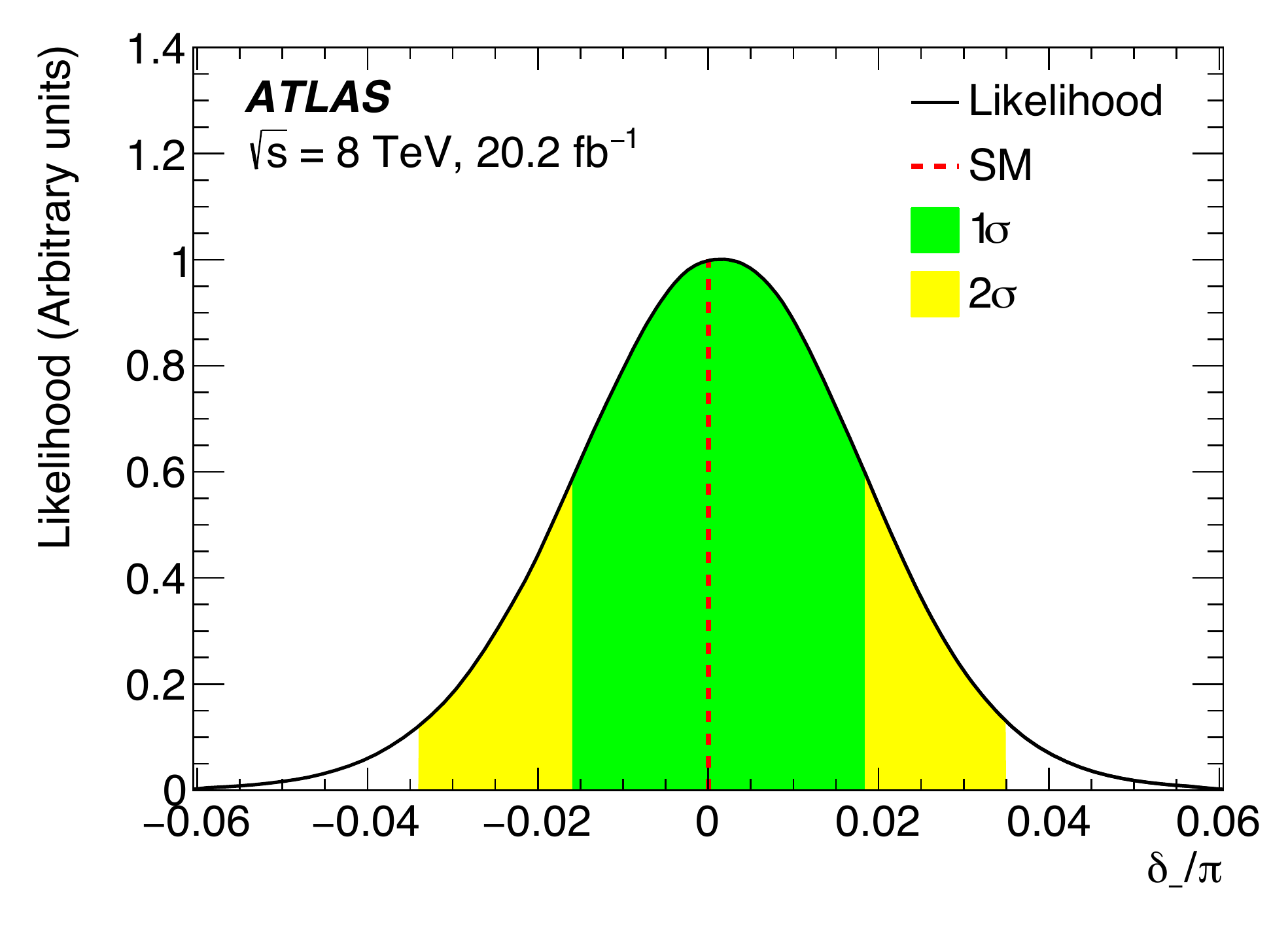}}
    }}
  \centerline{\makebox{
      \subfloat[$\delm$ vs \fu\ (stat. + syst.)]{\includegraphics[width=0.48\textwidth]{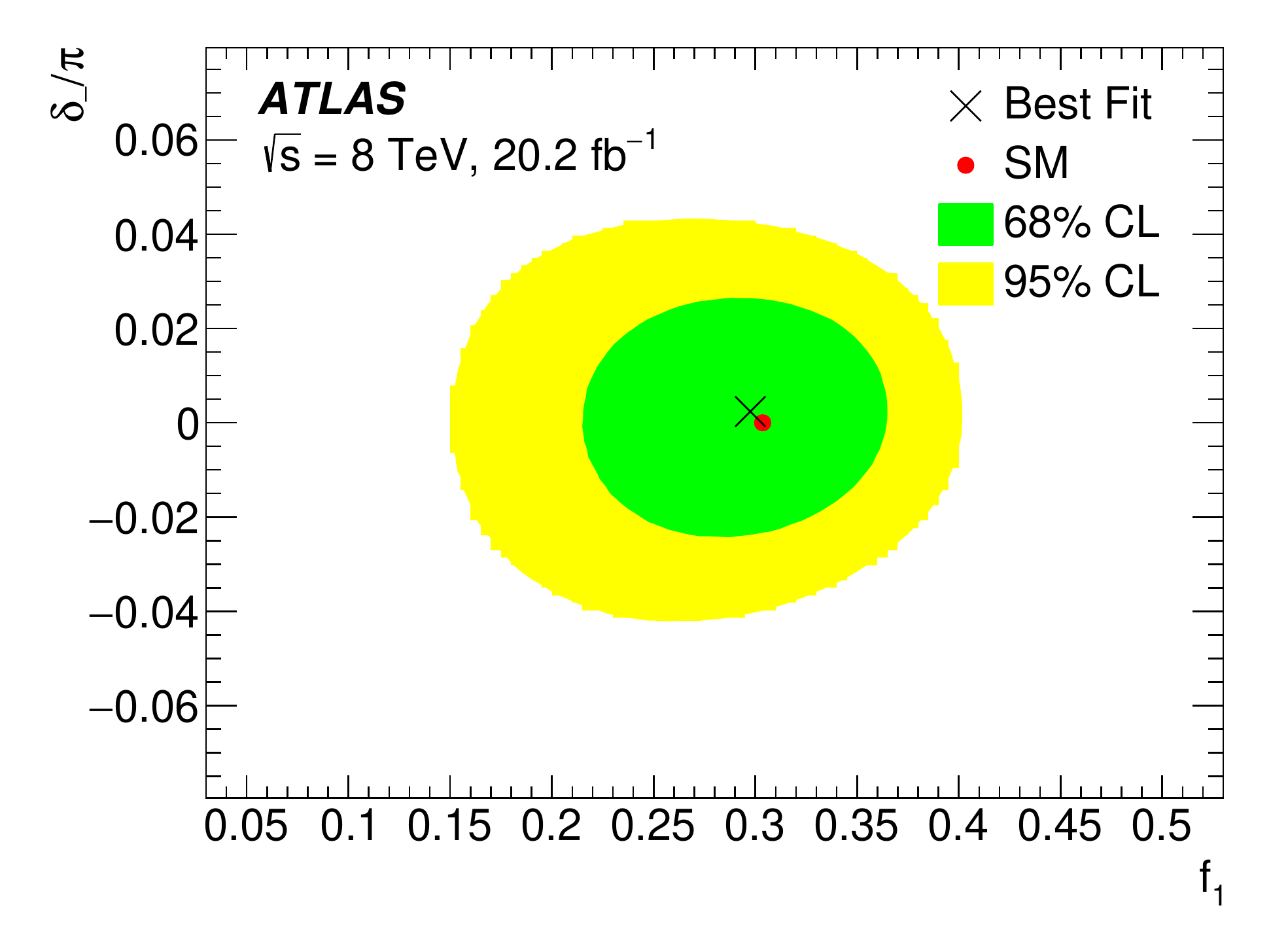}}
    }}
  \caption{The likelihood profiles for the parameters (a) \fu\ and (b) \delm\
    are shown. The black line indicates the evaluated likelihood in each bin of the profiled variable. The
    red dashed line represents the SM expectation. Additionally (c), the joint likelihood contour of \delm\ as a function of \fu\
    is shown. The red point represents the SM expectation while a
    black x mark indicates the observed value. The 68\% and 95\%~\text{CL} regions are shown in green and yellow, respectively.}
  \label{fig:resultFUPM10}
\end{figure}

For the parameters for which the analysis obtains point estimates rather than limits, i.e. the
fraction $\fu$ and the phase $\delm$ as discussed
in \Section{sec:intro}, likelihood profiles and a joint likelihood contour are
shown in \Figure{fig:resultFUPM10}. 
These parameters are measured to be
\begin{equation*}
  \begin{aligned}
    &\fu &=& \quad \cvfufullunc = \cvfutotaluncNoparenthesis \,,  \\ 
    &\delm &=& \quad \cvdelmfullunc = \cvdelmtotaluncNoparenthesis \,. 
  \end{aligned}
  \label{eq:result_parameters}
\end{equation*}
Correlations between the coefficients of \Figure{fig:coefficientsFromData_NP} are taken into account but do not lead to large correlations between these two parameters. 
The results are compatible with their SM expectations shown
in \Section{sec:triplediffrate}, and improve on the measurements from
double-differential angular decay rates done at \cmenergySeven\ by the
ATLAS Collaboration~\cite{Aad:2015yem}. 

The dependence of the parameters \fu\ and \delm\ on the top-quark
mass is evaluated using \tch, \Wt-channel, \sch, and \ttbar\ simulation samples with a
range of different top-quark masses. A linear dependence is found,
resulting from changes in acceptance at different masses, with a slope
of \fumassdep\ for \fu\ and consistent with zero for \delm. The uncertainty due to the top-quark mass dependence is not
included in the total systematic uncertainty since it has a negligible 
impact on the results.

\begin{figure}[!htb]
  \centering
  \centerline{\makebox{
      \subfloat[$|\vr/\vl|$ (stat. + syst.)]  {\includegraphics[width=0.48\textwidth]{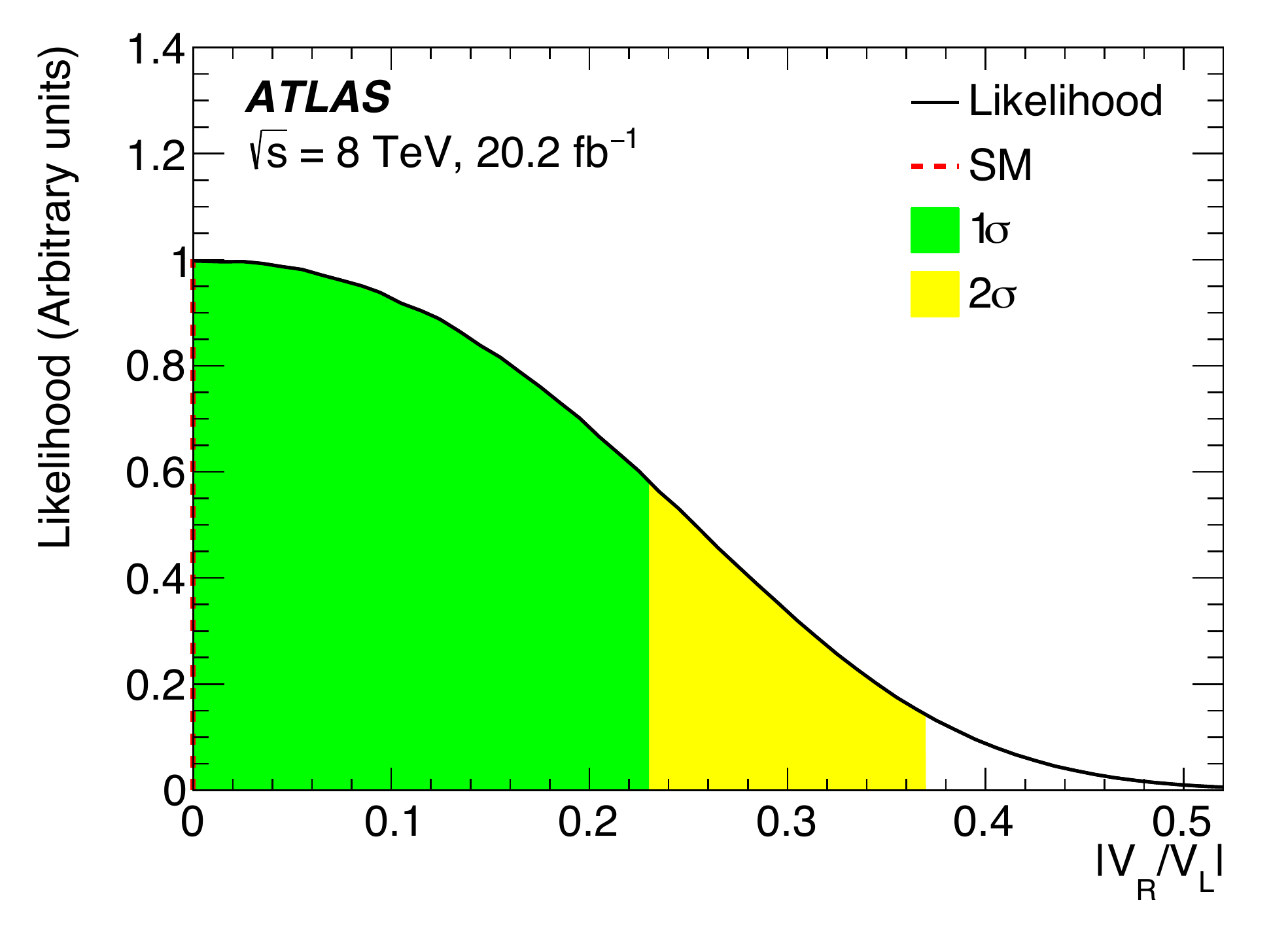}}
      \subfloat[$|\gl/\vl|$ (stat. + syst.)]  {\includegraphics[width=0.48\textwidth]{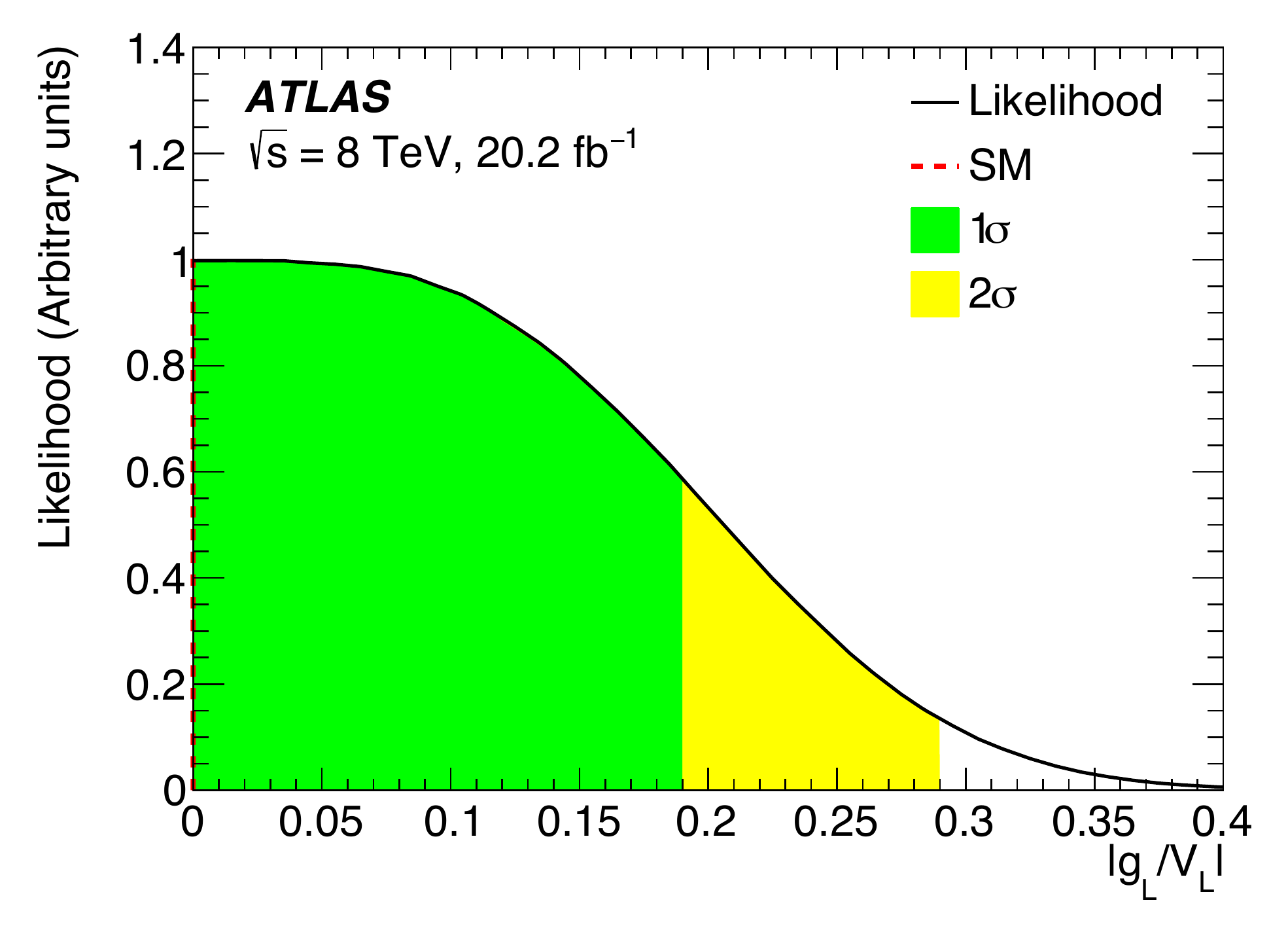}}
    }}
  \centerline{\makebox{
      \subfloat[$|\gl/\vl|$ vs. $|\vr/\vl|$ (stat. + syst.)] {\includegraphics[width=0.48\textwidth]{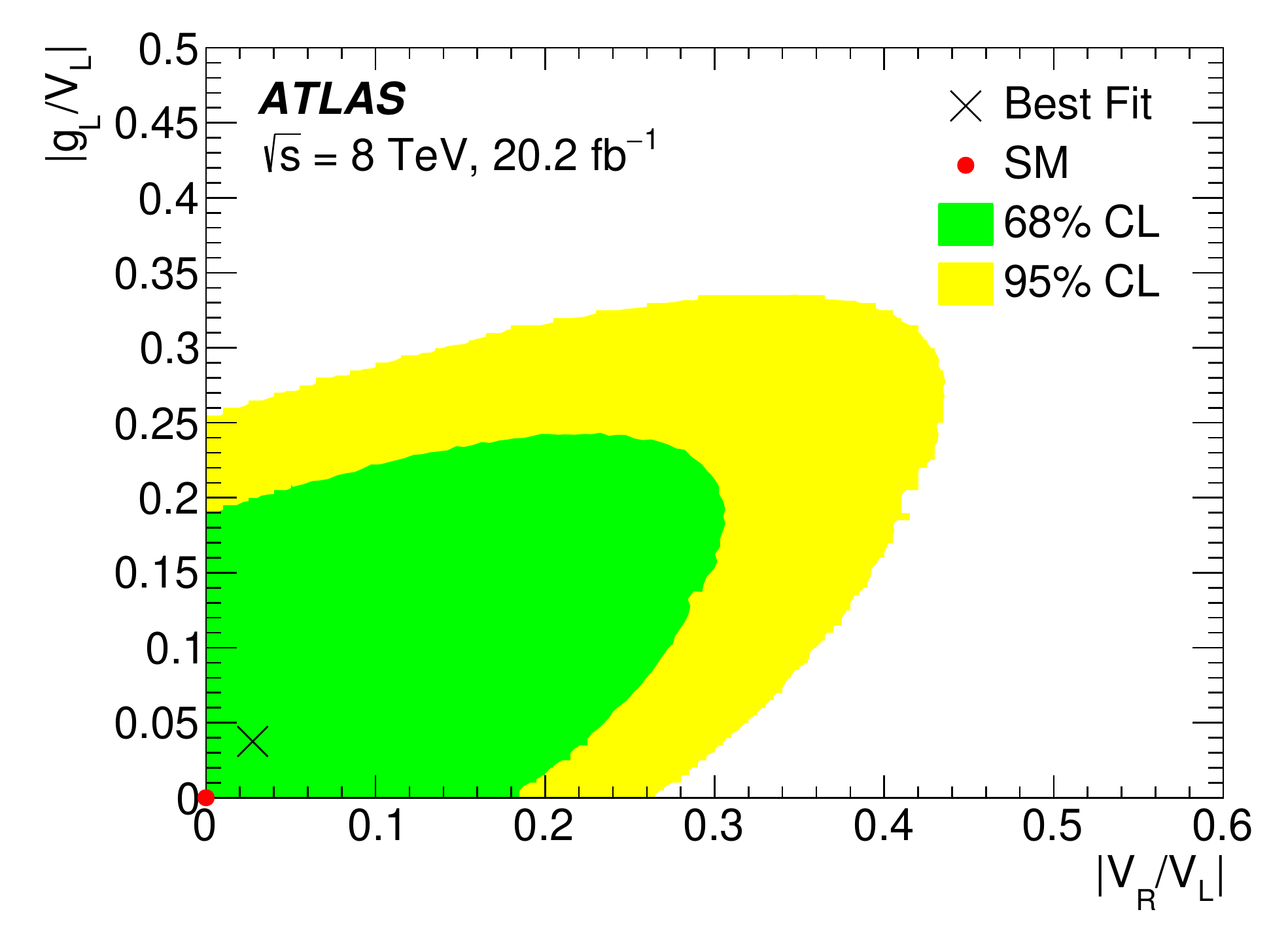}}
    }}
  \caption{The likelihood profiles for the parameters (a) $|\vr/\vl|$ and (b) $|\gl/\vl|$
     are shown. The black line indicates the evaluated likelihood in each bin of the profiled variable. The
    red dashed line, which overlaps the $y$-axis, represents the SM expectation. Additionally (c), the joint likelihood contour of $|\gl/\vl|$ as a function of $|\vr/\vl|$ is shown. The red point, which overlaps with the origin of
    the $x$- and $y$-axis, represents the SM expectation while a black
    x mark indicates the observed value. The 68\% and 95\%~\text{CL} regions are shown in green and yellow, respectively.}
  \label{fig:resultsREVR_REGL}
\end{figure}

\begin{figure}[!htb]
  \centering
  \centerline{\makebox{
      \subfloat[$\Re{\gr/\vl}$ (stat. + syst.)]  {\includegraphics[width=0.48\textwidth]{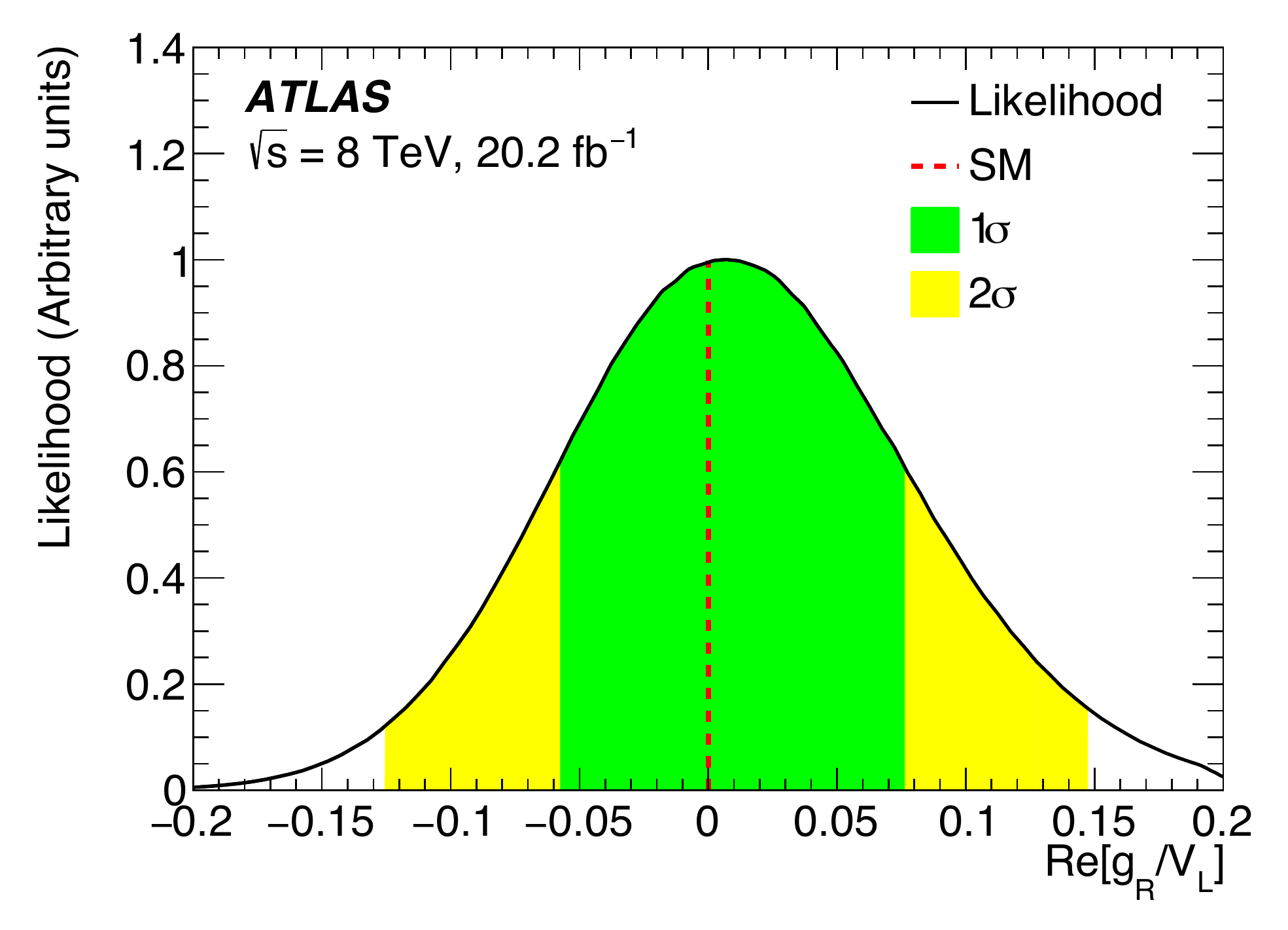}}
      \subfloat[$\Im{\gr/\vl}$ (stat. + syst.)]  {\includegraphics[width=0.48\textwidth]{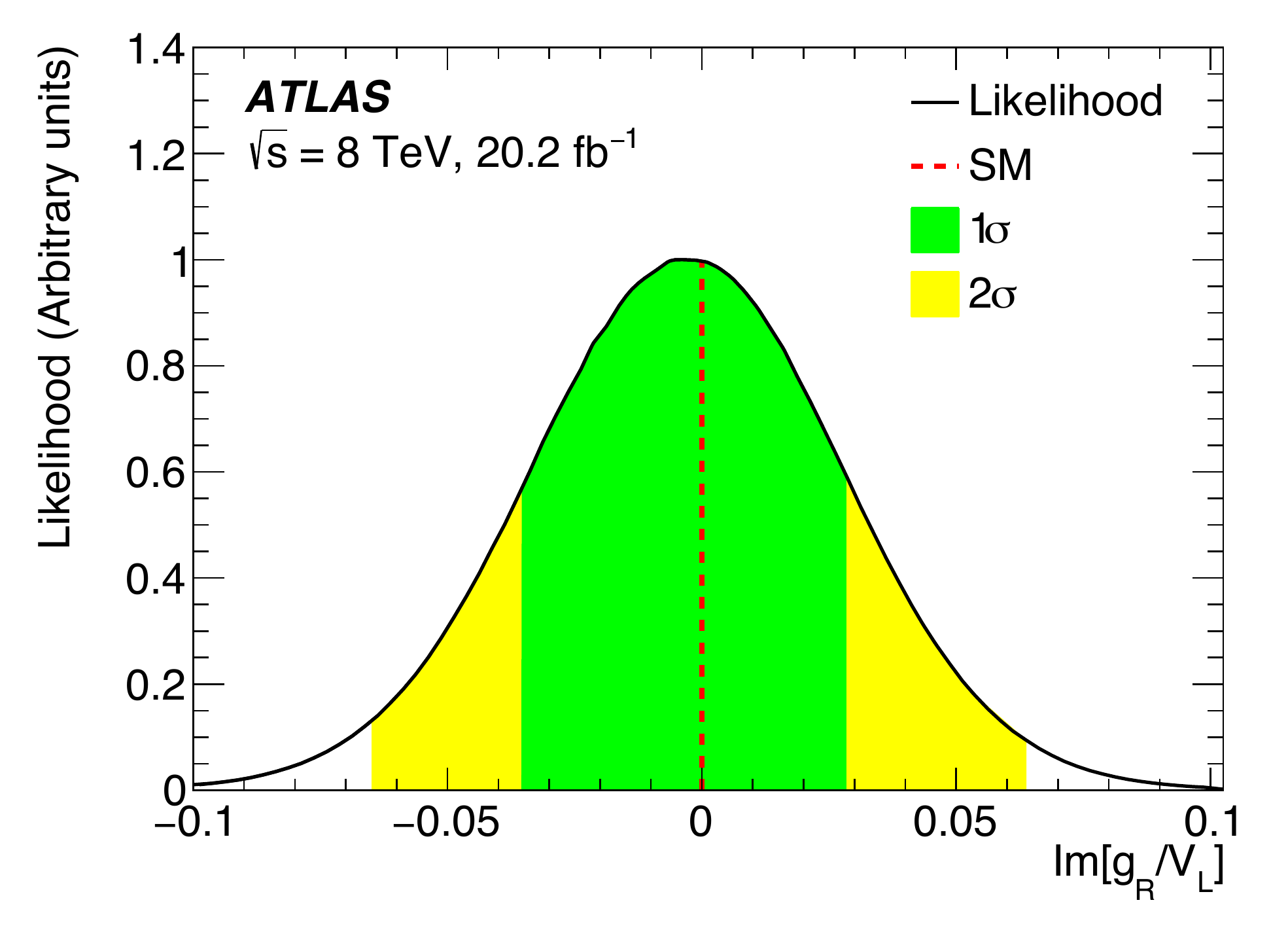}}
    }}
  \centerline{\makebox{
      \subfloat[$\Im{\gr/\vl}$ vs. $\Re{\gr/\vl}$ (stat. + syst.)] {\includegraphics[width=0.48\textwidth]{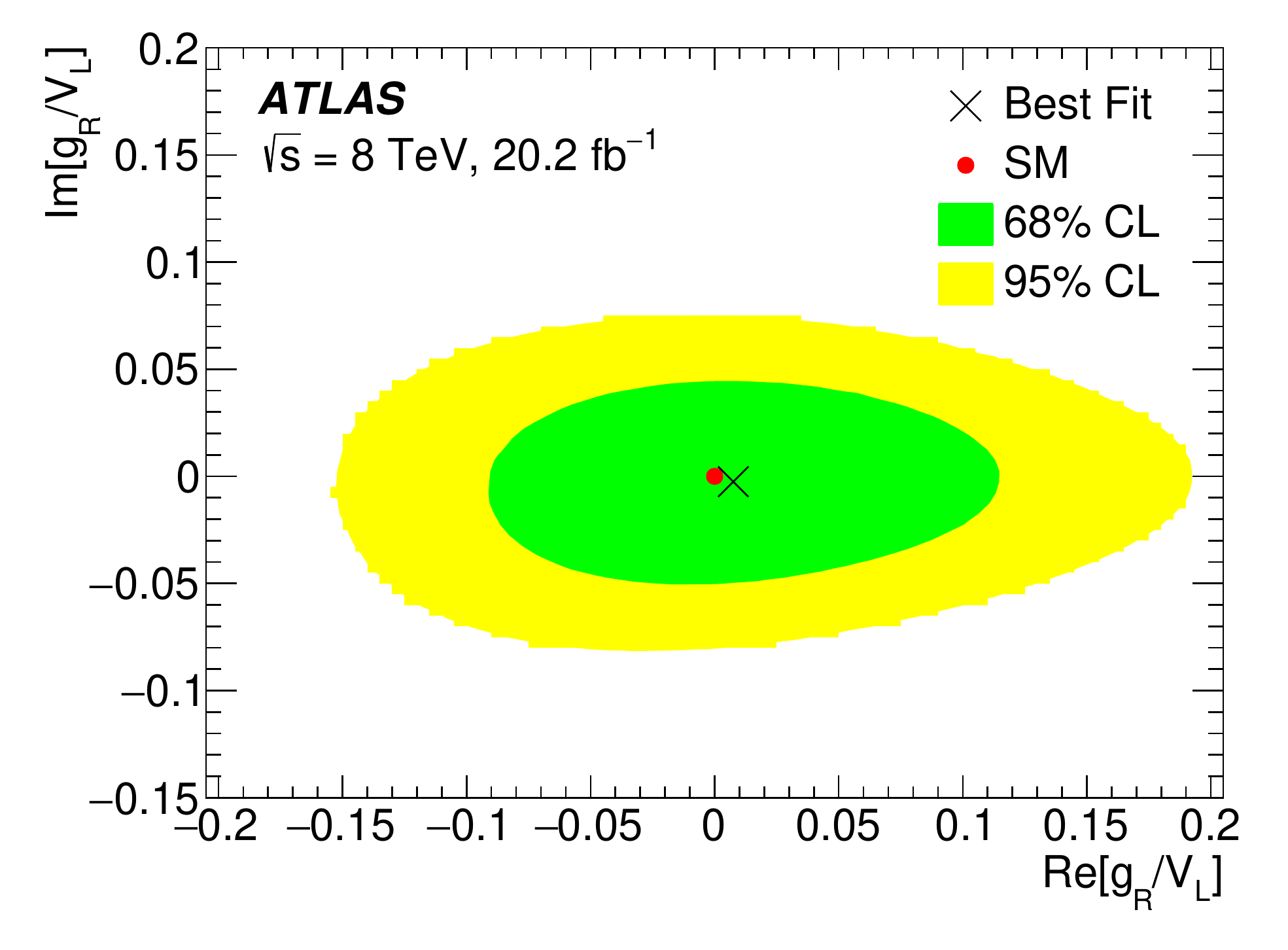}}
    }}
  \caption{The likelihood profiles for the parameters (a) $\Re{\gr/\vl}$ and (b) $\Im{\gr/\vl}$
     are shown. The black line indicates the evaluated likelihood in
     each bin of the profiled variable. The red dashed line represents
     the SM expectation. Additionally (c), the joint likelihood contour of $\Im{\gr/\vl}$ as a function of $\Re{\gr/\vl}$
    is shown. The red point represents the SM expectation while a
    black x mark indicates the observed value. The 68\% and 95\%~\text{CL} regions are shown in green and yellow, respectively.}
  \label{fig:resultsREGR_IMGR}
\end{figure}

The results for the generalised helicity fractions and phases can be interpreted in terms of anomalous couplings by propagating the statistical and systematic
uncertainties. 
Although a parameterisation of $P$ in terms of anomalous couplings, obtained from LO MC simulations, exists~\cite{Aguilar-Saavedra:2014eqa}, 
it is not included in this interpretation. 
Likelihood profiles and joint likelihood contours for these couplings are shown
in Figures~\ref{fig:resultsREVR_REGL} and~\ref{fig:resultsREGR_IMGR}. The 68\% contours represent the total 
uncertainty in the measurement. 
The normalised observables measured in this paper are sensitive to ratios of couplings, which are presented normalised to the dominant coupling in the SM, \vl.
The quantities \fup\ and \fzp\ depend most strongly on two different combinations of \vr\ and \gl, 
while the quantities $\fu (1-\fup)$ and \delm\ depend more strongly on \vl\ and \gr.
Since the likelihood is determined in terms of all of these quantities simultaneously, 
no assumptions need to be imposed on couplings in order to produce these distributions. 
In each case the measured values are consistent with the SM prediction, i.e. $\vr = \glr = 0$.

The bounds obtained on $\vr$ and $\gl$ are shown in \Figure{fig:resultsREVR_REGL}.
As this analysis yields no constraint on \delp , no constraint can be placed on the relative phase between $\vr$ and $\gl$.
Thus, only bounds on the magnitudes,
\begin{eqnarray*}
\left| \vr/\vl \right| < \vronesigma \qquad (68\%~\text{CL}) \,, \nonumber \\
\left| \vr/\vl \right| < \vrtwosigma \qquad (95\%~\text{CL}) \,, \nonumber \\
\end{eqnarray*}
and
\begin{eqnarray*}
\left| \gl/\vl \right| < \glonesigma \qquad (68\%~\text{CL}) \,, \nonumber \\
\left| \gl/\vl \right| < \gltwosigma \qquad (95\%~\text{CL}) \,, \nonumber \\
\end{eqnarray*}
are obtained.
Limits on these quantities have been obtained from $B$-meson decays~\cite{Grzadkowski:2008mf},
and from measurements of $W$ boson helicity fractions in top-quark
decays~\cite{Aad:2012ky, Chatrchyan:2013jna, Aaboud:2016hsq, Khachatryan:2016fky,Khachatryan:2014vma}, but all of those measurements can only place limits on combinations of couplings, and thus the quoted limits on individual couplings depend on the assumptions made about other couplings.

The propagation of the uncertainties to the $(\Re{\gr/\vl},\Im{\gr/\vl})$ space gives 
\begin{equation*}
  \begin{aligned}
    &\regrvl &=& \quad \cvregrfullunc = \cvregrtotaluncNoparenthesis \,, \\
    &\imgrvl &=& \quad \cvimgrfullunc = \cvimgrtotaluncNoparenthesis \,. 
  \end{aligned}
  \label{eq:result_couplings}
\end{equation*}

A linear dependence is found for the  coupling ratios on the top-quark
mass, which is evaluated with the top-quark mass-varied samples
mentioned before. 
A slope
of \regrmassdep\ is found for $\Re{\gr/\vl}$, while the slope is 
consistent with zero
for $\Im{\gr/\vl}$. 
Similarly to \fu\ and \delm, the uncertainty due to the top-quark mass dependence is not
included in the total systematic uncertainty since it has no significant impact on the results.

Confidence intervals are placed simultaneously on the values of the ratio of the anomalous couplings \gr\ and \vl\ at 95\%~CL, 
\begin{equation*}
  \regrvl\in\regrvllimits \quad \mathrm{and} \quad \imgrvl\in\imgrvllimits \,. 
  \label{eq:result_limits}
\end{equation*}

The best constraints on $\regr$ derive from measurements of the $W$
boson helicity fractions in top-quark pair decays, with $\regr \in [-0.02, 0.06]$ and $[-0.08,0.07]$, both at 95\%~CL, from ATLAS at \cmenergy~\cite{Aaboud:2016hsq} 
and from CMS at \cmenergySeven~\cite{Chatrchyan:2013jna}, respectively. However,
these limits use the measured single-top-quark production cross-section~\cite{Aad:2014fwa,Chatrchyan:2012ep} along with the
assumption that $\vl=1$, $\imgr = 0$, and either $\gl=0$ or $\vr=0$. 
Without these assumptions only a circular region in the complex $\gr$ plane within $0.0 \lesssim
\Re{\gr/\vl} \lesssim 0.8$ can be excluded by $W$ boson helicity
fractions measurements. The measurements presented here require no
assumptions in values of the other anomalous couplings, and on their own can exclude large values of $\Re{\gr/\vl}$.

Along these lines, from the double-differential angular
decay rates in \tch\ single-top-quark events in ATLAS at \cmenergySeven~\cite{Aad:2015yem},
confidence intervals are placed simultaneously on the coupling ratios, $\Re{\gr/\vl}
\in \regrvllimitstwoangle$ and $\Im{\gr/\vl} \in \imgrvllimitstwoangle$, at 95\%~CL,
assuming $\vr=\gl=0$. Furthermore, slightly better limits on the imaginary part of \gr\ are set
from asymmetries by ATLAS at \cmenergy, giving $\Im{\gr} \in \imgrvllimitspolarisation$~\cite{Aaboud:2017aqp}, at 95\%~CL,
assuming again $\vr=\gl=0$. The limits presented in this paper improve on both these results
and extend current constraints on $\gr$ to the whole
complex plane by simultaneously measuring 
information about $\Re{\gr/\vl}$ and
$\Im{\gr/\vl}$.

\FloatBarrier

\section{Conclusion}
\label{sec:conclusion}

The analysis presented in this paper uses the triple-differential decay 
rate in electroweak production and subsequent decay of single top
quarks to constrain the complex parameters of the effective Lagrangian that describes the properties
of the \Wtb\ vertex. An analysis of angular distributions of the decay
products of single top quarks produced in the \tch\ constrains these
parameters simultaneously. The analysis is based on \lumiInInvFb\ of $pp$
collision data at \cmenergy\ collected with the ATLAS 
detector at the LHC. The selected events contain one isolated electron or muon, large \MET, and exactly two jets, with one
of them identified as likely to contain a $b$-hadron. A cut-based
analysis is used to discriminate the signal events from background, and the electron and muon channels
are merged. An OSDE technique is used to perform an angular analysis of the triple-differential
decay rate in order to determine six observables simultaneously, i.e.~five generalised helicity fractions and phases, 
as well as the polarisation of the produced top quark.
Detector effects are deconvolved from data using Fourier techniques. 
The fraction $\fu$ of decays containing transversely polarised $W$ bosons is 
measured to be  $\fu=\cvfutotaluncNoparenthesisStandalone$.
The phase $\delm$ between amplitudes for transversely and longitudinally
polarised $W$ bosons recoiling against left-handed $b$-quarks, is measured to be 
$\delm = \cvdelmtotaluncNoparenthesis$, giving no indication of CP violation. 
The fractions of transverse and longitudinal $W$ bosons accompanied by right-handed 
$b$-quarks are also constrained at 95\% CL to $\fup < \fuptwosigma$
and $\fzp < \fzptwosigma$, respectively. The fractions \fu\ and \fup\ are related to the $W$ boson helicity
fractions ($\Fp$, $\Fz$, and $\Fm$), while the fraction \fzp, which is
previously unmeasured, separates \Fz\ into two components involving
left- and right-handed $b$-quarks. Based on these measurements, 95\% CL intervals are placed on the ratio of the 
complex coupling parameters \gr\ and \vl\ such that 
$\Re{\gr/\vl} \in \regrvllimits$ and $\Im{\gr/\vl} \in
\imgrvllimits$. Constraints at 95\% CL are also placed on the magnitudes of the
ratios $|\vr/\vl| < \vrtwosigma$ and $|\gl/\vl| < \gltwosigma$, and the polarisation of single top quarks in the \tch\ is constrained to be 
$P > \Ptwosigma$ (95\% CL). None of the above measurements make assumptions about the value of any of the 
other parameters or couplings and all of them are in agreement with
the SM expectations.

\section*{Acknowledgements}
\input{atlaslatex/acknowledgements/Acknowledgements}

\printbibliography

\newpage
\input{atlas_authlist}

\end{document}

%% file: paper-metadata.tex
\AtlasTitle{Analysis of the \boldmath{\Wtb}
  vertex from the measurement of triple-differential angular decay
  rates of single top quarks produced in the \boldmath{\tch} at \cmenergy\ with the ATLAS detector}

\author{The ATLAS Collaboration}

\PreprintIdNumber{CERN-EP-2017-089}

\AtlasJournal{JHEP 12 (2017) 017}
\AtlasDOI{10.1007/JHEP12(2017)017}

\AtlasAbstract{The electroweak production and subsequent decay of single top quarks in the \tch\
  is determined by the properties of the \Wtb\ vertex, which can be described by the complex parameters of an effective Lagrangian. 
  An analysis of a triple-differential decay rate in \tch\ production is used to
  simultaneously determine
  five generalised helicity fractions and phases, 
  as well as the polarisation of the produced top quark. The complex
  parameters are then constrained.
  This analysis is based on \lumiInInvFb\ of proton--proton
  collision data at a centre-of-mass energy of \lhcenergy\ collected with the ATLAS 
  detector at the LHC. 
  The fraction of decays containing transversely polarised $W$ bosons is 
  measured to be  $\fu=\cvfutotaluncNoparenthesisStandalone$.
  The phase between amplitudes for transversely and longitudinally
  polarised $W$ bosons recoiling against left-handed $b$-quarks is measured to be 
  $\delm = \cvdelmtotaluncNoparenthesis$, giving no indication of CP violation. 
  The fractions of longitudinal or transverse $W$ bosons accompanied by right-handed 
  $b$-quarks are also constrained.
  Based on these measurements, limits are placed at 95\% CL on the ratio of the 
  complex coupling parameters
  $\Re{\gr/\vl} \in \regrvllimits$ and $\Im{\gr/\vl} \in
  \imgrvllimits$. Constraints are also placed on the
  ratios $|\vr/\vl|$ and $|\gl/\vl|$. In addition, the polarisation of
  single top quarks in the \tch\ is constrained to be $P > \Ptwosigma$ (95\% CL). 
  None of the above measurements make assumptions about the value of any of the 
  other parameters or couplings and all of them are in agreement with the Standard
  Model.}

%% file: tables/EventYield_Merged_2-jetbin_CustomGroupList_combined_nominal_AllRegions_rounded.tex
\begin{tabular}{l | D{,}{{}\pm{}}{-1} | D{,}{{}\pm{}}{-1} | D{,}{{}\pm{}}{-1} | D{,}{{}\pm{}}{-1}} 
\hline
\hline
\multicolumn{1}{l|}{} & \multicolumn{1}{c|}{\multirow{2}{*}{Signal region}} & \multicolumn{1}{c|}{\multirow{2}{*}{\ttbar\ control region}} & \multicolumn{1}{c|}{\multirow{2}{*}{$W$+jets control region}} & \multicolumn{1}{c}{\multirow{2}{*}{$W$+jets validation region}} \\
\multicolumn{1}{l|}{Process} & \multicolumn{1}{c|}{} & \multicolumn{1}{c|}{} & \multicolumn{1}{c|}{} & \multicolumn{1}{c}{}\\
\hline
\textit{t}-channel & 4395,17 & 1688,12 & 11601,29 & 9306,27 \\ 
\textit{t$\bar{t}$}, \textit{Wt}, \textit{s}-channel & 2017,15 & 62864,77 & 48120,82 & 23937,61 \\ 
\textit{W}+heavy-jets & 1910,49 & 6898,65 & 45410,200 & 157260,480 \\ 
\textit{W}+light-jets & 87,31 & 218,38 & 3110,200 & 130900,1000 \\ 
\textit{Z}+jets, diboson & 157,7 & 1118,37 & 4734,77 & 17750,300 \\ 
Multijet & 375,13 & 862,27 & 8910,61 & 20140,120 \\ 
\hline
Total expected & 8941,64 & 73650,120 & 121890,310 & 359300,1200 \\ 
\hline
Data & \multicolumn{1}{c|}{8939} & \multicolumn{1}{c|}{73662} & \multicolumn{1}{c|}{121913} & \multicolumn{1}{c}{359320} \\ 
\hline
S/B & \multicolumn{1}{c|}{0.97} & \multicolumn{1}{c|}{0.02} & \multicolumn{1}{c|}{0.11} & \multicolumn{1}{c}{0.03} \\ 
\hline
\hline
\end{tabular}

%% file: tables/SystTable-PhysicsParams_and_CouplingRatios.tex
\begin{tabular}{lcccc}
  \hline \hline
  & \multicolumn{2}{|c}{Helicity parameters} & \multicolumn{2}{|c}{Coupling ratios} \\
  \cline{2-5}
  Source & \multicolumn{1}{|c}{$\sigma(\fu)$} & \multicolumn{1}{|c}{$\sigma(\delm)/\pi$} & \multicolumn{1}{|c}{$\sigma(\Re{\gr/\vl})$} & \multicolumn{1}{|c}{$\sigma(\Im{\gr/\vl})$} \\
  \hline \hline
  Statistical & \multicolumn{1}{|r}{0.022} & \multicolumn{1}{|r}{0.013} & \multicolumn{1}{|r}{0.030} & \multicolumn{1}{|r}{0.027} \\
  \hline \hline
  Jets   &  \multicolumn{1}{|r}{0.029} & \multicolumn{1}{|r}{0.007} & \multicolumn{1}{|r}{0.039} & \multicolumn{1}{|r}{0.009} \\
  Leptons & \multicolumn{1}{|r}{0.014} & \multicolumn{1}{|r}{0.002} & \multicolumn{1}{|r}{0.017} & \multicolumn{1}{|r}{<0.001} \\
  \MET\ & \multicolumn{1}{|r}{<0.001} & \multicolumn{1}{|r}{<0.001} & \multicolumn{1}{|r}{<0.001} &  \multicolumn{1}{|r}{<0.001} \\
  \hline
  Generator & \multicolumn{1}{|r}{0.027} & \multicolumn{1}{|r}{0.006} & \multicolumn{1}{|r}{0.030} & \multicolumn{1}{|r}{0.010} \\
  Parton shower and hadronisation & \multicolumn{1}{|r}{0.004} & \multicolumn{1}{|r}{0.003} & \multicolumn{1}{|r}{<0.001} & \multicolumn{1}{|r}{0.003} \\
  PDF variations & \multicolumn{1}{|r}{0.008} & \multicolumn{1}{|r}{0.004} & \multicolumn{1}{|r}{<0.001} & \multicolumn{1}{|r}{<0.001} \\
  \hline
  Background normalisation & \multicolumn{1}{|r}{<0.001} & \multicolumn{1}{|r}{<0.001} & \multicolumn{1}{|r}{<0.001} & \multicolumn{1}{|r}{<0.001} \\
  \QCD\ normalisation & \multicolumn{1}{|r}{<0.001} & \multicolumn{1}{|r}{<0.001} & \multicolumn{1}{|r}{<0.001} & \multicolumn{1}{|r}{<0.001} \\
  $W$+jets shape & \multicolumn{1}{|r}{0.015} & \multicolumn{1}{|r}{0.005} & \multicolumn{1}{|r}{0.007} & \multicolumn{1}{|r}{0.009} \\
  Luminosity & \multicolumn{1}{|r}{<0.001} & \multicolumn{1}{|r}{<0.001} & \multicolumn{1}{|r}{<0.001} & \multicolumn{1}{|r}{<0.001} \\
  \hline
  MC sample sizes & \multicolumn{1}{|r}{0.009} & \multicolumn{1}{|r}{0.006} & \multicolumn{1}{|r}{<0.001} & \multicolumn{1}{|r}{0.013} \\
  Other & \multicolumn{1}{|r}{<0.001} & \multicolumn{1}{|r}{<0.001} & \multicolumn{1}{|r}{<0.001} & \multicolumn{1}{|r}{<0.001} \\
  \hline\hline
  Total systematic uncertainty & \multicolumn{1}{|r}{0.044} & \multicolumn{1}{|r}{0.010} & \multicolumn{1}{|r}{0.061} & \multicolumn{1}{|r}{0.017} \\
  \hline\hline
  \textbf{Total} & \multicolumn{1}{|r}{0.049} & \multicolumn{1}{|r}{0.017} & \multicolumn{1}{|r}{0.068} & \multicolumn{1}{|r}{0.032} \\
  \hline \hline
\end{tabular}

%% file: atlaslatex/acknowledgements/Acknowledgements.tex

We thank CERN for the very successful operation of the LHC, as well as the
support staff from our institutions without whom ATLAS could not be
operated efficiently.

We acknowledge the support of ANPCyT, Argentina; YerPhI, Armenia; ARC, Australia; BMWFW and FWF, Austria; ANAS, Azerbaijan; SSTC, Belarus; CNPq and FAPESP, Brazil; NSERC, NRC and CFI, Canada; CERN; CONICYT, Chile; CAS, MOST and NSFC, China; COLCIENCIAS, Colombia; MSMT CR, MPO CR and VSC CR, Czech Republic; DNRF and DNSRC, Denmark; IN2P3-CNRS, CEA-DSM/IRFU, France; SRNSF, Georgia; BMBF, HGF, and MPG, Germany; GSRT, Greece; RGC, Hong Kong SAR, China; ISF, I-CORE and Benoziyo Center, Israel; INFN, Italy; MEXT and JSPS, Japan; CNRST, Morocco; NWO, Netherlands; RCN, Norway; MNiSW and NCN, Poland; FCT, Portugal; MNE/IFA, Romania; MES of Russia and NRC KI, Russian Federation; JINR; MESTD, Serbia; MSSR, Slovakia; ARRS and MIZ\v{S}, Slovenia; DST/NRF, South Africa; MINECO, Spain; SRC and Wallenberg Foundation, Sweden; SERI, SNSF and Cantons of Bern and Geneva, Switzerland; MOST, Taiwan; TAEK, Turkey; STFC, United Kingdom; DOE and NSF, United States of America. In addition, individual groups and members have received support from BCKDF, the Canada Council, CANARIE, CRC, Compute Canada, FQRNT, and the Ontario Innovation Trust, Canada; EPLANET, ERC, ERDF, FP7, Horizon 2020 and Marie Sk{\l}odowska-Curie Actions, European Union; Investissements d'Avenir Labex and Idex, ANR, R{\'e}gion Auvergne and Fondation Partager le Savoir, France; DFG and AvH Foundation, Germany; Herakleitos, Thales and Aristeia programmes co-financed by EU-ESF and the Greek NSRF; BSF, GIF and Minerva, Israel; BRF, Norway; CERCA Programme Generalitat de Catalunya, Generalitat Valenciana, Spain; the Royal Society and Leverhulme Trust, United Kingdom.

The crucial computing support from all WLCG partners is acknowledged gratefully, in particular from CERN, the ATLAS Tier-1 facilities at TRIUMF (Canada), NDGF (Denmark, Norway, Sweden), CC-IN2P3 (France), KIT/GridKA (Germany), INFN-CNAF (Italy), NL-T1 (Netherlands), PIC (Spain), ASGC (Taiwan), RAL (UK) and BNL (USA), the Tier-2 facilities worldwide and large non-WLCG resource providers. Major contributors of computing resources are listed in Ref.~\cite{ATL-GEN-PUB-2016-002}.

%% file: atlas_authlist.tex
\begin{flushleft}
{\Large The ATLAS Collaboration}

\bigskip

M.~Aaboud$^\textrm{\scriptsize 137d}$,
G.~Aad$^\textrm{\scriptsize 88}$,
B.~Abbott$^\textrm{\scriptsize 115}$,
J.~Abdallah$^\textrm{\scriptsize 8}$,
O.~Abdinov$^\textrm{\scriptsize 12}$$^{,*}$,
B.~Abeloos$^\textrm{\scriptsize 119}$,
S.H.~Abidi$^\textrm{\scriptsize 161}$,
O.S.~AbouZeid$^\textrm{\scriptsize 139}$,
N.L.~Abraham$^\textrm{\scriptsize 151}$,
H.~Abramowicz$^\textrm{\scriptsize 155}$
% ...
\bigskip
\\
% ...
$^{8}$ Department of Physics, The University of Texas at Arlington, Arlington TX, United States of America\\
% ...
$^{12}$ Institute of Physics, Azerbaijan Academy of Sciences, Baku, Azerbaijan\\
% ...
$^{88}$ CPPM, Aix-Marseille Universit{\'e} and CNRS/IN2P3, Marseille, France\\
% ...
$^{115}$ Homer L. Dodge Department of Physics and Astronomy, University of Oklahoma, Norman OK, United States of America\\
% ...
$^{119}$ LAL, Univ. Paris-Sud, CNRS/IN2P3, Universit{\'e} Paris-Saclay, Orsay, France\\
% ...
$^{137}$ $^{(a)}$ Facult{\'e} des Sciences Ain Chock, R{\'e}seau Universitaire de Physique des Hautes Energies - Universit{\'e} Hassan II, Casablanca; $^{(b)}$ Centre National de l'Energie des Sciences Techniques Nucleaires, Rabat; $^{(c)}$ Facult{\'e} des Sciences Semlalia, Universit{\'e} Cadi Ayyad, LPHEA-Marrakech; $^{(d)}$ Facult{\'e} des Sciences, Universit{\'e} Mohamed Premier and LPTPM, Oujda; $^{(e)}$ Facult{\'e} des sciences, Universit{\'e} Mohammed V, Rabat, Morocco\\
% ...
$^{139}$ Santa Cruz Institute for Particle Physics, University of California Santa Cruz, Santa Cruz CA, United States of America\\
% ...
$^{151}$ Department of Physics and Astronomy, University of Sussex, Brighton, United Kingdom\\
% ...
$^{155}$ Raymond and Beverly Sackler School of Physics and Astronomy, Tel Aviv University, Tel Aviv, Israel\\
% ...
$^{161}$ Department of Physics, University of Toronto, Toronto ON, Canada\\
% ...
$^{a}$ Also at Department of Physics, King's College London, London, United Kingdom\\
$^{b}$ Also at Institute of Physics, Azerbaijan Academy of Sciences, Baku, Azerbaijan\\
$^{c}$ Also at Novosibirsk State University, Novosibirsk, Russia\\
$^{d}$ Also at TRIUMF, Vancouver BC, Canada\\
$^{e}$ Also at Department of Physics {\&} Astronomy, University of Louisville, Louisville, KY, United States of America\\
% ...
$^{*}$ Deceased
\end{flushleft}
